\documentclass{aa}  

\usepackage{graphicx}
\usepackage{txfonts}
\def\gapprox{\;\rlap{\lower 2.5pt            
 \hbox{$\sim$}}\raise 1.5pt\hbox{$>$}\;}       
\def\lapprox{\;\rlap{\lower 2.5pt            
 \hbox{$\sim$}}\raise 1.5pt\hbox{$<$}\;} 
\begin{document}
   \title{Supersonic turbulence in 3D isothermal flow collision}

   \subtitle{}

   \author{Doris Folini\inst{1} \and  Rolf Walder\inst{1} \and Jean M. Favre\inst{2}}

   \institute{\'{E}cole Normale Sup\'{e}rieure, Lyon, CRAL, UMR CNRS 5574, 
           Universit\'{e} de Lyon, France\\
           \email{doris.folini@ens-lyon.fr}
           \and
           Swiss National Supercomputing Center, CSCS Lugano, Switzerland;
             }

   \date{Received ... ; accepted ...}

  \abstract
  {Large scale supersonic bulk flows are present in a wide range of
    astrophysical objects, from O-star winds to molecular clouds,
    galactic sheets, accretion, or $\gamma$-ray bursts. Associated
    flow collisions shape observable properties and internal physics
    alike. Our goal is to shed light on the interplay between large scale
    aspects of such collision zones and the characteristics of the
    compressible turbulence they harbor. Our model setup is as simple
    as can be: 3D hydrodynamical simulations of two head-on colliding,
    isothermal, and homogeneous flows with identical upstream
    (subscript $\mathrm{u}$) flow parameters and Mach numbers $2 <
    M_{\mathrm{u}} < 43$.\newline The turbulence in the collision zone
    is driven by the upstream flows, whose kinetic energy is partly
    dissipated and spatially modulated by the shocks confining the
    zone. Numerical results are in line with expectations from
    self-similarity arguments. The spatial scale of modulation grows
    with the collision zone. The fraction of energy dissipated at the
    confining shocks decreases with increasing $M_{\mathrm{u}}$. The
    mean density is $\rho_{\mathrm{m}} \approx 20 \rho_{\mathrm{u}}$,
    independent of $M_{\mathrm{u}}$. The root mean square Mach number
    is $M_{\mathrm{rms}} \approx 0.25 M_{\mathrm{u}}$.  Deviations
    toward weaker turbulence are found as the collision zone thickens
    and for small $M_{\mathrm{u}}$. The density probability function
    is not log-normal.\newline The turbulence is inhomogeneous, weaker
    in the center of the zone than close to the confining shocks.  It
    is also anisotropic: transverse to the upstream flows
    $M_{\mathrm{rms}}$ is always subsonic. We argue that uniform,
      head-on colliding flows generally disfavor turbulence that is
    at the same time isothermal, supersonic, and isotropic. The
    anisotropy carries over to other quantities like the density
    variance - Mach number relation. Line-of-sight effects thus exist.
    Structure functions differ depending on whether they are computed
    along a line-of-sight perpendicular or parallel to the upstream
    flow. Turbulence characteristics generally deviate markedly from
    those found for uniformly driven, supersonic, isothermal
    turbulence in 3D periodic box simulations. We suggest that this
    should be kept in mind when interpreting turbulence
    characteristics derived from observations. Our
    simulations show that even a simple model setup results in a
    richly structured interaction zone. The robustness of our findings
    toward more realistic setups remains to be tested.  }

   \maketitle
%

%
%
\section{Introduction}
\label{sec:intro}
Shock-bound interaction zones are ubiquitous in astrophysics. They
occur, for example, in binary star systems~\citep{stevens-et-al:92,
  myasnikov-zhekov:98, parkin-pittard:10}, the jets of high-energy
objects~\citep{panaitescu-et-al:99, kaiser-et-al:00, fan-wei:04}, or
galaxy formation and cosmology~\citep{anninos-norman:96,
  kang-et-al:05, aegertz-et-al:09, klar-muecket:12}. The idea that
such zones may also play a crucial role in the context of molecular
clouds and star formation stimulated extensive research in recent
years, on their large scale properties as well as on the turbulence they
harbor. The present paper stresses the combined view, the interplay
between large scale aspects of the collision zone and the characteristics
of the turbulence in its interior.

Substantial progress in understanding supersonic turbulence and its
characteristics has come from 3D periodic box simulations. These
simulations meanwhile cover an impressive range of physics from pure
isothermal hydrodynamics to the inclusion of magnetic fields,
radiative cooling, self-gravity, chemistry, or relativistic
flows~\citep[][]{stone-ostriker-gammie:98,
  maclow-klessen-burkert:98, maclow:99, padoan-nordlund:99,
  boldyrev-et-al1:02, boldyrev-et-al2:02, kritsuk-norman:04,
  padoan-et-al:04, kritsuk-et-al:06, gazol-et-al:07, kritsuk-et-al:07,
  padoan-et-al:07, federrath-et-al:08, schmidt-et-al:09,
  federrath-et-al:10, glover-et-al:10, gazol-kim:10, price-et-al:11,
  seifried-et-al:11, kritsuk-et-al:11, kritsuk-et-al2:11,
  konstandin-et-al:12, molina-et-al:12, downs:12, gazol-kim:13,
  federrath-klessen:13}. Turbulence in such simulations is either
left to decay or is forced.  Typical ways of forcing are continuous
energy input in each grid cell or occasional energy input at randomly
selected grid cells, to mimic energy input into the interstellar
medium by, for example, collimated or spherical outflows or supernova
explosions~\citep{li-nakamura:06, de-avillez-bretschwerdt:07,
  wang-et-al:10, moraghan-et-al:13}. Only rather recently have
analytical results been presented that demonstrate the existence of an
intermediate scaling range and derive scaling relations for
compressible turbulence\citep[][]{aluie:11, galtier-banerjee:11,
  banerjee-galtier:13}.

For isothermal hydrodynamics, on which the present paper concentrates,
3D periodic box studies reveal the crucial role of the form of the energy
input for the characteristics of the turbulence. The wave length at
which the energy is put into the system sets the largest spatial scale
of the turbulent structures~\citep{maclow:99, ballesteros-maclow:02}.
Purely compressible forcing leads to much sharper density contrasts
and wider density PDFs than purely solenoidal
forcing~\citep{federrath-et-al:09, federrath-et-al:10}. Likewise,
stronger density contrasts result if more energy is put into the
system~\citep{maclow:99}.

Less explored is the turbulence harbored within flow collision zones
and its interplay with the overall properties of such zones.  It has
long been known that isothermal head-on colliding flows are
unstable~\citep[e.g.,][]{vishniac:94, blondin-marks:96}. The role of
additional physics, like radiative cooling, self-gravity, or the
inclusion of magnetic fields, has been investigated since by a number
of studies~\citep{walder-folini:98, hennebelle-perault:99,
  walder-folini:00, audit-hennebelle:05, folini-walder:06,
  vazquez-semadeni-et-al:06, vazquez-semadeni-et-al:07,
  hennebelle-audit:07, hennebelle-et-al:08, inoue-inutsuka:09,
  heitsch-et-al2:09, folini-walder-favre:10,
  vazquez-semadeni-et-al:10, klessen-hennebelle:10,
  ntormousi-et-al:11, gong-ostriker:11, zrake-macfadyen:11,
  inoue-inutsuka:12, zrake-macfadyen-2:12, zrake-macfadyen:12,
  inoue-fukui:13}. Of interest in the context of the present paper is
the observation that the very early evolution of the collision zone
can be much more violent ~\citep{heitsch-et-al:06,
  vazquez-semadeni-et-al:06} than the situation later on, when a still
unstable but less violent, statistically stationary situation
develops~\citep{walder-folini:00, folini-walder:06,
  audit-hennebelle:10}. It is this later stage of the evolution we are
interested in, and there in particular in the characteristics of the
supersonically turbulent interior of the collision zone and its
interplay with the confining shocks, which modulate the energy input
that drives the turbulence.

The paper is a follow up of~\citet{folini-walder:06} (FW06), a 2D
study of head-on colliding isothermal flows. FW06 pointed out the
coupling between the turbulence within the collision zone and the
confining shocks of this zone. For the upstream Mach numbers
considered ($5 < M_{\mathrm{u}} < 90$), the fraction of upstream
kinetic energy that passes the confining shocks without getting
dissipated (thus is available for driving the turbulence in the
collision zone) scales with the root mean square Mach number of the
turbulence. In addition, FW06 derived approximate self-similarity
relations for collision zone mean quantities and checked them against
numerical results.

Here we extend this earlier study by considering isothermal head-on
colliding flows in 3D instead of 2D and by investigating further
turbulence characteristics. In particular, the following questions
shall be addressed. What level of turbulence, what root mean square
Mach number $M_{\mathrm{rms}}$, can be reached within the interaction
zone for a given upstream Mach number $M_{\mathrm{u}}$?  How
homogeneous and isotropic is the turbulent interior of the interaction
zone? What do established characteristics of supersonic turbulence
look like, in particular the density PDF, the density variance-Mach
number relation, and the structure functions? Finally, we contemplate on
potential implications of our results for real astrophysical objects,
for their observation, and on how additional physics may alter the
presented results.

We stress that the setup studied here is highly idealized. In reality,
there exist no strictly isothermal, uniform, or precisely head-on
colliding flows of equal Mach number and density.  The study
highlights, however, the wealth of phenomena that can result already
from such a setup - and, likewise, indicates characteristics beyond
reach. And it offers a somewhat complementary, thus potentially
interesting, view on supersonic turbulence, as compared to the more
wide spread (and often equally idealized) 3D periodic box perspective.

The structure of the paper is as follows. In
Sect.~\ref{sec:runs_and_tools} we present the numerical method and the
physical simulation setup. Results follow in Sect.~\ref{sec:results}
and are discussed in Sect.~\ref{sec:discussion}. Summary and
conclusions are given in Sect.~\ref{sec:conc}
\section{Physical model and numerical method}
\label{sec:runs_and_tools}
The physical model and numerical method used are the same as
in~FW06 but extended now to 3D. We refer to this paper for a
detailed description and restrict ourselves here to the most
important facts and information specific to the 3D
simulations.
\subsection{Physical model problem}
\label{sec:phys_model}
We consider a 3D, plane-parallel, infinitely extended (yz-direction),
isothermal, shock compressed slab. Two high Mach-number flows,
oriented parallel (left flow, subscript $l$) and anti-parallel (right
flow, subscript $r$) to the x-direction, collide head-on. The
resulting high-density interaction zone we denote by CDL for
`cold dense layer' to remain consistent with notation used in previous
papers~\citep{walder-folini:96,walder-folini:98, folini-walder:06}. We
investigate this system within the frame of Euler equations, together
with a polytropic equation of state. For the polytropic exponent, we
choose $\gamma = 1.000001$.  This value guarantees that jump
conditions and wave speeds of a Mach-90 shock are within 0.01 per cent
of the isothermal values.

We only consider symmetric settings, where the left and right
colliding flow have identical density and velocity (subscript $u$ for
upstream): $\rho_{\mathrm{l}} = \rho_{\mathrm{r}} \equiv
\rho_{\mathrm{u}}$ and $|v_{\mathrm{l}}| = |v_{\mathrm{r}}| \equiv
v_{\mathrm{u}}$.

We express velocities in units of the isothermal sound speed $a$,
densities in terms of the upstream density $\rho_{\mathrm{u}}$, and
lengths in units of $\mathrm{Y}$, the y-extent of the computational
domain we used. This artificial choice is necessary as there is no
natural time-independent length scale to the problem (see
Sect.~\ref{sec:results}). The y-extent of the computational domain is
the same in all our simulations and is identical with the z-extent,
so $\mathrm{Y} = \mathrm{Z}$.
\subsection{Numerical method}
\label{sec:num_meth}
Our results are computed with the hydrodynamic code from the A-MAZE
code-package~\citep{amaze:00, folini-et-al:03, melzani-et-al:13}. We
use the multidimensional high-resolution finite-volume-integration
scheme developed by~\citet{colella:90} on the basis of a Cartesian
mesh.  In all our simulations we use a version of the scheme that is
(formally) second order accurate in space and time for smooth flows.
We combine this integration scheme with the adaptive mesh algorithm
by~\citet{berger:85}. A coarse mesh is used for the upwind flows, a
finer mesh for the CDL. The meshes adapt automatically to the
increasing spatial extension of the CDL.

As described in~FW06, we have our CDL moving in x-direction
at Mach 20-40 to avoid alignment effects of strong
shocks, a well known problem not particular to our
scheme~\citep{colella-woodward:84, quirk:94, jasak-weller:95}. We rely
on the MILES approach to mimic the high-wavenumber end of the inertial
subrange~\citep[e.g.,][]{boris-et-al:92, porter-woodward:94}, but see
also~\citet{garnier-et-al:99} and~\citet{domaradzki:10} for a
discussion of limitations.
\subsection{Numerical settings of simulations}
\label{sec:num_settings}
The y- and z-extent of the computational domain (directions
perpendicular to the upstream flows) are identical in all our
simulations, $\mathrm{Y} = \mathrm{Z}$. The x-extent is 200 times
larger. Boundary conditions in x-direction are `supersonic inflow'.
Periodic boundary conditions are used in y- and z-direction.

The runs performed differ in their upstream Mach-number
$M_{\mathrm{u}}$, with $2 \lapprox M_{\mathrm{u}} \lapprox 45 $, and
in their discretization, with 128 or 256 cells in $y-$ and
$z-$direction on the finest level of refinement, which only covers the
CDL. Runs are labled with M\_R, where M is the upwind
Mach-number and R indicates the refinement: 128 cells (R=1) or 256
cells (R=2). The runs are listed in Table~\ref{tab:list_of_runs}.

At $t=0$, no CDL is present in our simulations. Random density
perturbations of up to 2\% are added initially to the flow left of the
interface separating both flows, up to a distance $\mathrm{Y}/2$
from the interface. This initialization results in a fast development
of turbulence in the CDL without imposing any artificial mode.
\begin{table*}[ht]
  \caption{List of performed simulations. Individual columns denote: the 
    upstream Mach number ($M_{\mathrm{u}}$); the size of the CDL at the end of 
    each simulation in scaled units 
    ($\ell_{\mathrm{end}}$, see Sect.~\ref{sec:data_anal}) and with respect 
    to the transverse domain size $Y$ ($\ell^{\mathrm{end}}_{\mathrm{cdl}} / \mathrm{Y}$); 
    the average shock area (left and right shock) with respect 
    to the transverse domain size ($\mathrm{s}_{\mathrm{sh}} / Y^{2}$); the driving 
    efficiency or fraction of upstream kinetic energy density that passes the confining shocks 
    of the CDL unthermalized ($f_{\mathrm{eff}}$, Eq.~\ref{eq:exp_feff}); the 
    ratio of the CDL mean density to the upstream density 
    ($\rho_{\mathrm{m}} / \rho_{\mathrm{u}}$, Eq.~\ref{eq:exp_rho} 
    and Fig.~\ref{fig:dens_mach}); the root mean square Mach number
    of the CDL, as well as its transverse (y-z-direction) and parallel (x-direction)
    components, and the ratio of the parallel and transverse component 
    ($M_{\mathrm{rms}}$, $M_{\mathrm{rms,\perp}}$, $M_{\mathrm{rms,\parallel}}$, 
    $M_{\mathrm{rms,\parallel}} / M_{\mathrm{rms,\perp}}$, Eq.~\ref{eq:exp_mach} 
    and Fig.~\ref{fig:dens_mach}); the density variance - Mach number parameter 
    $b = \sigma(\rho) / (\rho_{\mathrm{m}} M_{\mathrm{rms}})$, 
    as well as its transverse (y-z-direction) and
    parallel (x-direction) components 
    ($b$, $b_{\mathrm{\perp}}$, $b_{\mathrm{\parallel}}$, Eq.~\ref{eq:dens-machnumber});
    the CDL density variance $\sigma(\rho)$ normalized by the CDL mean density 
    ($\sigma(\rho) / \rho_{\mathrm{m}}$).
    All columns except $\ell_{\mathrm{end}}$  and $\ell^{\mathrm{end}}_{\mathrm{cdl}} / \mathrm{Y}$ contain 
    averages over $\ell \in [11,12]$. Values in parentheses in column
    $f_{\mathrm{eff}}$ are maximum values reached (see Fig.~\ref{fig:feff}).}
\begin{center}
\begin{tabular}{lcccccccccccccc} 
\hline
\hline 
 label                                                              & 
 $M_{\mathrm{u}}$                                                   & 
 $\ell_{\mathrm{end}}$                                              & 
 $\frac{\ell^{\mathrm{end}}_{\mathrm{cdl}}}{\mathrm{Y}}$            & 
 $\frac{\mathrm{s}_{\mathrm{sh}}}{Y^{2}}$                           & 
 $f_{\mathrm{eff}}$                                                 &
 $\frac{\rho_{\mathrm{m}}}{\rho_{\mathrm{u}}}$                      & 
 $M_{\mathrm{rms}}$                                                 & 
 $M_{\mathrm{rms,\perp}}$                                           & 
 $M_{\mathrm{rms,\parallel}}$                                       & 
 $\frac{M_{\mathrm{rms,\parallel}}}{M_{\mathrm{rms,\perp}}}$        &  
 $b$                                                                &  
 $b_{\mathrm{\perp}}$                                               &  
 $b_{\mathrm{\parallel}}$                                           &
 $\frac{\sigma(\rho)}{\rho_{\mathrm{m}}}$                           \\
\hline 
R2\_2  &  2.71 & 13 & 1.47 & 1.1 & 0.09 (0.15) &  9 &  0.33 & 0.17 & 0.28 & 1.63 & 0.34 & 0.65 & 0.40 & 0.11 \\ 
R4\_2  &  4.06 & 13 & 0.88 & 1.1 & 0.20 (0.22) & 15 &  0.68 & 0.34 & 0.58 & 1.72 & 0.32 & 0.64 & 0.37 & 0.26 \\ 
R5\_2  &  5.42 & 18 & 0.81 & 1.2 & 0.27 (0.32) & 21 &  0.99 & 0.47 & 0.87 & 1.85 & 0.30 & 0.63 & 0.34 & 0.29 \\ 
R7\_2  &  6.78 & 14 & 0.57 & 1.4 & 0.39 (0.45) & 23 &  1.4  & 0.60 & 1.24 & 2.08 & 0.28 & 0.64 & 0.31 & 0.38 \\ 
R8\_2  &  8.15 & 17 & 0.65 & 1.6 & 0.53 (0.56) & 23 &  1.8  & 0.69 & 1.66 & 2.42 & 0.26 & 0.69 & 0.28 & 0.47 \\ 
R11\_2 & 10.9  & 24 & 0.88 & 2.1 & 0.65 (0.67) & 24 &  2.5  & 0.77 & 2.35 & 3.05 & 0.23 & 0.73 & 0.24 & 0.56 \\ 
R16\_2 & 16.3  & 22 & 0.94 & 3.1 & 0.80 (0.82) & 22 &  4.0  & 0.80 & 3.89 & 4.89 & 0.16 & 0.80 & 0.16 & 0.63 \\ 
R22\_2 & 21.7  & 34 & 1.46 & 3.9 & 0.84 (0.86) & 21 &  5.4  & 0.81 & 5.37 & 6.64 & 0.12 & 0.81 & 0.12 & 0.65 \\ 
R27\_2 & 27.1  & 15 & 0.73 & 4.7 & 0.87 (0.89) & 20 &  7.0  & 0.77 & 6.94 & 9.01 & 0.09 & 0.82 & 0.09 & 0.63 \\ 
R33\_2 & 32.4  & 36 & 1.72 & 5.3 & 0.88 (0.92) & 20 &  8.5  & 0.75 & 8.47 & 11.3 & 0.07 & 0.84 & 0.07 & 0.63 \\ 
R43\_2 & 43.4  & 37 & 1.88 & 6.0 & 0.88 (0.93) & 20 & 11.5  & 0.72 & 11.5 & 16.1 & 0.05 & 0.85 & 0.05 & 0.61 \\ 
\hline 
R11\_1 & 10.9  & 23 & 0.95 & 2.3 & 0.71 (0.72) & 20 &  2.7  & 0.59 & 2.63 & 4.44 & 0.18 & 0.83 & 0.19 & 0.49 \\ 
R22\_1 & 21.7  & 18 & 0.92 & 3.6 & 0.82 (0.85) & 19 &  5.8  & 0.59 & 5.78 & 9.75 & 0.09 & 0.85 & 0.09 & 0.51 \\ 
R33\_1 & 32.4  & 17 & 0.86 & 4.4 & 0.83 (0.85) & 19 &  9.1  & 0.59 & 9.07 & 15.3 & 0.06 & 0.85 & 0.06 & 0.50 \\ 
R43\_1 & 43.4  & 16 & 0.82 & 4.9 & 0.84 (0.86) & 20 & 12.4  & 0.59 & 12.4 & 21.0 & 0.04 & 0.87 & 0.04 & 0.52 \\ 
\hline
\end{tabular}
\end{center}
\label{tab:list_of_runs}
\end{table*}
\begin{figure*}[tp]
\centerline{
\includegraphics[height=8.7cm]{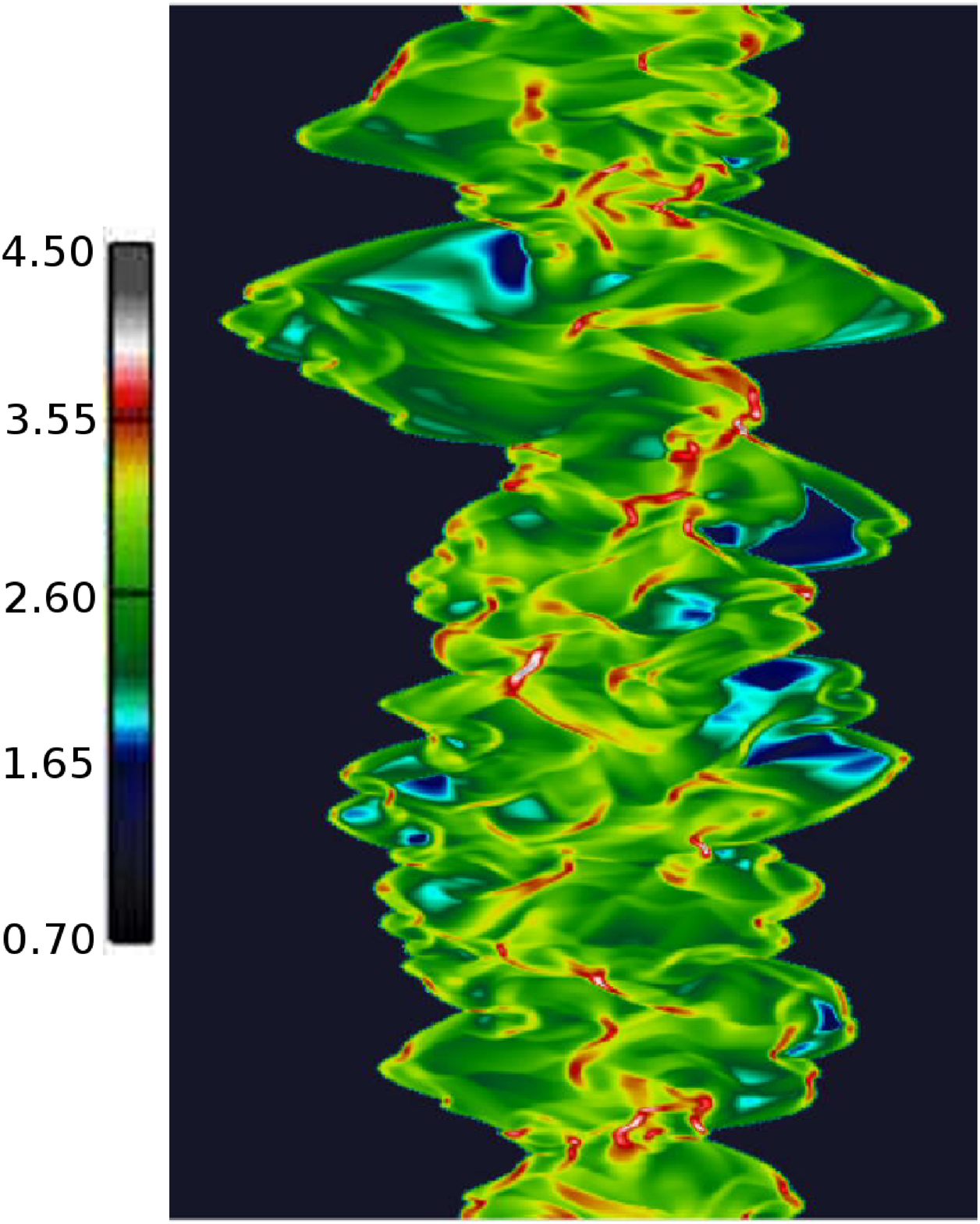}
\hspace{.5cm}
\includegraphics[height=8.4cm]{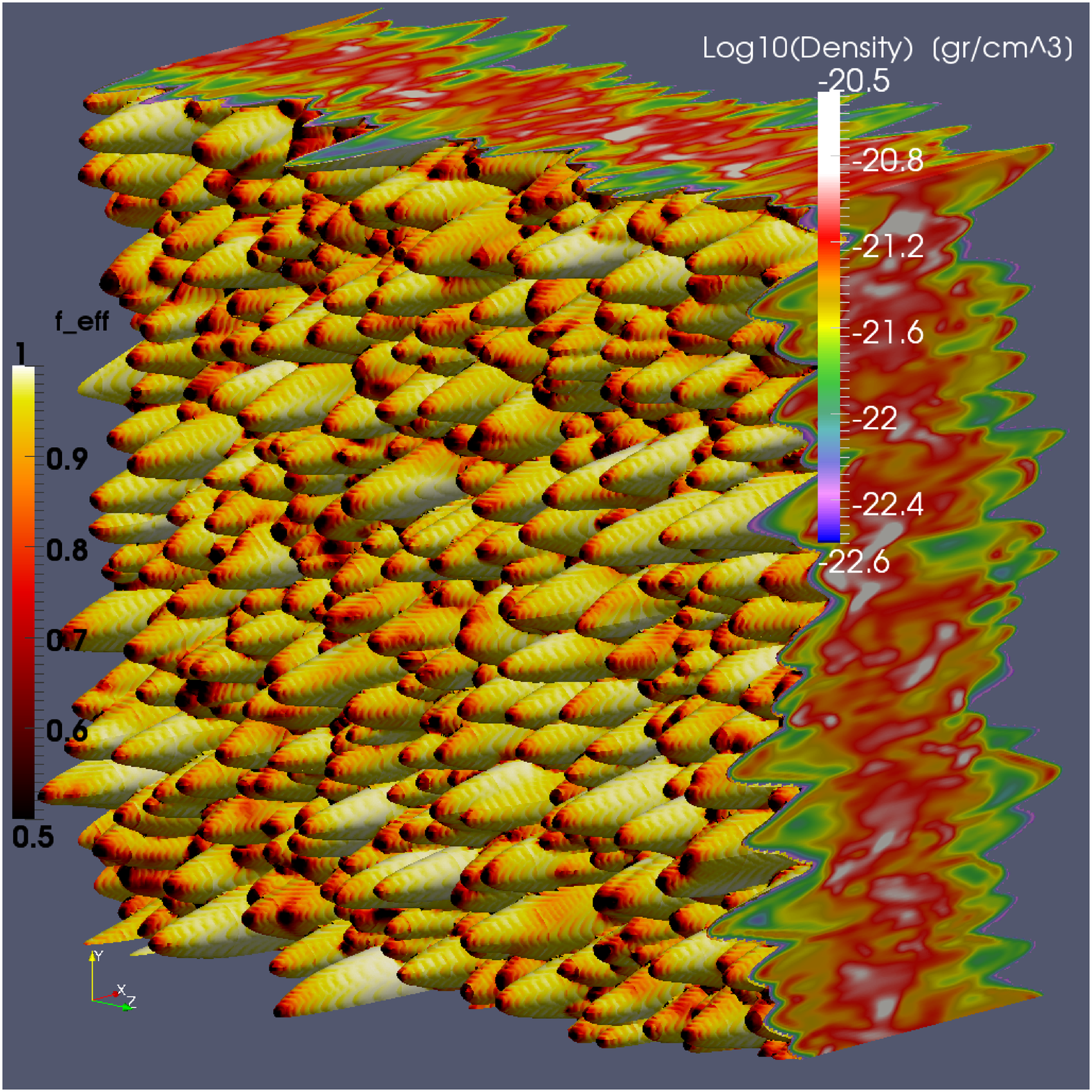}
}
\caption{Different geometry of confining shocks. In 2D (periodic in
  y-direction, infinite in z-direction), the confining shocks
  resemble corrugated sheets ({\bf left panel}). In 3D
  (periodic in y- and z-direction) they evoke the idea of
  cardboard egg wrappings ({\bf right panel}). Both simulations have
  $M_{\mathrm{u}}=21.7$. Color coded is density (log10($\rho$);
particles/cm$^{3}$ in 2D; g/cm$^{3}$ in 3D) and, right panel only,
the driving efficiency $f_{\mathrm{eff}}$ at the shock
surface.}
\label{fig:eggwrapping}
\end{figure*}
\subsection{Data analysis}
\label{sec:data_anal}
The relevant quantity for the time evolution of CDL quantities is the
average x-extension of the CDL, $\ell_{\mathrm{cdl}} \equiv V /
(\mathrm{Y} \mathrm{Z})$ with $V$ the volume of the CDL (see
Sect.~\ref{sec:results} or the 2D case in~FW06).
The average CDL extension does, however, not need to be a strictly
monotonically increasing function of time, as the CDL compression
fluctuates slightly with time.

A proxy to $\ell_{\mathrm{cdl}}$ that does not show such fluctuations
can be defined through the monotonically increasing CDL mass.
Expressing $\ell_{\mathrm{cdl}}$ in terms of the CDL mass
$m_{\mathrm{cdl}}$ and mean density $\rho_{\mathrm{m}}$ as
$\ell_{\mathrm{cdl}} = m_{\mathrm{cdl}} /(\rho_{\mathrm{m}}
\mathrm{Y}^{2})$, where we used $\mathrm{Y} = \mathrm{Z}$, we can
replace $\rho_{\mathrm{m}}$ by $\rho_{\mathrm{u}}$ and normalize with
$\mathrm{Y}$ to obtain a monotonically increasing, dimensionless
quantity, $\ell \equiv m_{\mathrm{cdl}} /(\rho_{\mathrm{u}} \mathrm{Y}^{3})$.

All simulations were advanced till at least $\ell=12$, some much
further. Except for the lowest Mach number simulations, $\ell \approx 12$
corresponds to within 10\% to a spatial extent of $\mathrm{Y}/2$, i.e.,
the x-extension of the CDL is about half the domain size in
yz-direction. Data analysis is done off-line, on previously dumped data
sets. Temporal averages, unless otherwise stated, are taken
over $\ell \in [11,12]$. This interval comprises at least three data
sets.

Part of the data analysis was done with
VisIt\footnote{https://wci.llnl.gov/codes/visit/}, combined with
python scripting. We developed a reader for VisIt that can cope with
our block structure adaptive mesh refinement data, stored in hdf5
format.  VisIt proved particularly useful for the slice wise analysis
of the data (see Sect.~\ref{sec:2d_slices}) and for the numerical
determination of the driving efficiency. The latter involves local
decomposition of the upstream flow into components perpendicular and
parallel to the confining shocks and integration over these shocks
(see Appendix~\ref{app:feff}). We use VisIt to create an iso-surface
representative of the confining shock, defined by a density threshold
slightly greater than $\rho_{\mathrm{u}}$, and then exploit VisIts
capabilities of providing surface normals to iso-surfaces. An example
of the reconstructed shock surface, colored in driving efficiency
$f_{\mathrm{eff}}$ (see Eq.~\ref{eq:exp_feff}), is shown in
Fig.~\ref{fig:eggwrapping}, right panel.
\section{Results}
\label{sec:results}
The section is structured along two main perspectives: the view on the
CDL as an entity and the view on its interior structure. In
Sect.~\ref{sec:mean_quantities}, we concentrate on approximate scaling
relations for some CDL mean quantities in 1D to 3D. In
Sect.~\ref{sec:velo_aniso} we focus on the anisotropy and
inhomogeneity of the CDL interior. 2D slices are used to analyze how
mean quantities change with distance from the CDL center.
Sect.~\ref{sec:density_pdf} is dedicated to density PDFs,
Sect.~\ref{sec:struc_func} to structure functions. The physical
results are complemented by considerations on numerical
resolution in Sect.~\ref{sec:num_effects}.
\subsection{CDL Mean quantities: approximate scaling relations}
\label{sec:mean_quantities}
Within the frame of Euler equations, i.e., in the absence of viscous
dissipation, and making some further assumptions as detailed below,
approximate self-similar scaling relations for CDL mean quantities can
be derived and expressed in terms of upstream flow parameters. The
relations are qualitatively different in 1D as compared to 2D and 3D
(see FW06). In particular, in 1D the mean density increases with the
upstream Mach number squared. In 2D / 3D the CDL mean density is
expected to be limited, the root mean square Mach number should
increase linearly with the upstream Mach number.
\subsubsection{Analytical scaling relations for CDL mean quantities}
\label{sec:anal-scaling}
Looking first at the 1D case and denoting the density and velocity of
the CDL by $\rho_{\mathrm{1d}}$ and $v_{\mathrm{1d}}$, those of the
left and right upwind flows by $\rho_{\mathrm{i}}$ and
$v_{\mathrm{i}}$ ($i=l,r$), and the isothermal sound speed by $a$, it
can be shown that~\citep[e.g.,][]{courant-friedrichs} in the rest frame
of the CDL
\begin{eqnarray}
\label{eq:self_sim1} 
\rho_{\mathrm{1d}} / \rho_{\mathrm{i}} & = & M_{\mathrm{i}}^\mathrm{2} + 1 
                         \approx M_{\mathrm{i}}^2,\\
\label{eq:self_sim2}
v_{\mathrm{1d}} & = & 0.
\end{eqnarray}
The approximation holds for large Mach-numbers.  A relation between
characteristic length and time scales of the solution, the
self-similarity variable $\kappa_{\mathrm{1d}}$, can be obtained as
the ratio between the spatial extension $\ell_{\mathrm{1d}}$ of the
CDL and the time $\tau$ needed to accumulate the corresponding column
density ${ N }_{\mathrm{1d}}$:
\begin{equation}
 N_{\mathrm{1d}} = \rho_{\mathrm{1d}} \ell_{\mathrm{1d}},
\label{eq:mass_column}
\end{equation}
\begin{equation}
 N_{\mathrm{1d}} = \tau \left( \rho_l v_l + \rho_r v_r\right),
\label{eq:mass_cons}
\end{equation}
and using $\rho_\mathrm{l}/\rho_\mathrm{r} =
M^{2}_\mathrm{r}/M^{2}_\mathrm{l}$
\begin{equation}
\kappa_{\mathrm{1d}} \equiv \frac{\ell_{\mathrm{1d}}}{\tau}
  = a \frac{M_l + M_r}{M_l \cdot M_r}.
\label{eq:1d_kappa}
\end{equation}
For strong shocks and symmetric settings ($\mathrm{l} = \mathrm{r}$)
the above relations reduce to $\rho_{\mathrm{1d}} / \rho_{\mathrm{u}} = M_{\mathrm{u}}^{2}$
and $\kappa_{\mathrm{1d}} =2a/M_{\mathrm{u}}$.

For the 2D case, FW06 derived scaling relations
for five CDL mean quantities: mean density $\rho_{\mathrm{m}}$; root
mean square Mach number $M_{\mathrm{rms}}$; driving energy $\dot{\cal
  E}_{\mathrm{drv}}$, i.e., the kinetic energy density entering the CDL
per time and unit length in the y-direction; $\dot{\cal
  E}_{\mathrm{diss}}$, the energy density dissipated per time within an
average column of length $\ell_{\mathrm{cdl}}$; the change per time of
the average kinetic energy energy density contained within such an average column, $\dot{\cal
  E}_{\mathrm{kin}}$.

Extension to 3D is straightforward. The only derivation we repeat here
(see Appendix~\ref{app:feff}) is that of the driving energy $\dot{\cal
  E}_{\mathrm{drv}}$, which involves an integral over the confining
shocks.  Part of the total (left plus right flow) upwind kinetic
energy flux density, ${\cal F}_{\mathrm{e_{\mathrm{kin}},u}} =
\rho_{\mathrm{u}}v_{\mathrm{u}}^{3}$, is thermalized at these shocks.
The remaining part, $\dot{\cal E}_{\mathrm{drv}}$, drives the
turbulence in the CDL. In analogy with the 2D case, we write
$\dot{\cal E}_{\mathrm{drv}} = f_{\mathrm{eff}} {\cal
  F}_{\mathrm{e_{\mathrm{kin}},u}}$, where the driving efficiency
$f_{\mathrm{eff}}$ is a function of the upwind Mach-number alone.  The
formal dependencies, involving upstream quantities $ \rho_{\mathrm{u}}$
and $M_{\mathrm{u}}$, as well as parameters $\beta_{\mathrm{i}}$ and
$\eta_{\mathrm{i}}$ (to be determined), are as in the 2D case:
\begin{eqnarray}
\label{eq:exp_rho}
\rho_{\mathrm{m}}        & = & \eta_{\mathrm{1}} \rho_{\mathrm{u}} M_{\mathrm{u}}^{\beta_{\mathrm{1}}} 
                           =   \eta_{\mathrm{1}} \rho_{\mathrm{u}},\\
\label{eq:exp_mach}
M_{\mathrm{rms}}         & = & \eta_{\mathrm{2}} M_{\mathrm{u}}^{\beta_{\mathrm{2}}}
                           =   \eta_{\mathrm{1}}^{-1/2} M_{\mathrm{u}},\\
\label{eq:exp_kappa}
\kappa_{\mathrm{3d}}     & = & \ell_{\mathrm{cdl}}/\tau 
                           =   2 \eta_{\mathrm{1}}^{-1} a M_{\mathrm{u}},\\
\label{eq:exp_feff}
f_{\mathrm{eff}}         & = & (1 - \eta_{\mathrm{3}} M_{\mathrm{u}}^{\beta_{\mathrm{3}}}),\\
\label{eq:exp_edrv}
\dot{\cal E}_{\mathrm{drv}}  & = & \rho_{\mathrm{u}} a^{3} M_{\mathrm{u}}^{3} 
                              (1 - \eta_{\mathrm{3}} M_{\mathrm{u}}^{\beta_{\mathrm{3}}}),\\
\label{eq:exp_ekin}
\dot{\cal E}_{\mathrm{kin}}  & = & \rho_{\mathrm{u}} a^{3} M_{\mathrm{u}}^{3} 
                               \; \eta_{\mathrm{2}}^{2},\\
\label{eq:exp_ediss}
\dot{\cal E}_{\mathrm{diss}} & = & \rho_{\mathrm{u}} a^{3} M_{\mathrm{u}}^{3} 
                              (1 -\eta_{\mathrm{3}} M_{\mathrm{u}}^{\beta_{\mathrm{3}}} 
                                 - \eta_{\mathrm{2}}^{2}).
\end{eqnarray}
Here, $\beta_{\mathrm{1}} = 0$ and $\beta_{\mathrm{2}} = 1$ from
analytical considerations, whereas $\beta_{\mathrm{3}}$ as well as
$\eta_{\mathrm{i}}$, $i=1,2,3$, can be determined only from numerical
simulations and turn out to be different in 2D and 3D.  The 3D
numerical results, to be presented in Sect.~\ref{sec:num-scaling},
confirm $\beta_{\mathrm{1}} = 0$ and $\beta_{\mathrm{2}} = 1$, and
yield $\beta_{\mathrm{3}} \approx -1.17$, $\eta_{\mathrm{1}} \approx
25$, $\eta_{\mathrm{2}} \approx 0.2 $, and $\eta_{\mathrm{3}} \approx
5$ as best fit values.

The above relations imply that the mean density of the CDL does not
depend on the upstream Mach number $M_{\mathrm{u}}$, in strong
contrast to the 1D case where $\rho_{\mathrm{m}} \propto
M_{\mathrm{u}}^{2}$. The CDL root mean square Mach number should
increase linearly with $M_{\mathrm{u}}$. Also: the larger
$M_{\mathrm{u}}$, the larger $f_{\mathrm{eff}}$, i.e., the more energy
passes the confining shocks of the CDL unthermalized

We stress the four basic assumptions made in deriving the above
relations: a) a simple Mach-number dependency of $\rho_{\mathrm{m}}$,
$v_{\mathrm{rms}}$, and $f_{\mathrm{eff}}$; b) CDL density and
velocity are uncorrelated; c) we have high Mach-numbers in the sense
that $\eta_{\mathrm{2}}^{2} M_{\mathrm{u}}^{2 \beta_{\mathrm{2}}} >>
1$ or $M_{\mathrm{rms}}^{2} >> 1$; d) $v_{\mathrm{diss}} \propto
v_{\mathrm{rms}}$. For a discussion of the validity of these
assumptions we refer to FW06.
\begin{figure}[tbp]
\vspace{-0.6cm}
\centerline{\includegraphics[width=9.0cm,height=5.5cm]{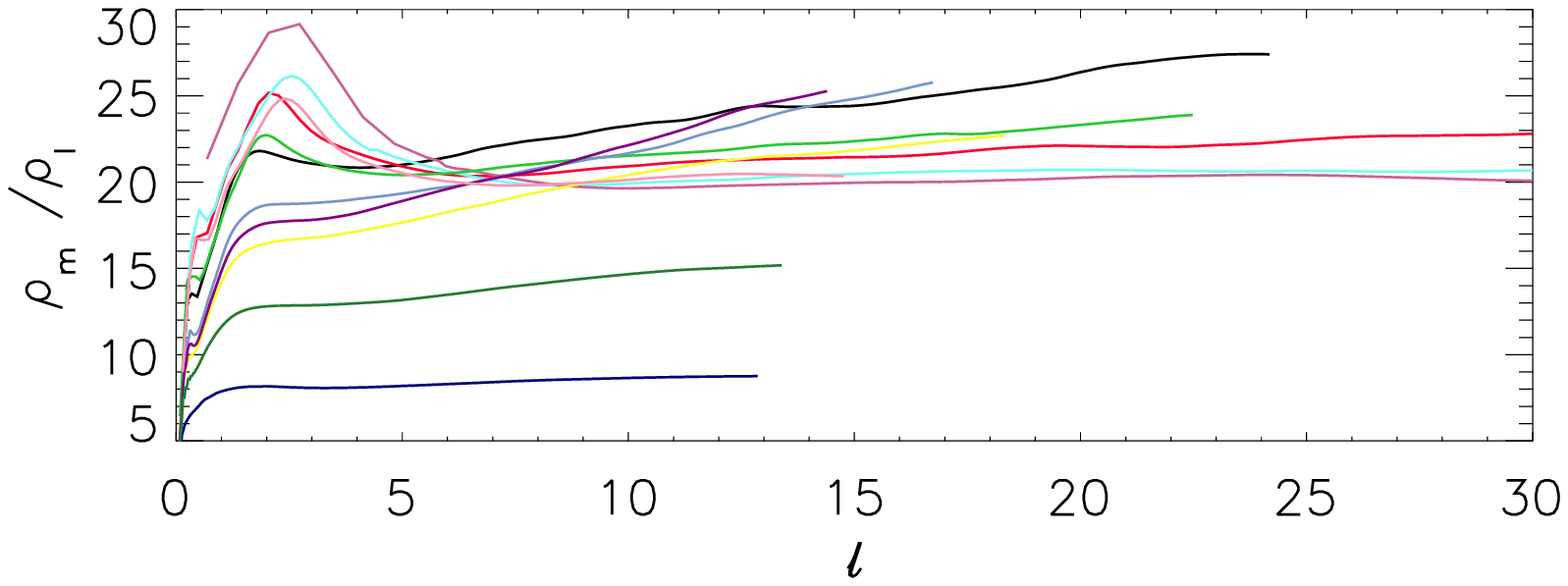}}
\vspace{-0.6cm}
\centerline{\includegraphics[width=9.0cm,height=5.5cm]{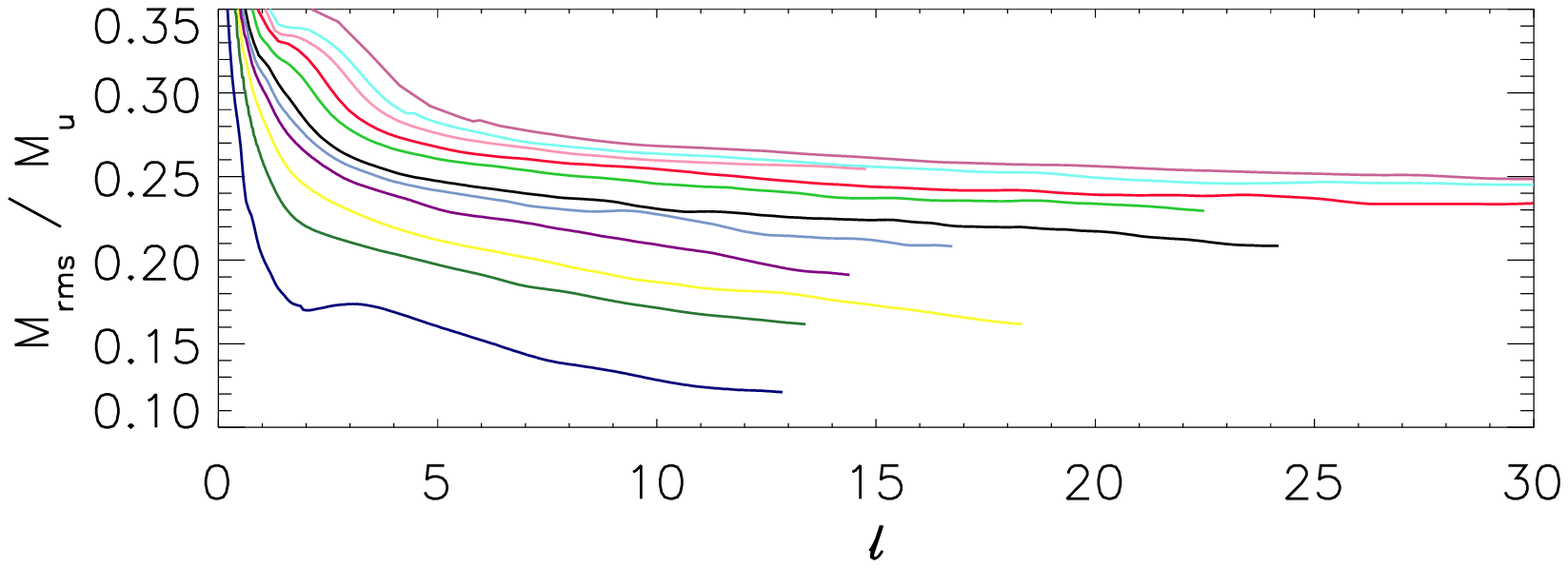}}
\vspace{-0.6cm}
\centerline{\includegraphics[width=9.0cm,height=5.5cm]{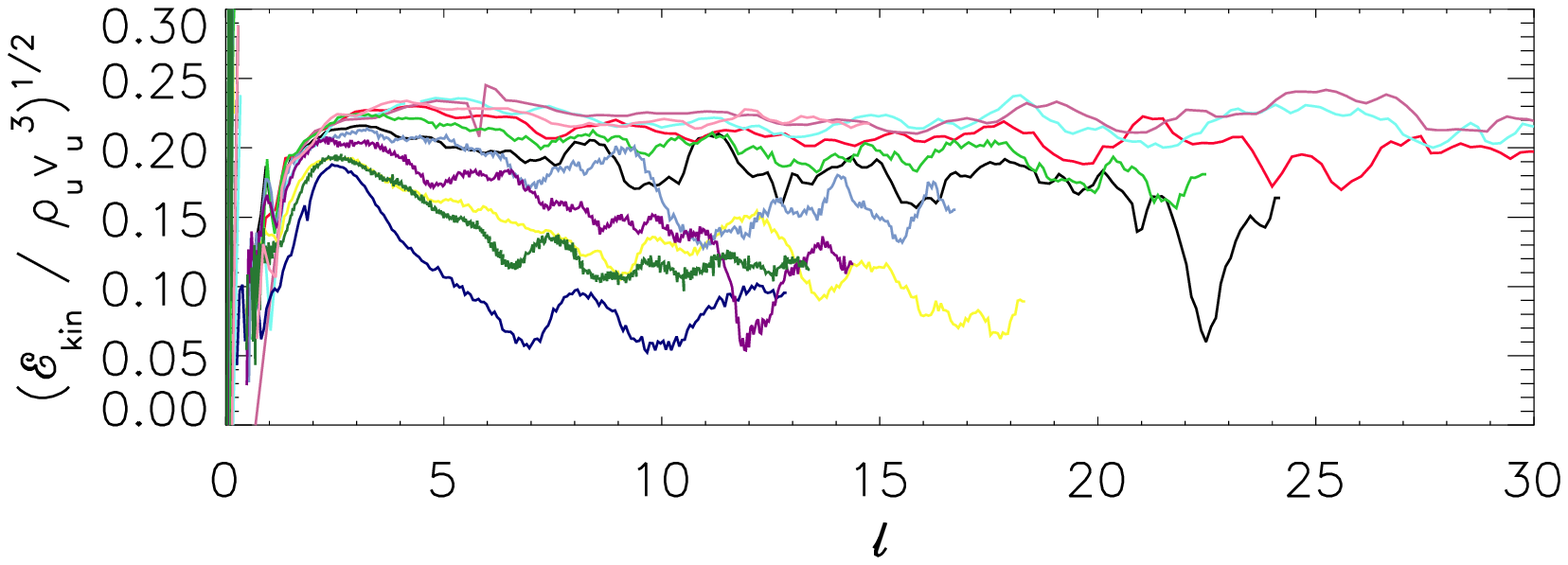}}
\caption{Mean density $\rho_{\mathrm{m}}$ ({\bf top panel}), root
         mean square Mach number $M_{\mathrm{rms}}$ ({\bf middle panel}), and
         $(\dot{\cal E}_{\mathrm{kin}} / (\rho_{\mathrm{u}} a^{3} M_{\mathrm{u}}^{3}))^{1/2}$  
         ({\bf bottom panel}) as
         function of the monotonically increasing CDL size $\ell$ for 
         simulations R*\_2. Individual curves denote $M_{\mathrm{u}} = 2$ 
         (dark blue), 4 (dark green), 5 (yellow), 7 (purple), 8 (light blue), 11 (black), 16 (green), 
         22 (red), 27 (pink), 33 (cyan), and 43 (magenta).
}
\label{fig:dens_mach}
\end{figure}
\begin{figure}[tbp]
\centerline{\includegraphics[width=8.5cm,height=5cm]{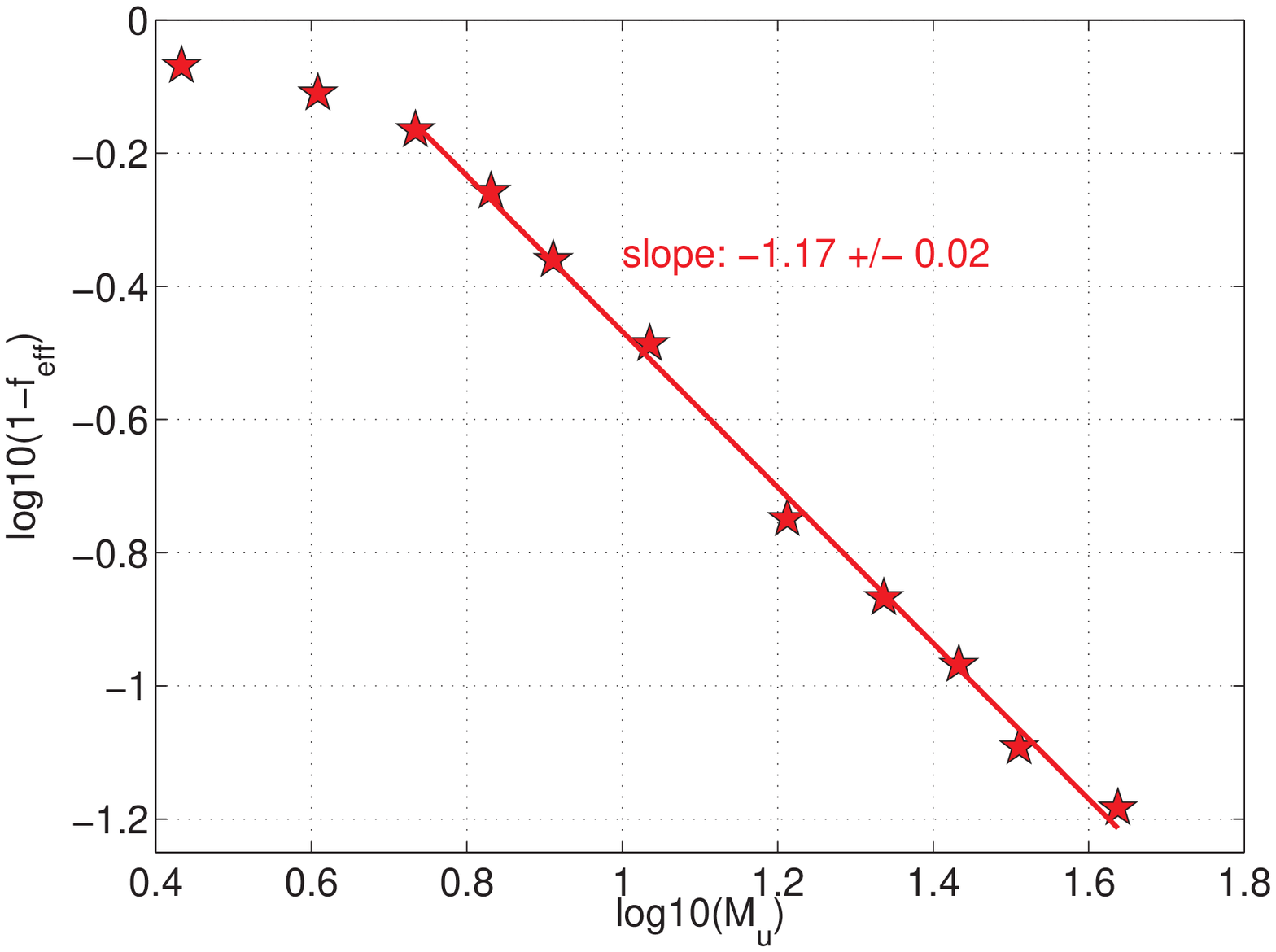}}
\centerline{\includegraphics[width=8.5cm,height=5cm]{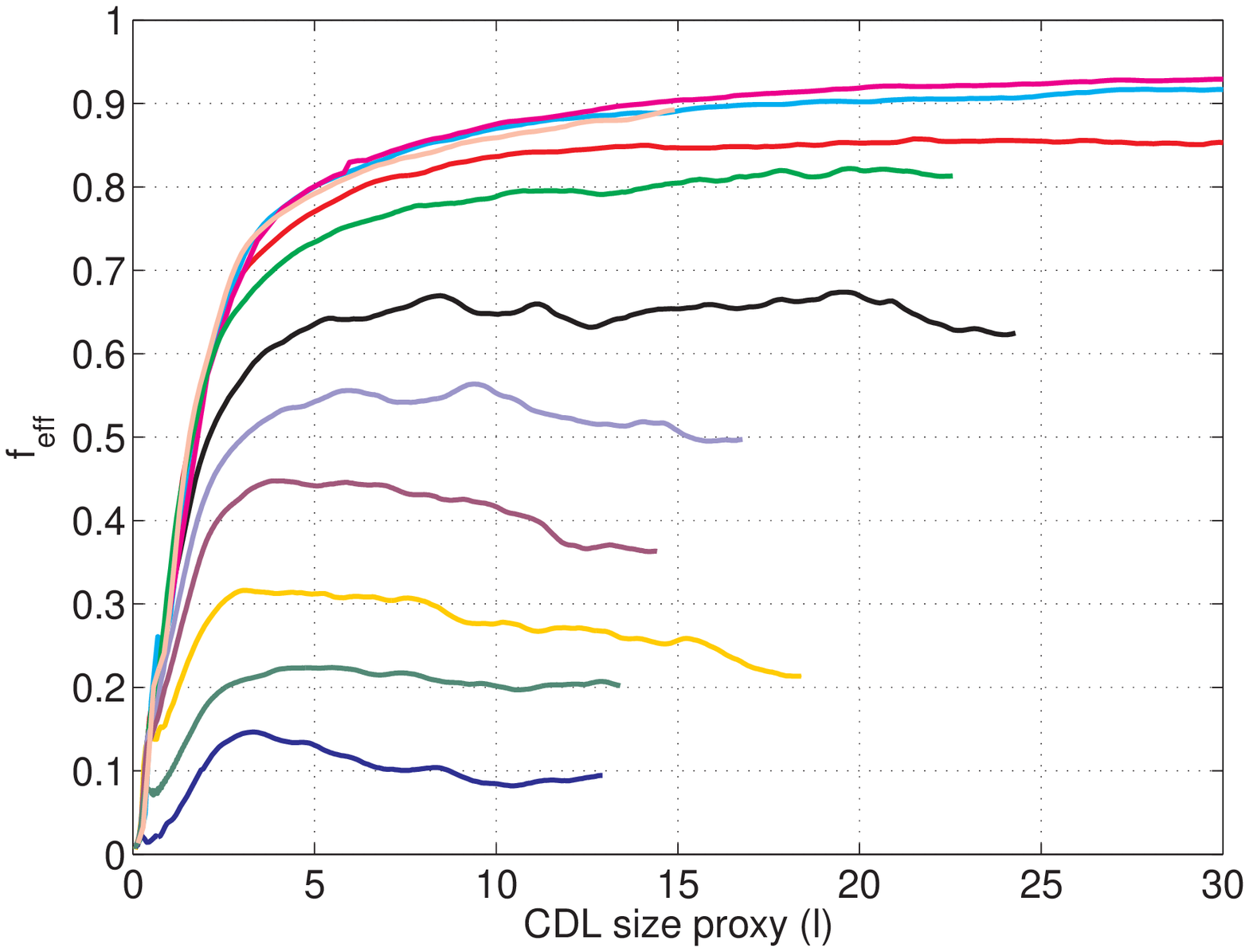}}
\caption{Driving efficiency of 3D slabs, same simulations and color
  coding as in Fig.~\ref{fig:dens_mach}.  {\bf Top panel:} Comparison
  of numerical results with expected scaling relation
  $\beta_{\mathrm{3}} \mathrm{log}_{\mathrm{10}} (M_{\mathrm{u}})
  \propto \mathrm{log}_{\mathrm{10}}(1 - f_{\mathrm{eff}})$. Each star
  denotes the maximum driving efficiency of one simulation. {\bf
    Bottom panel:} Evolution of $f_{\mathrm{eff}}$ as function of CDL
  size.}
\label{fig:feff}
\end{figure}
\subsubsection{3D scaling relations: numerical results}
\label{sec:num-scaling}
The analytical relations in Eq.~\ref{eq:exp_rho} to~\ref{eq:exp_ediss}
basically state that CDL mean quantities can be expressed in terms of
upstream quantities and that the corresponding relations do not change
with time / CDL size. Comparing the analytical predictions with our
numerical results thus means checking for two things: do the relations
hold across simulations, i.e., across widely varying upstream Mach
numbers, and do the relations hold independent of time / CDL size.

We first focus on the comparison of analytical and numerical results
with regard to the first question, the validity of the analytical
relations across different upstream Mach numbers. In doing so, we
tacitly assume that $\beta{\mathrm{i}}$ and $\eta_{\mathrm{i}}$ do not
change with time / CDL size.

Looking first at $\rho_{\mathrm{m}}$, Fig.~\ref{fig:dens_mach}, top
panel, and ignoring for the moment the obvious drift with increasing
CDL thickness, and concentrating instead on CDL thicknesses $6 \le
\ell \le 12$, the density compression ratios can be seen to lie mostly
in a range of $20 \le \eta_{\mathrm{1}} \le 25$. This narrow range is
remarkable given the range of upstream Mach numbers ($2.7 \le
M_{\mathrm{u}} \le 43.4$) and the associated 1D compression ratios ( $7
< \rho_{\mathrm{1d}}/\rho_{\mathrm{u}} < 1900$). The only clear
exceptions here are simulation R2\_2 and R4\_2, whose upstream Mach
numbers allow at most for 1D compression factors of about 7 and 16,
respectively, roughly the values attained also in the 3D case (dark
blue and dark green curves in Fig.~\ref{fig:dens_mach}).  The
analytically predicted value of $\beta_{\mathrm{1}}=0$ is confirmed by
the numerical results in that the scatter among the different curves
in Fig.~\ref{fig:dens_mach}, top panel, cannot be improved by any
other choice for $\beta_{\mathrm{1}}$.

For the root mean square Mach number (Fig.~\ref{fig:dens_mach}, middle
panel), we find $\eta_{\mathrm{2}} = M_{\mathrm{rms}} / M_{\mathrm{u}}
= 0.25$ to within about 10\% for all simulations except those with
very low upstream Mach number ($M_{\mathrm{u}} \le 7$). As in the case
of density, changing the predicted dependence of $M_{\mathrm{rms}}$ on
$M_{\mathrm{u}}$ does not reduce the scatter of the curves, unless low
Mach number simulations are included, in which case
$\beta_{\mathrm{2}}=1.1$ or even 1.2 gives a smaller spread among
curves than the predicted value of $\beta_{\mathrm{2}}=1$.  Using
Eq.~\ref{eq:exp_mach}, the above range for $\eta_{\mathrm{2}}$ implies
for $\eta_{\mathrm{1}} = \eta_{\mathrm{2}}^{-2}$ a value in the range
from 13 to 20, somewhat smaller but roughly consistent with the
numerically obtained density compression ratios discussed above.
Better consistency is achieved if $\eta_{\mathrm{2}}$ is determined
using Eq.~\ref{eq:exp_ekin} instead of Eq.~\ref{eq:exp_mach}. From
Fig.~\ref{fig:dens_mach}, bottom panel, and neglecting again low Mach
number simulations, we find $\eta_{\mathrm{2}} \approx 0.2$, which
corresponds to $\eta_{\mathrm{1}} \approx 25$. We ascribe the
ambiguity in the numerical value of $\eta_{\mathrm{2}}$ to slight
violations of the underlying assumptions in deriving
Eq.~\ref{eq:exp_rho} to~\ref{eq:exp_ediss}, in the case of low Mach
number simulations especially the assumption that
$M_{\mathrm{rms}}>>1$.
 
Numerical results regarding the remaining scaling relations, those
involving the driving efficiency $f_{\mathrm{eff}}$ and associated
parameters $\eta_{\mathrm{3}}$ and $\beta_{\mathrm{3}}$, are shown in
Fig.~\ref{fig:feff}. The top panel relates the maximum driving
efficiency achieved in each simulation with its upstream Mach number.
For $M_{\mathrm{u}} \ge 5$, an excellent linear relation ship between
log($M_{\mathrm{u}}$) and log($1 - f_{\mathrm{eff}}$) is found with
slope $\beta_{\mathrm{3}}=-1.17 \pm 0.02$ and $\eta_{\mathrm{3}} = 5$.

Changing perspective and looking at the evolution with CDL thickness
(or time) in Fig.~\ref{fig:dens_mach}, all simulations show a
persistent and clear decrease of the root mean square Mach number with
time and an associated increase of the density compression ratio, in
clear disagreement with the expected scaling relations. In units
$M_{\mathrm{rms}} / M_{\mathrm{u}}$, the decrease is about the same
for all simulations, around 0.02 to 0.03 (between about 10\% and 20\%)
for a doubling of the CDL thickness $\ell$ from 6 to 12.  Relative
changes are more pronounced for simulations with smaller  $M_{\mathrm{u}}$.

\begin{figure}[tp]
\centerline{\includegraphics[width=9.0cm,height=5cm]{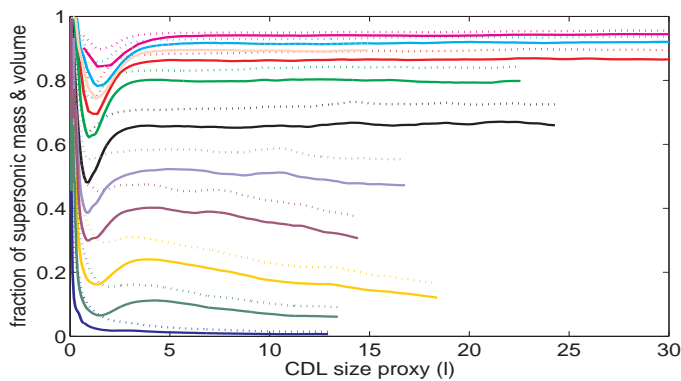}}
\caption{Supersonic mass fraction (solid) and volume fraction (dotted),
color coding as in Fig.~\ref{fig:dens_mach}.}
\label{fig:frac_supersonic}
\end{figure}

The same effect was observed in the 2D case and there attributed to
the increasing role of numerical dissipation with increasing CDL
size~(FW06). We favor the same explanation given there also for the 3D
case. The volume (or mass) fraction of the CDL that is subsonic
remains about constant with time (Fig.~\ref{fig:frac_supersonic}). In
absolute terms, the subsonic parts of the CDL grow as the CDL grows.
Energy loss in these parts, dominated by numerical dissipation /
viscous dissipation in terms of the MILES approach, grows in
proportion - while $\dot{\cal E}_{\mathrm{drv}} \approx
\mathrm{const.}$. The effect leads to an overall decay of CDL
turbulence.  It is more severe for smaller $M_{\mathrm{u}}$, as a
larger fraction of the CDL is subsonic
(Fig.~\ref{fig:frac_supersonic}), thus in violation of the assumptions
made in deriving the analytical scaling relations in
Sect.~\ref{sec:anal-scaling}.

Similar drifts with time exist for the driving efficiency,
Eq.~\ref{eq:exp_feff} and Fig.~\ref{fig:feff} (bottom panel).  However,
considering again the period $6 \le \ell \le 12$, the drifts with time
of $f_{\mathrm{eff}}$ can be to either larger or smaller values.
Apparently, the larger $M_{\mathrm{u}}$, the larger the CDL has to
grow before the driving efficiency reaches its peak value. 

In summary, we conclude from the above findings that the numerical
results essentially confirm the validity of the self-similar
analytical relations Eqs.~\ref{eq:exp_rho} to~\ref{eq:exp_ediss}
across different upstream Mach numbers, although the numerical results
display a systematic drift with increasing CDL size, which we
attribute to the effect of numerical dissipation / viscous dissipation.
\subsection{Inhomogeneity and anisotropy of CDL}
\label{sec:velo_aniso}
The driving of the turbulence within the CDL is highly anisotropic and
inhomogeneous: energy is injected only at the two confining shocks of
the CDL and only in the form of flows directed parallel or
anti-parallel to the x-axis. The amount of kinetic energy entering the
CDL, as well as its decomposition into compressible and solenoidal
components, depends on the orientation of the confining shock with
respect to the upstream flow and thus changes with position along the
confining shock. It is thus to be expected that the turbulence within
the CDL is neither homogeneous, nor isotropic.  Viewing angle effects
in observational data are a logical consequence.
\begin{figure}[tp]
\vspace{-0.3cm}
\centerline{\includegraphics[width=9.0cm,height=5.5cm]{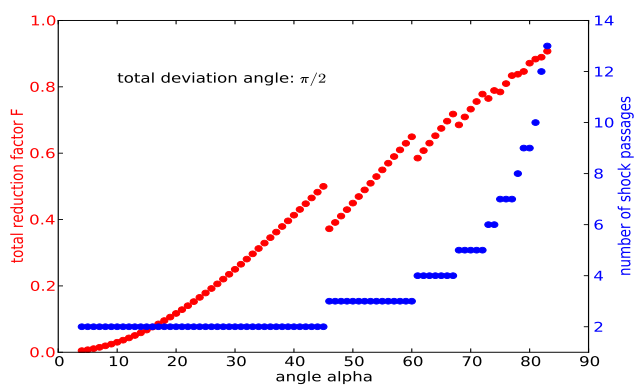}}
\caption{Deviating a highly supersonic, isothermal flow by (at least)
  $\beta_{\mathrm{tot}} = \pi/2$ through $n$ (right y-axis) subsequent
  shock passages at angle $\alpha$ (x-axis; in degrees; angle between
  flow direction and surface normal of shock) results in a total
  reduction of the Mach number by a factor $F_{\mathrm{tot}}$ (left
  y-axis).}
\label{fig:isotropization}
\end{figure}
\subsubsection{Mach number anisotropy}
\label{sec:mach_aniso}
From Table~\ref{tab:list_of_runs} it can be taken that the CDL root
mean square Mach number is highly anisotropic.  Root mean square Mach
numbers in direction transverse to the upstream flow (yz-plane,
$M_{\mathrm{rms,\perp}}$) are subsonic in all our experiments, with
concrete Mach numbers ranging from about 0.2 to 0.8.  In the direction
parallel to the upstream flow, supersonic root-mean-square Mach
numbers $M_{\mathrm{rms,\parallel}}$ are found from R7\_2 onwards. The
ratio $M_{\mathrm{rms,\parallel}} / M_{\mathrm{rms,\perp}}$ ranges
from around 2 to over 11.

Indeed, analytical considerations show the difficulty associated with
substantially deviating isothermal head-on colliding flows, a
necessary condition to reach isotropic conditions, while retaining
supersonic flow conditions. To illustrate the point, look at a uniform
flow hitting an oblique shock. Let $\alpha$ denote the angle between
the flow direction and the surface normal of the shock. Decomposing
the upstream and downstream flow velocities, $\mathbf{v}_{\mathrm{u}}$
and $\mathbf{v}_{\mathrm{d}}$, into components parallel and
perpendicular to the shock, assuming $M_{\mathrm{u}}>>1$, and applying
isothermal shock jump conditions $v_{\mathrm{d}\perp} =
v_{\mathrm{u}\perp} M_{\mathrm{u}\perp}^{-2}$, one can write
\begin{eqnarray}
\label{eq:v_up}
\mathbf{v}_{\mathrm{u}} & = & \mathbf{v}_{\mathrm{u\parallel}} + \mathbf{v}_{\mathrm{u\perp}}
                          =  v_{\mathrm{u}} \mathrm{sin} \alpha \; \mathbf{e}_{\mathrm{\parallel}} + 
                             v_{\mathrm{u}} \mathrm{cos} \alpha \; \mathbf{e}_{\mathrm{\perp}} \\
\label{eq:v_down}
\mathbf{v}_{\mathrm{d}} & = & \mathbf{v}_{\mathrm{d\parallel}} + \mathbf{v}_{\mathrm{d\perp}}
                          =  v_{\mathrm{u}} \mathrm{sin} \alpha \; \mathbf{e}_{\mathrm{\parallel}} + 
                             v_{\mathrm{u}} \mathrm{cos} \alpha \; 
                             M_{\mathrm{u}}^{-2} \mathrm{cos}^{-2} \alpha \; \mathbf{e}_{\mathrm{\perp}}.
\end{eqnarray}
These relations can be recast in the form of the Mach number reduction
factor $F \equiv v_{\mathrm{d}} / v_{\mathrm{u}}$ and the deviation
angle $\beta$ between the upstream and the downstream flow direction,
\begin{eqnarray}
\label{eq:reduction_F}
F & \equiv & \frac{v_{\mathrm{d}}}{v_{\mathrm{u}}} 
    =   (1 - \mathrm{cos}^{2}\alpha - \mathrm{cos}^{-2}\alpha \; M_{\mathrm{u}}^{-4})^{1/2}
    \approx  \mathrm{sin}\alpha \\
\label{eq:deviation_beta}
\beta & = & \mathrm{arccos} \left( 
            \frac{\mathbf{v}_{\mathrm{u}} \cdot \mathbf{v}_{\mathrm{d}}}{v_{\mathrm{u}} v_{\mathrm{d}}}
            \right) 
         \approx  \mathrm{arccos}(F) = \frac{\pi}{2} - \alpha,
\end{eqnarray}
where the approximations hold for large $M_{\mathrm{u}}$ and angles
$\alpha$ not too close to either $0$ or $\pi/2$. From
Eqs.~\ref{eq:reduction_F} and~\ref{eq:deviation_beta} it is apparent
that large deviation angles $\beta$ come at the price of
substantial reduction factors $F$, at least for single shock passages.
Multiple 'grazing' shock passages (large $\alpha$) allow, in
principle, for large total deviation angles $\beta_{\mathrm{tot}}$
while retaining an overall reduction factor $F_{\mathrm{tot}}$ close
to one. For $n$ subsequent shock passages at the same angle $\alpha$
one has $\beta_{\mathrm{tot}} = n \; (\pi/2 - \alpha)$ and
$F_{\mathrm{tot}} = (\mathrm{sin}\alpha)^{n}$. The situation is
illustrated in Fig.~\ref{fig:isotropization} for $\beta_{\mathrm{tot}}
\ge \pi/2$. For example, two subsequent shock passages at $\alpha =
\pi/4$ allow for a total deviation of $\pi/2$ while retaining about
half of the initial upstream Mach number ($F_{\mathrm{tot}} \approx 0.5$),
while seven passages at an angle of close to $80^{\circ}$ are needed
to retain 80\% of $M_{\mathrm{u}}$.  From
Fig.~\ref{fig:isotropization} it can also be taken that already one
shock passage at an angle $\alpha < 30^{\circ}$ reduces $M_{\mathrm{u}}$ by at
least 75\%. 

It is this later situation, even a few shock passages at small angles
$\alpha$, which we believe to be responsible for the low transverse
Mach numbers in our 3D slabs, despite the generally rather 'grazing'
angles at the confining shocks.
\begin{figure}[tbp]
\centerline{\includegraphics[width=9.0cm,height=5cm]{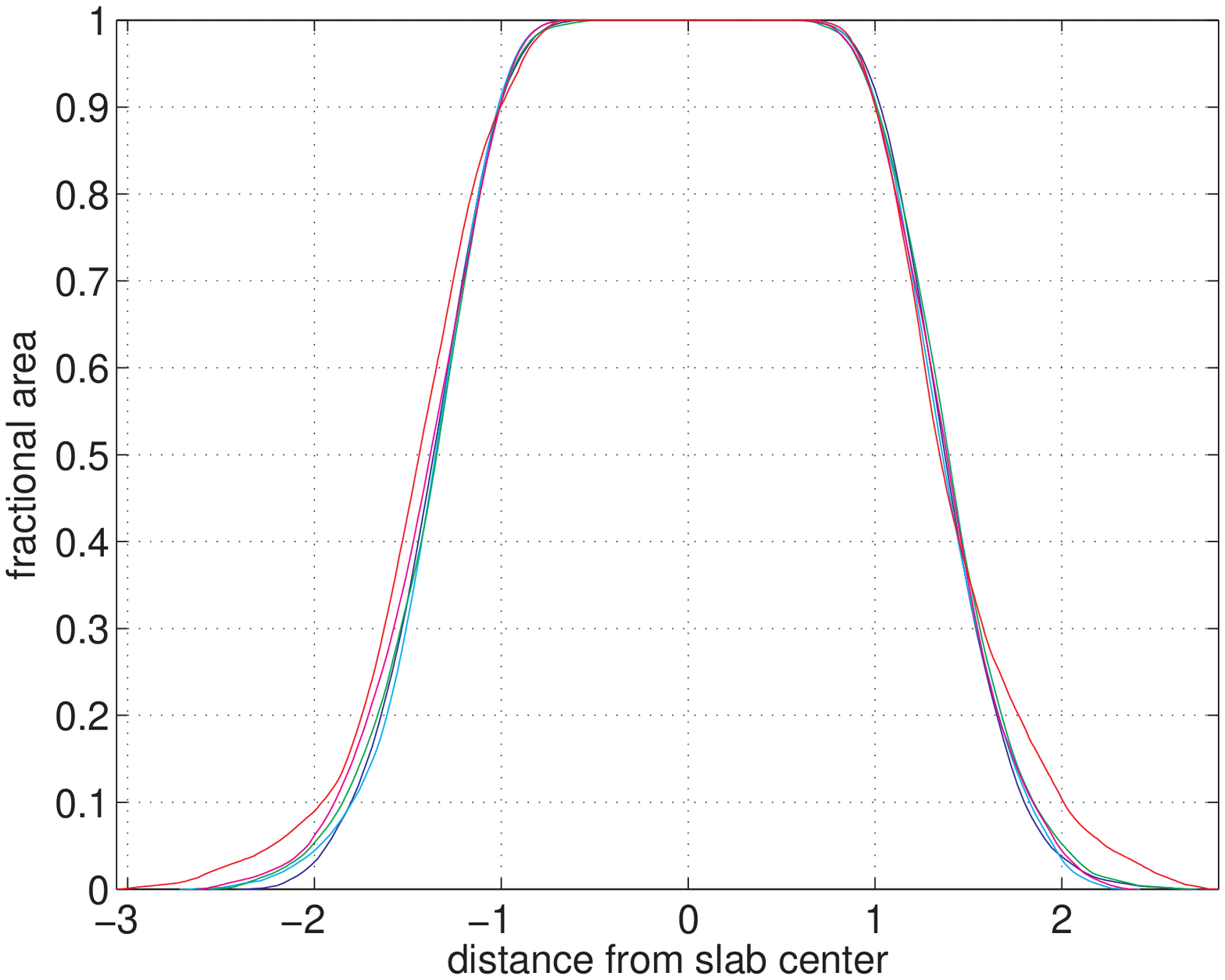}}
\centerline{\includegraphics[width=9.0cm,height=5cm]{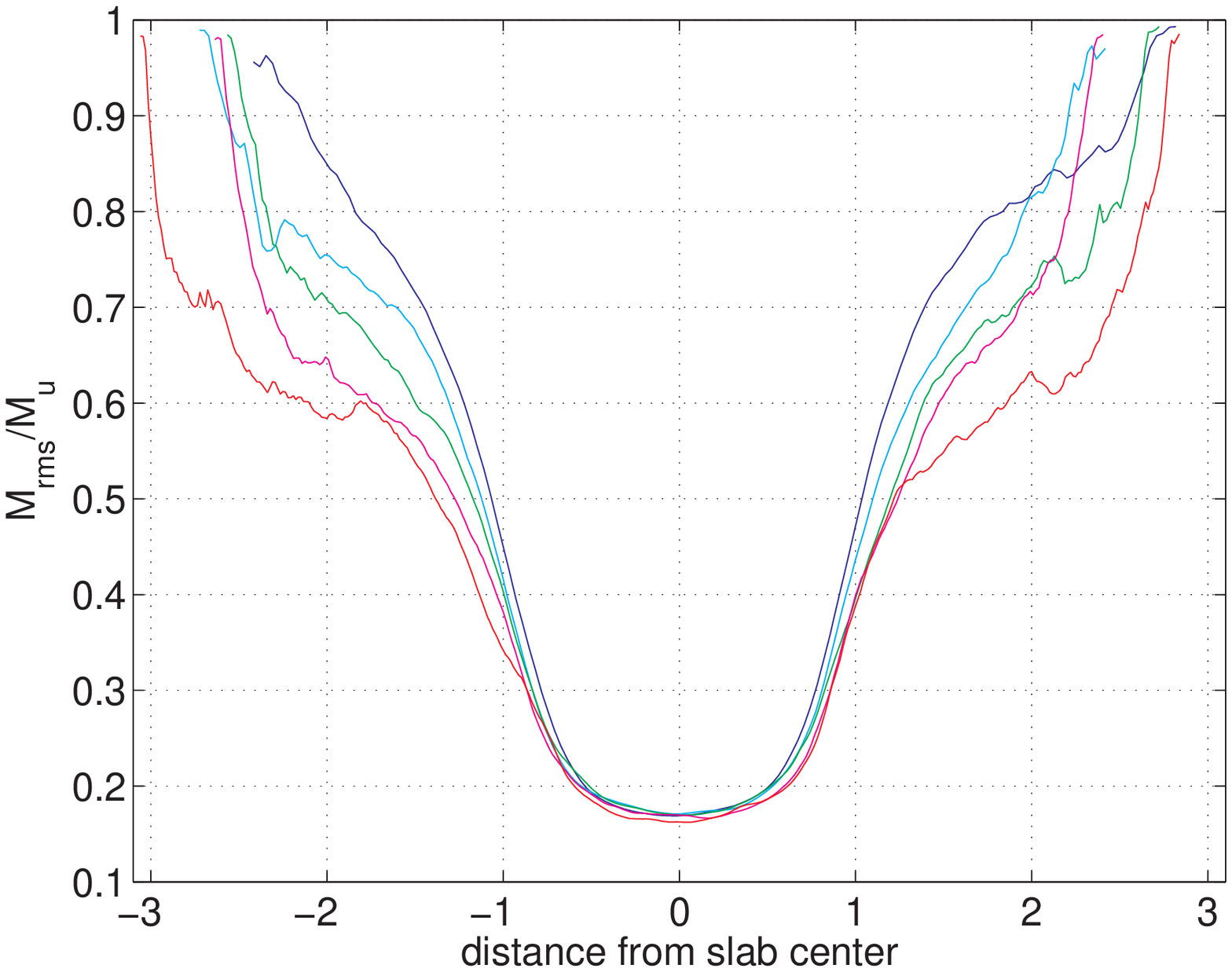}}
\caption{2D slice (yz-plane) averages as function of (scaled)
  x-coordinate for R22\_2 for different times ($\ell=12, 17, 23, 29,$
  and $34$, for the blue, cyan, green, magenta, and red curves,
  respectively). {\bf Top panel:} Fractional 2D CDL area $F(x)$. {\bf
    Bottom panel:} Root mean square Mach number in units of $M_{\mathrm{u}}$. For
  details see text, Sect.~\ref{sec:2d_slices}.}
\label{fig:2d_slice_selfsim}
\end{figure}
\subsubsection{Density variance - Mach number relation}
\label{sec:densmach}
This anisotropy of $M_{\mathrm{rms}}$ carries over to the
density variance - Mach number relation,
\begin{equation}
\frac{\sigma(\rho)}{\rho_{\mathrm{m}} M_{\mathrm{rms}}} = b.
\label{eq:dens-machnumber}
\end{equation}
From 3D periodic box simulations of driven, isothermal, supersonic,
hydrodynamical turbulence a value of $b \in [1/3,1]$ is found,
depending on the nature of the forcing, purely compressible ($b=1$) or
purely solenoidal ($b=1/3$)~\citep{padoan-nordlund-jones:97,
  passot-vazquez:98, federrath-et-al:10}.

Using Eq.~\ref{eq:dens-machnumber} for our CDL with $M_{\mathrm{rms}}$
the total root-mean-square Mach number, we find $b \in [0,1/3]$ (see
Table~\ref{tab:list_of_runs}), a range hardly overlapping with 3D
periodic box results. Higher $b$-values are obtained for the CDL if
only the transverse component $M_{\mathrm{rms,\perp}}$ is considered:
around 0.6 to 0.7 for small $M_{\mathrm{u}}$ and up to 0.85 for large
$M_{\mathrm{u}}$.  In the context of turbulence in flow collision
zones, the density variance - Mach number relation thus seems rather
indicative of the viewing angle (line-of-sight along
$M_{\mathrm{rms,\parallel}}$ or $M_{\mathrm{rms,\perp}}$) than of the
driving of the turbulence.
\subsubsection{Role of distance to confining shocks: 2D slices}
\label{sec:2d_slices}
CDL quantities are not only anisotropic, but they also differ between
regions close to the confining shocks, the only place where energy is
injected into the CDL, and areas close to the central plane of the
CDL. What 'close to the confining shocks' really means may be debated.
One interpretation could be to study flow properties as a function of
distance from the confining shock.  This we will not attempt here, as
our spatial resolution is likely too coarse anyway to capture the
details of such 'boundary layer flows'.

Instead, we study how averages over 2D slices (yz-plane) of the CDL
vary with the position of the slice from the CDL center. Technically
speaking, we take 2D slices (yz-plane) at different positions along
the x-axis, identify those cells within a given slice that belong to
the CDL (and not the upstream flow) by checking for $M_{\mathrm{rms}}
< M_{\mathrm{u}}$, and compute average quantities for each such 2D CDL
area $A(x)$.  Introducing the fractional 2D CDL area $F(x)=A(x)/(YZ)$,
with $YZ$ the yz-cross-section of the computational domain, slices
close to the CDL center are characterized by $F(x)=1$, whereas slices
close to outermost tips of the confining shocks have $F(x) \approx 0$.
To plot the resulting 2D averages, we use scaled x-units, which we
define by mapping the core part of the CDL with $F(x) > 0.9$ onto the
interval [-1,1] in scaled x-units. We expect this mapping to make
sense because of (approximate) self-similarity. This is confirmed by
Fig.~\ref{fig:2d_slice_selfsim}, top panel: the curves for the
fractional 2D CDL area $F(x)$ as a function of scaled x-units
(simulation R22\_2) fall essentially on top of each other, although
the CDL size increases from $\ell \approx 12$ to $\ell \approx 34$.
Similar figures are obtained for other simulations.
\begin{figure}[tbp]
\centerline{\includegraphics[width=9.0cm,height=4.5cm]{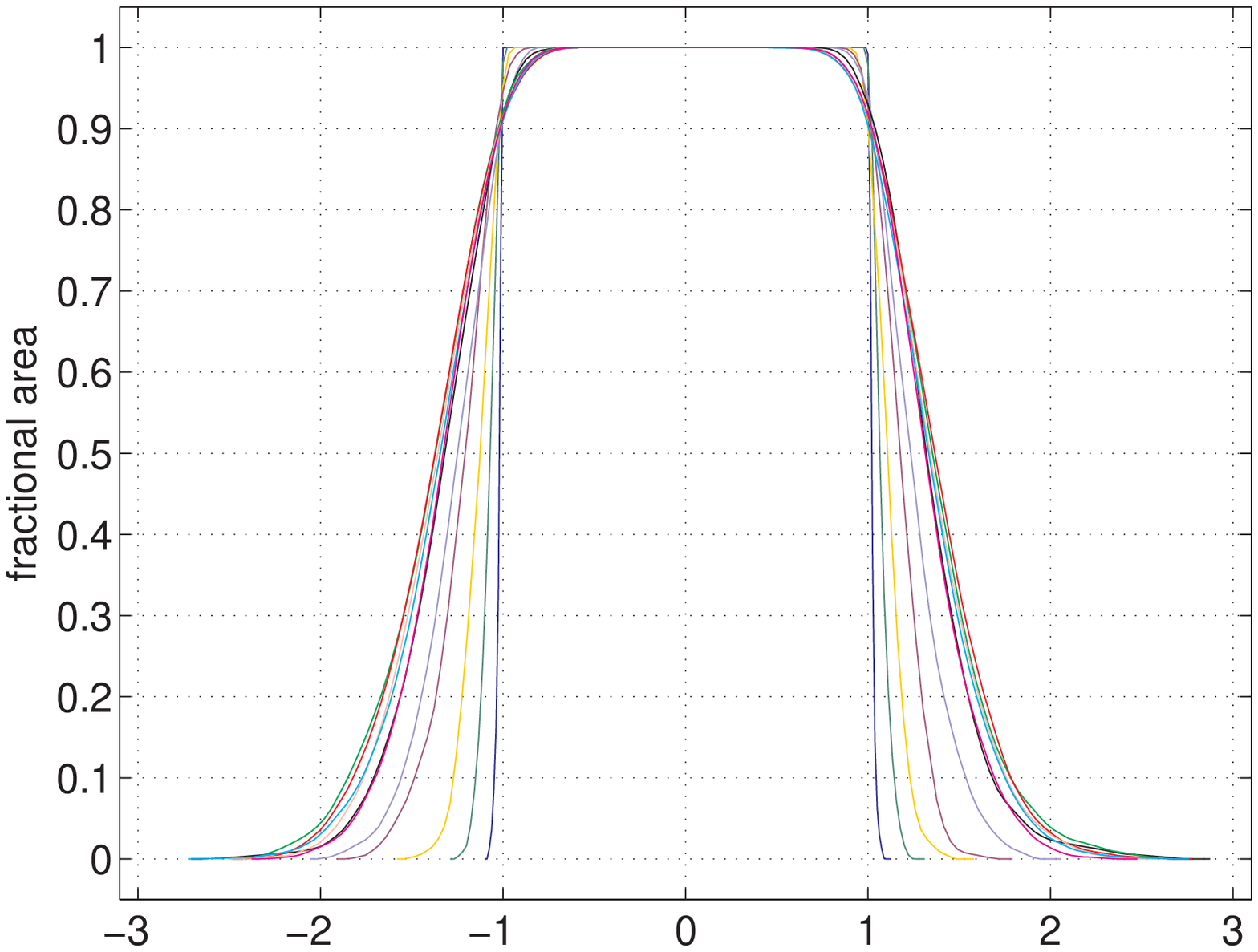}}
\centerline{\includegraphics[width=9.0cm,height=4.5cm]{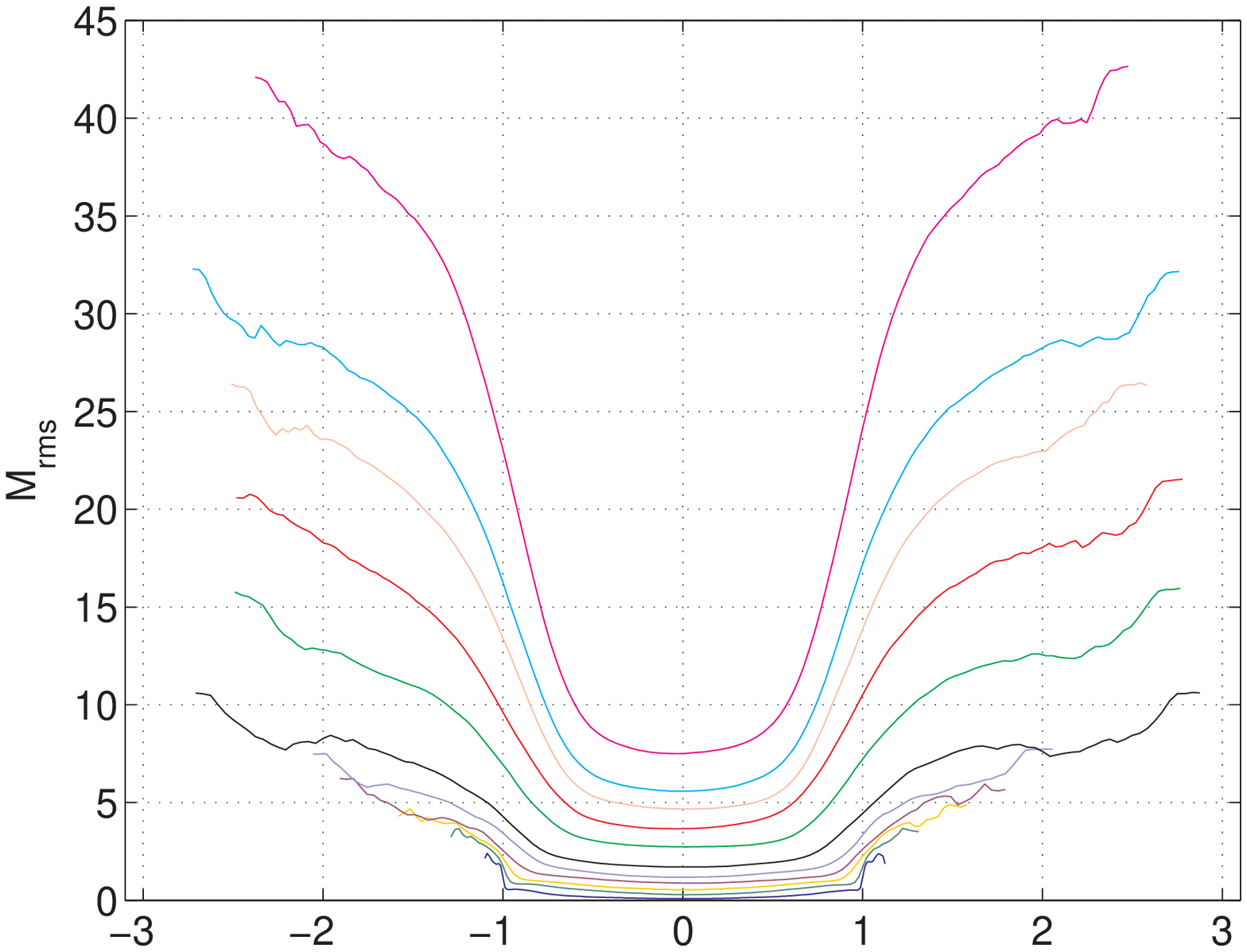}}
\centerline{\includegraphics[width=9.0cm,height=4.5cm]{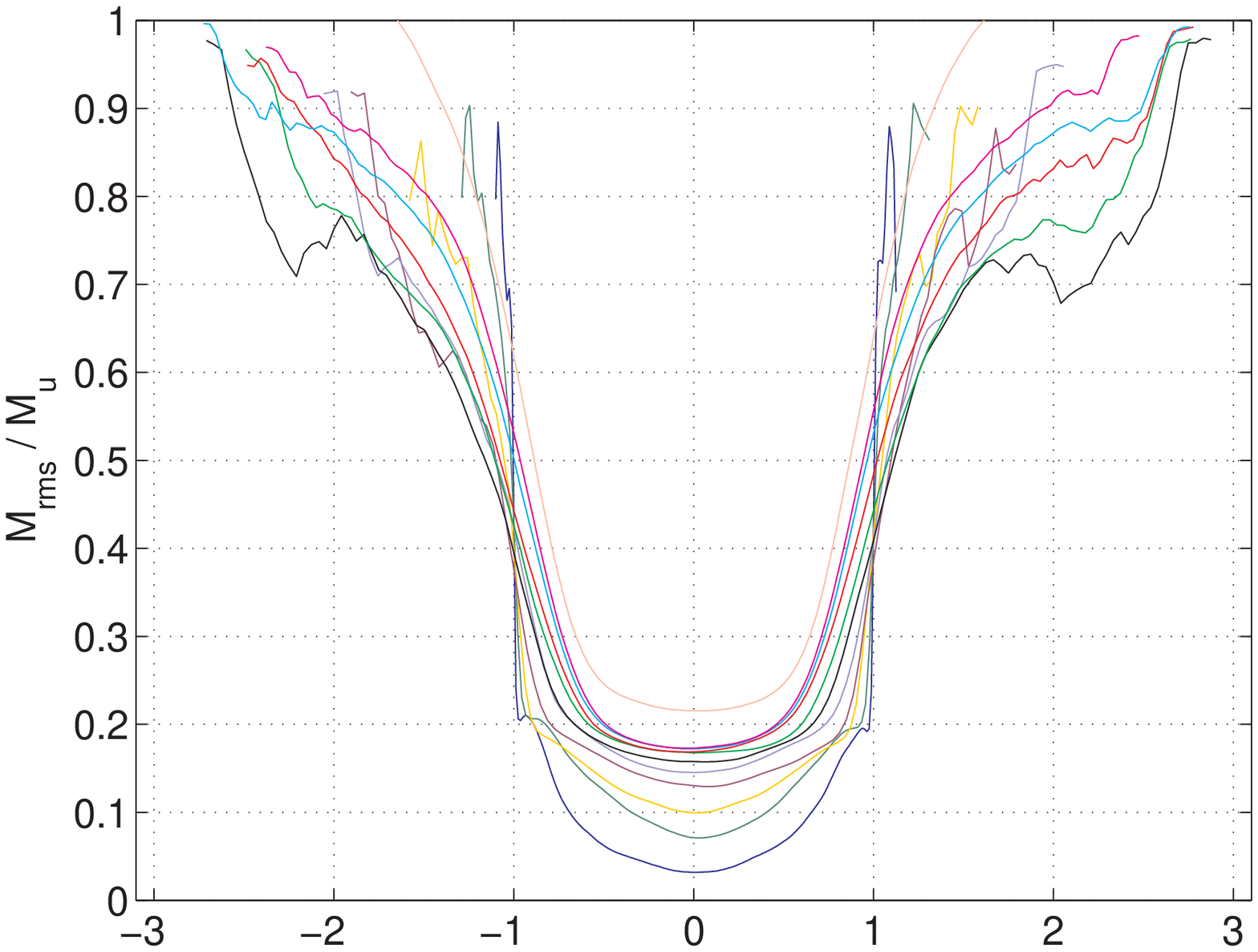}}
\centerline{\includegraphics[width=9.0cm,height=4.5cm]{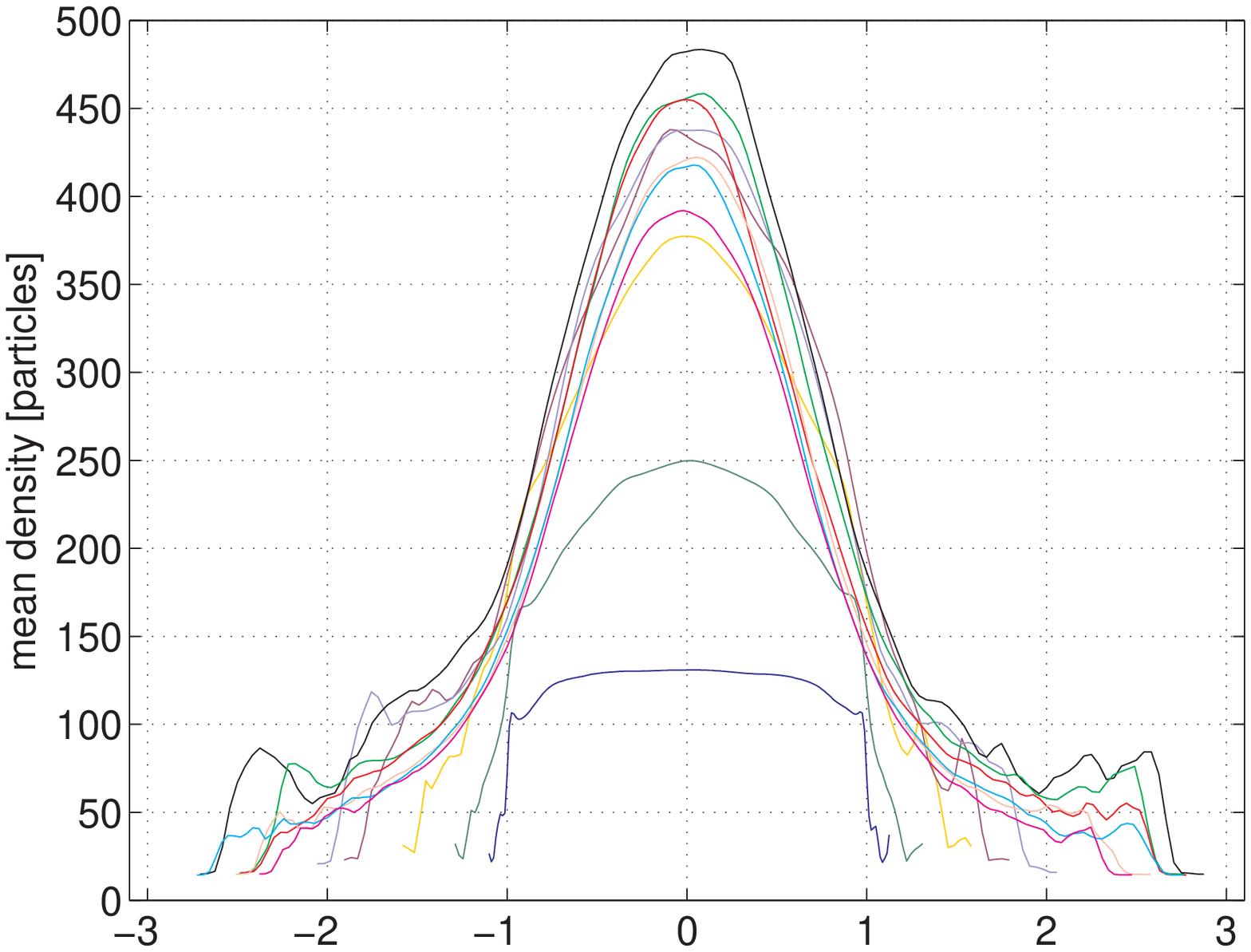}}
\centerline{\includegraphics[width=9.0cm,height=4.5cm]{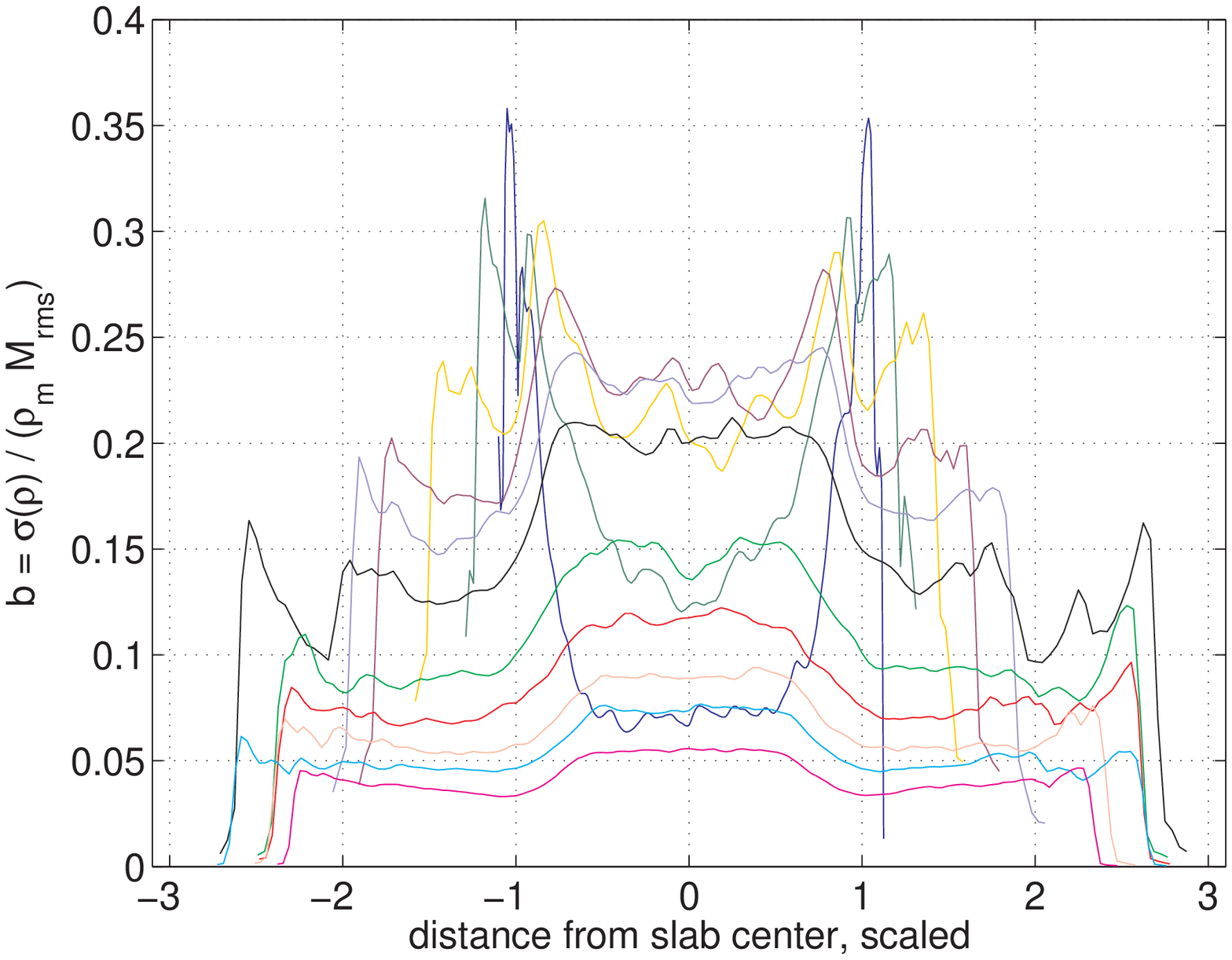}}
\caption{2D slices, average quantities for different runs
  (color coding as in Fig.~\ref{fig:dens_mach}) at CDL size $\ell \approx
  12$.  {\bf From top to bottom:} fractional
  CDL area $F(x)$, root mean square Mach number $M_{\mathrm{rms}}$,
  scaled root mean square Mach number
  $M_{\mathrm{rms}}/M_{\mathrm{u}}$, mean density $\rho_{\mathrm{m}}$, and
  $b = \sigma(\rho) / (\rho_{\mathrm{m}} M_{\mathrm{rms}})$.}
\label{fig:2d_slice_compo}
\end{figure}

Denoting parts where $F(x) < 1$ as 'CDL boundary layer' and parts with
$F(x) = 1$ as 'CDL core region', Fig.~\ref{fig:2d_slice_selfsim}, top
panel, further shows that the relative widths of the 'CDL boundary
layer' as compared to the 'CDL core region' remain essentially
constant throughout the simulation. The boundary layer does not
increase with respect to the core region or vice versa, the spatial
extension of each part obeys approximate self-similarity again.
Within the 'CDL boundary layer', $M_{\mathrm{rms}}$ generally
decreases with time for any given fixed relative x-position
(Fig.~\ref{fig:2d_slice_selfsim}, bottom panel). A possible
explanation for the decrease could be that as the CDL gets wider,
individual wiggles of the confining shocks extend farther and comprise
larger volumes ('get fatter'), the surface to volume ratio decreases.
In a 2D cut half way between $F(x)=0.9$ and $F(x) \approx 0$, the CDL
patches get larger for larger CDL thickness $\ell$. The distance from
the patch center to its boundary - the confining shock where energy is
injected - gets larger. The patch area (where turbulence must be driven)
divided by the patch circumference (where energy is injected) gets
larger.

Passing from one simulation at different times to comparison of
different simulations for the same CDL size, as shown in
Fig.~\ref{fig:2d_slice_compo} for different quantities, reveals again
the prominent role of the upstream Mach number. Larger upstream Mach
numbers result in larger 'CDL boundary layers', relative to the 'CDL
core region' (Fig.~\ref{fig:2d_slice_compo}, first panel). The root
mean square Mach number in the 'CDL core region' is slightly lower
than the CDL mean (around 15\% to 20\% of $M_{\mathrm{u}}$, second and
third panel of Fig.~\ref{fig:2d_slice_compo}). The density
distribution is more peaked around the CDL center than the
$M_{\mathrm{rms}}$ distribution (Fig.~\ref{fig:2d_slice_compo}, fourth
panel). Consistent with CDL mean values of
Sect.~\ref{sec:mean_quantities}, the largest densities are reached for
intermediate upstream Mach numbers, not for the highest. A tentative
interpretation here is that low upstream Mach numbers, on the one
hand, cannot generate the necessary compression, whereas high upstream
Mach numbers generate too Much turbulence, leading again to somewhat
lower mean densities. Numerical limitations (resolution) to maximum
compression ratios may also play a role (see Sect.~\ref{sec:num_effects}).
Finally, the density variance-Mach number relation shows less and less
dependence on the slice position with increasing $M_{\mathrm{u}}$
(Fig.~\ref{fig:2d_slice_compo}, fifth panel)

In summary, the analysis in terms of 2D slices illustrates the
inhomogeneity of the CDL turbulence, the CDL core region is less
turbulent than the CDL boundary layer. The relative weight of core and
boundary layer is independent of the CDL extension.
\begin{figure}[tbp]
\centerline{\includegraphics[width=9.cm,height=5cm]{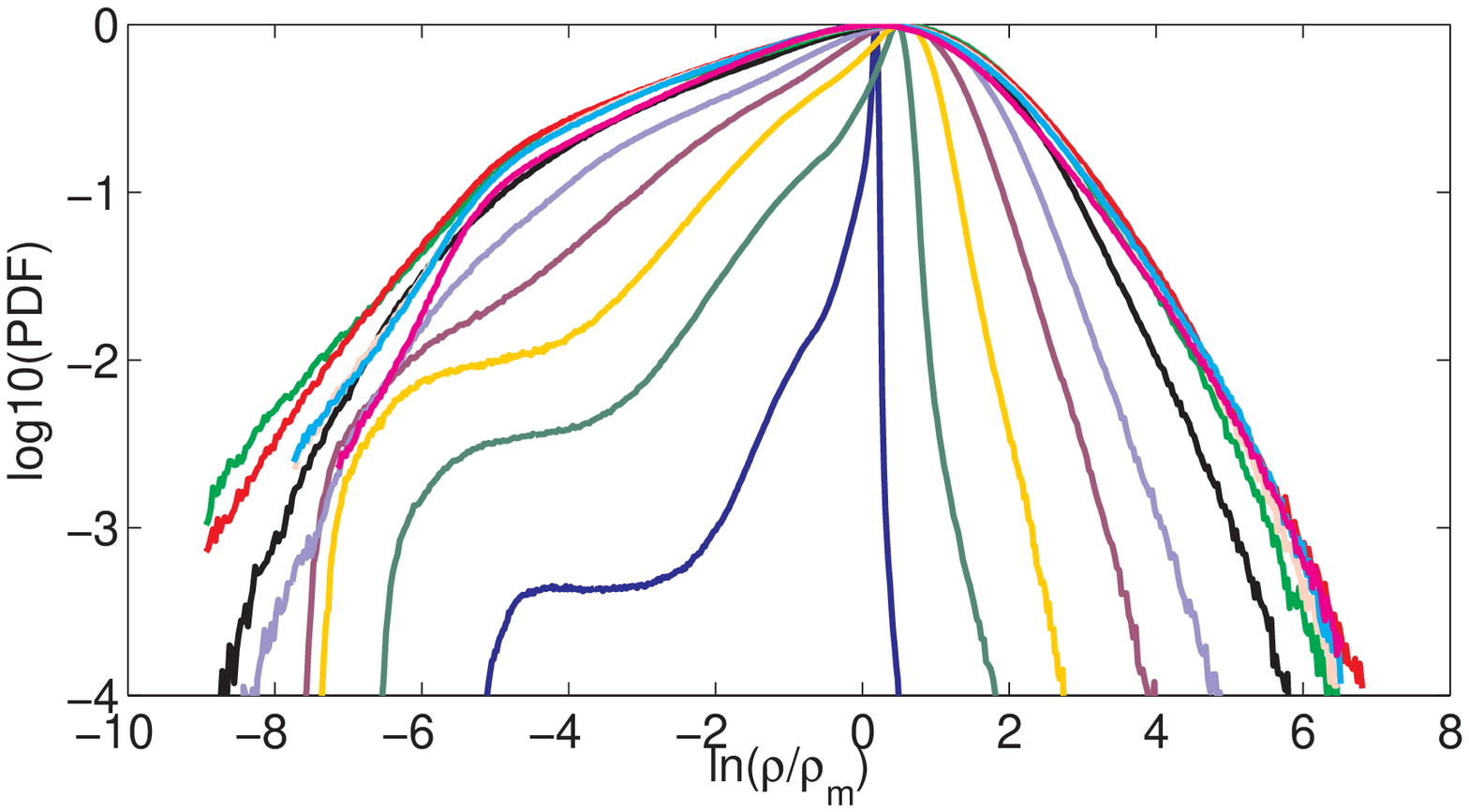}}
\centerline{\includegraphics[width=9.cm,height=5cm]{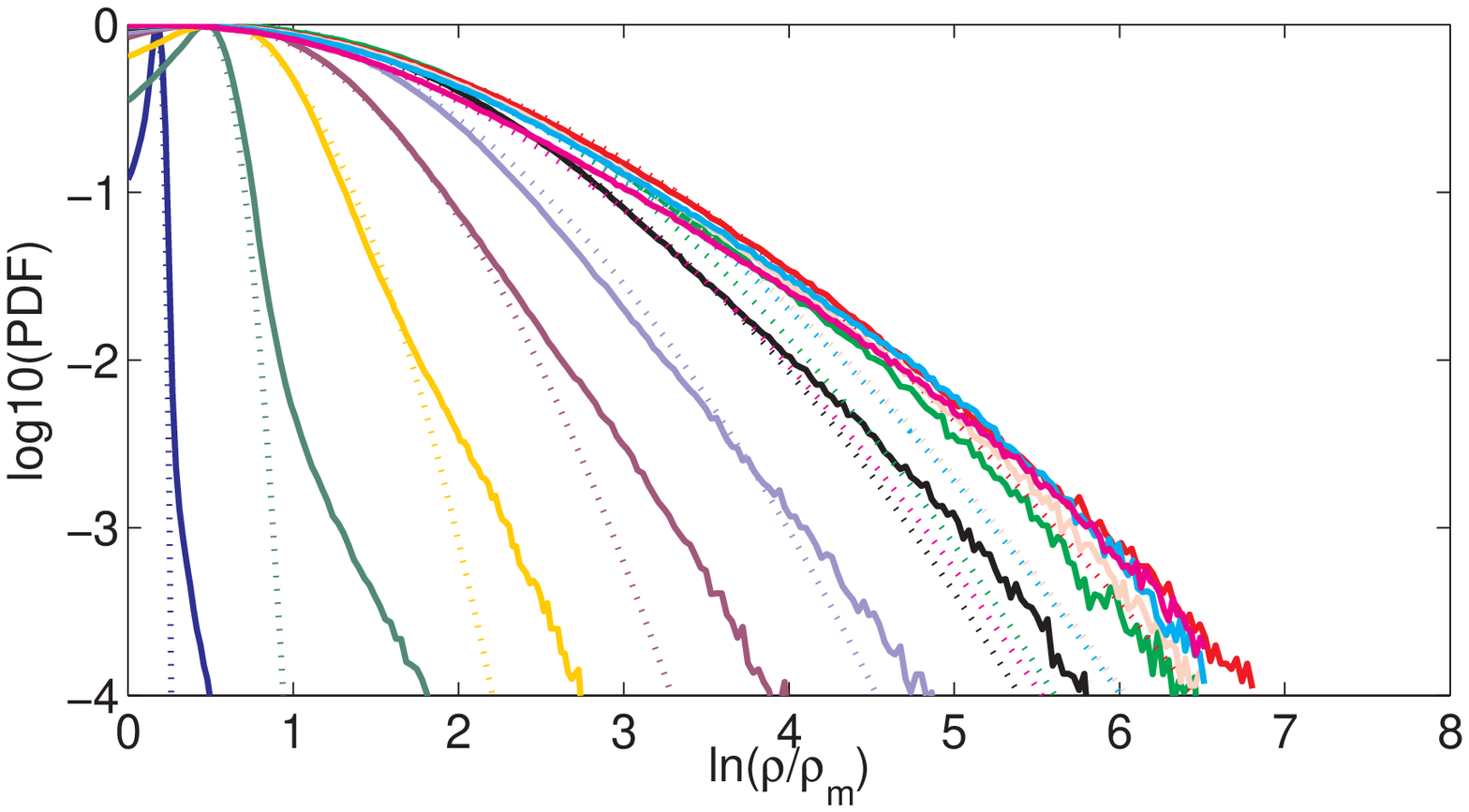}}
\caption{{\bf Top panel:} Volume-weighted PDFs of $s = ln(\rho /
  \rho_{m})$ for the different runs, color coding as in
  Fig.~\ref{fig:dens_mach}.  Each curve is an average over three data
  sets between $\ell \approx 11$ and $\ell \approx 12$.  {\bf Bottom
    panel:} Centrally (at peak value of PDF) mirrored PDF with
  Gaussian fitting (dotted line).}
\label{fig:densitypdf}
\end{figure}
\subsection{Density probability distribution function}
\label{sec:density_pdf}
The probability distribution function (PDF) of turbulent density
fluctuations is of key interest in connection with star formation and
as such extensively covered in the
literature~\citep[e.g.,][]{padoan-nordlund:02, krumholz-mckee:05,
  hennebelle-chabrier:11}.  A number of studies suggest that the
density PDF of isothermal, isotropic, and homogeneous supersonic
turbulence is log-normal, i.e., $s = \mathrm{ln}(\rho / \rho_{m})$ follows a
Gaussian distribution~\citep[e.g.,][]{vazquez-semadeni:94,
  passot-vazquez:98}. The PDF may deviate from a log-normal
distribution if additional physics is included, e.g., self-gravity, or
if subsonic root mean square Mach numbers result from solenoidal
forcing~\citep{federrath-et-al:09, kitsionas-et-al:09}.

The PDFs of $\mathrm{ln}(\rho / \rho_{m})$ we find for our different
runs also clearly deviate from a log-normal distribution
(Fig.~\ref{fig:densitypdf}, top panel).
Each PDF is the average over three snapshots of the CDL with seizes 
$\ell \in [11,12]$. 

To the right of their maxima, the PDFs can be approximately captured
by a log-normal distribution, but typically have a fat tail
(Fig.~\ref{fig:densitypdf}, bottom panel).  The fit (dotted line) was
obtained by first mirroring the right half of the PDF at its peak,
then determining the FWHM $w$ of the resulting distribution, from
which $\sigma$ can be obtain using $w = 2(2 ln(2))^{1/2} \sigma$ and
assuming that the distribution is normal. 

To the left of their maxima, the PDFs show a more intricate structure.
For runs with $M_{\mathrm{rms}}<1$, the peak value decreases sharply
before a plateau region is formed.  This plateau region is likely
associated with grid cells close to or even within the confining
shocks of the CDL, which on numerical grounds are about 3 cells wide.
Except for this plateau region, the shape of the PDFs at these low
Mach numbers resembles density PDFs found
by~\citet{kitsionas-et-al:09} for 3D periodic box simulations with purely
solenoidal forcing at subsonic root mean square Mach numbers. As
we go to higher Mach numbers, the decline of the PDF to the left of
its peak value becomes shallower, reaching about a constant slope for
runs with $M_{\mathrm{u}} > 11$. The plateau region of the low Mach
number runs turns into a steep slope for the high Mach number runs.

In summary, despite having purely isothermal conditions and no
additional physics, like for example self-gravity, the density PDFs of
our CDL always remain far from a log-normal distribution, even if the
turbulence is highly supersonic (large $M_{\mathrm{rms}}$).
\subsection{Structure functions}
\label{sec:struc_func}
\begin{figure*}[tbp]
\centerline{
\includegraphics[width=5.0cm,height=5.0cm]{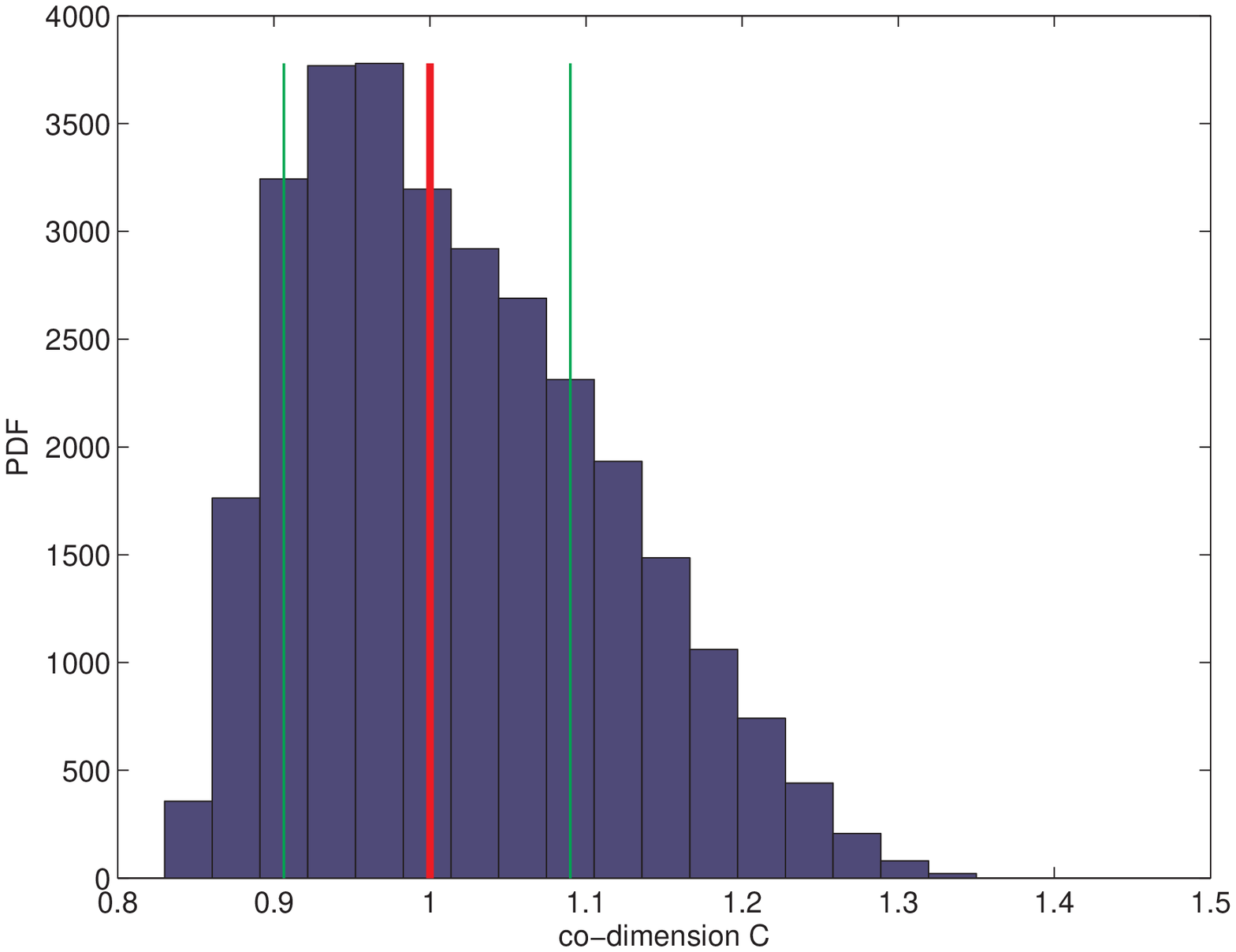}
\includegraphics[width=5.0cm,height=5.0cm]{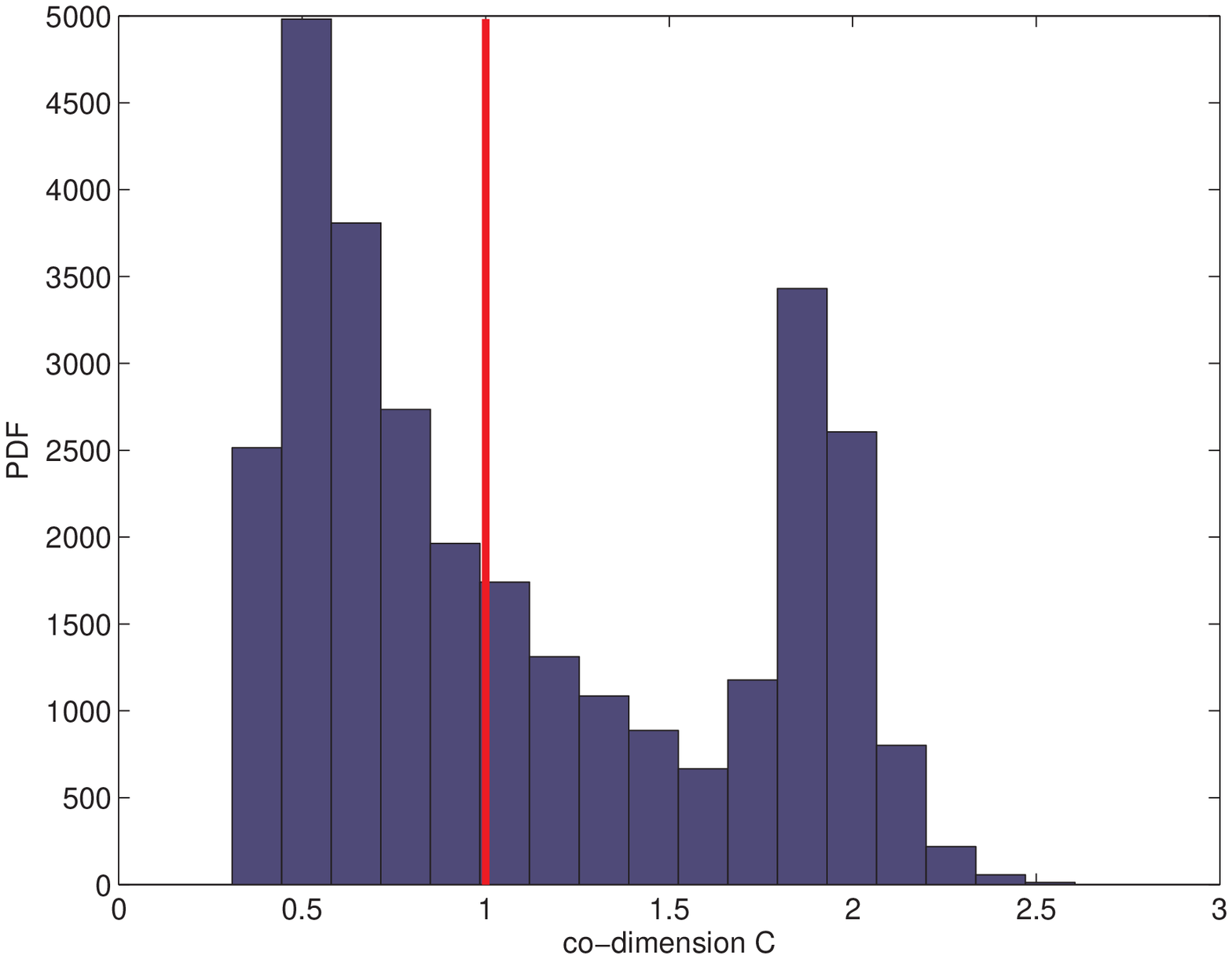}
\includegraphics[width=5.0cm,height=5.0cm]{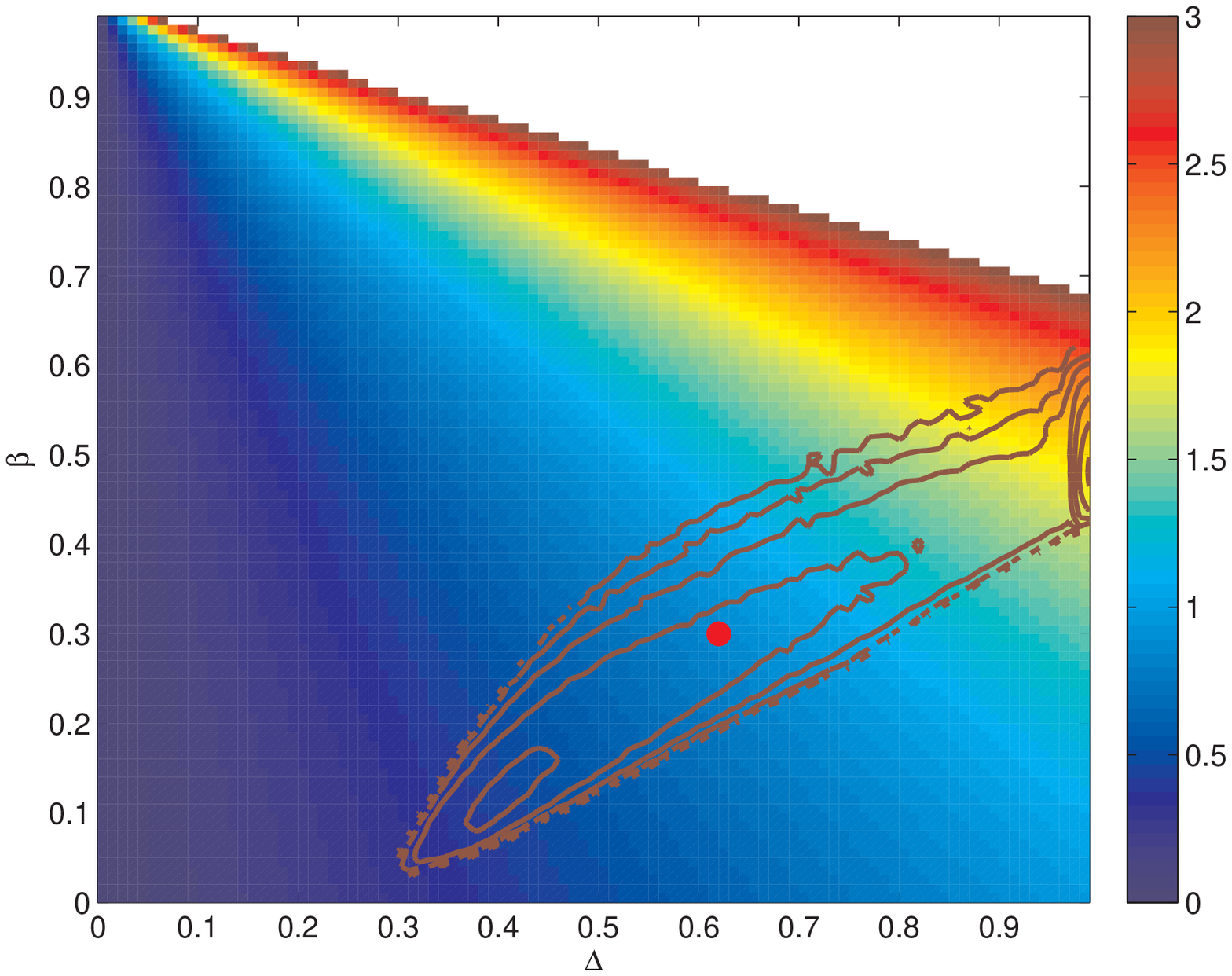}
}
\caption{A 5\% random perturbation of $Z_{p}$ (30\,000 realizations,
  at the example of $Z_{p}$ from~\citet{boldyrev:02}, $p=1$ to $p=5$)
  causes smearing of co-dimensions $C$ derived from Eq.~\ref{eq:ess}
  ({\bf left panel}). A double peak distribution results if $C$ is
  computed from Eq.~\ref{eq:dubrulle} ({\bf middle panel}).  The
  analytical value $C=1$ is shown as a red line, green lines in the
  left panel indicate where the 2/3 largest values in the histogram
  are located. In the {\bf right panel}, the 2D histogram (log10
  contours, spacing 0.5, spanning three orders of magnitude) of
  co-dimensions from perturbed $Z_{p}$ (500\,000 realizations) is
  over-plotted on the co-dimension $C=\Delta / (1-\beta)$ (color
  coded) in the $\Delta$-$\beta$ plane (Eq.~\ref{eq:dubrulle}). Two
  sharp peaks can be distinguished, at $\Delta \approx 0.4$ and $\beta
  \approx 0.1 $ ($C \approx 0.44$), as well as at $\Delta \approx 1.0$
  and $\beta \approx 0.5$ ($C \approx 1.9$).  The red dot is the
  analytical value by~\citet{boldyrev:02}, $\Delta=2/3$, $\beta=1/3$, $C=1$.}
\label{fig:ess_challenge}
\end{figure*}
Since the seminal work of~\citet{kolmogorov:41} (K41), velocity
structure functions have become an established tool for the
characterization of isotropic, homogeneous, incompressible, and
stationary turbulence. By contrast, the understanding of anisotropic,
inhomogeneous, compressible, non-stationary, or bound turbulence is
only in its infancy~\citep[][]{kurien-sreenivasan:00,
  biferale-et-al:08, byrne-et-al:11, hansen-et-al:11,
  biferale-et-al:12, grauer-et-al:12}. Characterizing our CDL data in
terms of structure functions thus is rather exploratory. We
nevertheless consider presentation of associated results worthwhile,
to make a few first steps in this direction, to compare our setup
with results from 3D periodic box simulation, and to illustrate the
potential role of line-of-sight effects, the latter also in view of
published, observation based structure functions for molecular clouds
whose physical interpretation continues to be
debated~\citep[e.g.,][]{padoan-et-al:03, gustafsson-et-al:06,
  hily-blant-et-al:08}.
\subsubsection{Working with low resolution CDL data}
\label{sec:sf_basics}
Anticipating quantitative values given in Sect.~\ref{sec:sf_num_comp},
here we first want to carve out two implications of the low resolution
of our CDL data: the compelling use of extended self-similarity and
why we determine the co-dimension of the most dissipative structure by
using the one parameter expression of Eq.~\ref{eq:ess} instead of the
two parameter expression of Eq.~\ref{eq:dubrulle}. We also introduce
some notation used later on.

The structure function of order $p$ is a scalar function of distance
$r$, defined as
\begin{equation}
S_{p}(r) \equiv < | \mathbf{u}(\mathbf{x} + \mathbf{r}) - \mathbf{u}(\mathbf{x})  |^{p} >,
\label{eq:struc_func_def}
\end{equation}
where $<\dots>$ denotes the average over all positions $\mathbf{x}$
within the sample and over all directed distances $\mathbf{r}$ from
each position. Structure functions are termed longitudinal
($S_{p}^{\parallel}$) if only orientations $\mathbf{u} \parallel
\mathbf{r}$ enter the average, and perpendicular ($S_{p}^{\perp}$) if
only $\mathbf{u} \perp \mathbf{r}$ enters. In the inertial range, both
can be well approximated by power laws with scaling exponents $\zeta_{p}$
\begin{equation}
S_{p}(r) \propto r^{\zeta_{p}}.
\label{eq:spdef}
\end{equation}

The numerical determination of $\zeta_{p}$ is difficult if the
inertial range is small, as is the case for our data (see
Sect.~\ref{sec:sf_num_comp}). The extended self-similarity (ESS)
hypothesis~\citep{benzi-et-al:93} can remedy the situation to some
extent. It basically states that the self-similarity of the velocity
field, if it exists, extends beyond the usual inertial range, into the
dissipation range. This extended range can be explored by considering
not absolute scaling exponents $\zeta_{p}$ but ratios $Z_{p} =
\zeta_{p} / \zeta_{3}$.

ESS scaling exponents $Z_{p}$ can be related to the co-dimension $C =
3-D$ of the most dissipative structures (dimension $D$),
\begin{equation}
Z_{p} = 
\frac{\zeta_{p}}{\zeta_{3}} = 
\frac{p}{9} + C \left ( 1 - \left ( 1 - \frac{2}{3C} \right)^{p/3}\right ).
\label{eq:ess}
\end{equation}
For 1D vortex filaments $C=2$~\citep{she-leveque:94},
for sheet-like structures $C=1$~\citep{boldyrev:02}. Intermediate
values of $C$ are found in numerical
simulations~\citep{padoan-et-al:04}. From multi-phase 3D periodic box
simulations~\citet{kritsuk-norman:04} find $D=2.3$, which is, as
pointed out by the authors, surprisingly close to observationally
determined fractal dimensions of molecular clouds~\citep[about 2.3,
e.g.,][]{elmegreen-falgarone:96, roman-duval-et-al:10}.  Roughly
speaking, velocity structure functions link the dimension of the most
dissipative turbulent structures to (observable) velocity differences.

Behind the one-parameter ($C$) view of Eq.~\ref{eq:ess} is the
two-parameter ($\Delta$ and $\beta$) expression for $Z_{p}$
by~\citet{dubrulle:94},
\begin{equation}
Z_{p} = 
(1 - \Delta)\frac{p}{3} + \frac{\Delta}{1-\beta}\left( 1 - \beta^{p/3} \right),
\label{eq:dubrulle}
\end{equation}
with $C=\Delta / (1-\beta)$. For $\Delta=0$, the K41 law is recovered.
For incompressible turbulence, \citet{she-leveque:94} postulated
$\Delta=2/3$ and with $\beta=2/3$ got $C=2$. For supersonic
turbulence, \citet{boldyrev:02} kept $\Delta=2/3$ but chose
$\beta=1/3$, giving $C=1$. There is, however, no reason why
$\Delta=2/3$ should also apply in the supersonic case, and why one may
collapse in this way the two-parameter model of Eq.~\ref{eq:dubrulle}
into the one-parameter model of Eq.~\ref{eq:ess}\footnote{For a good
  discussion, see~\citet{schmidt-et-al:09}, who
  follow~\citet{frisch:95} and~\citet{pan-et-al:08}.}.

Using Eq.~\ref{eq:dubrulle} to determine $C$ thus seems preferential.
However, in an application like ours, another factor comes into play:
Eq.~\ref{eq:dubrulle} is much more sensitive to uncertainties in the
$Z_{p}$ than Eq.~\ref{eq:ess}.  Fig.~\ref{fig:ess_challenge}
illustrates the point. We start with theoretical values $Z_{p}$
from~\citet{boldyrev:02} for $p=1$ to $p=5$. We randomly perturb each
$Z_{p}$ by at most 5\% (our accuracy requirement for $Z_{p}$, see
Appendix~\ref{app:sf_ess_comp}), thus obtaining 30\,000 perturbed data
sets. For each data set, we determine $C$ as best fit to either
Eq.~\ref{eq:ess} or Eq.~\ref{eq:dubrulle} (best fit = smallest root
mean square error between analytical $Z_{p}$, $p=1..5$, and $Z_{p}$ of
perturbed data set).  If Eq.~\ref{eq:ess} is used, the 5\%
perturbation (uncertainty) of $Z_{p}$ results in a smearing of $C$,
with $C$ mostly in a range 0.9 to 1.1 (Fig.~\ref{fig:ess_challenge},
left panel). Using Eq.~\ref{eq:dubrulle} results in a distribution with
two peaks at $C \approx 0.44$ and $C \approx 1.9$
(Fig.~\ref{fig:ess_challenge}, middle panel). A similar dichotomy was
reported by~\citet{kritsuk-et-al2:07} and by~\citet{schmidt-et-al:09}.
Seen in the $\Delta$-$\beta$-plane (Fig.~\ref{fig:ess_challenge},
right panel) the 5\% uncertainty in $Z_{p}$ translates into a whole
band of ($\Delta$, $\beta$) pairs and associated co-dimensions
($\Delta \approx 0.4$ and $\beta \approx 0.1 $ for $C \approx 0.44$;
$\Delta \approx 1.0$ and $\beta \approx 0.5$ for $C \approx 1.9$).

In view of these results, whose deeper discussion is beyond the scope
of this paper, we decided to use the more robust one parameter
expression Eq.~\ref{eq:ess} to compute the co-dimension $C$ from the
ESS exponents $Z_{p}$.
\subsubsection{Computing structure functions, ESS scaling exponents, and co-dimensions from CDL data}
\label{sec:sf_num_comp}
To compute $S_{p}(r)$ we use data at $\ell \approx 12$. Only points
$\mathbf{x}$ within the CDL may contribute, the number of points
contributing to $S_{p}(r)$ therefore decreases with $r$. When a
directed distance $\mathbf{r}$ leaves the CDL it is forbidden to
re-enter. The computation, detailed in Appendix~\ref{app:sf}, is
repeated in four different ways.  Spherical averaging in
Eq.~\ref{eq:struc_func_def} ('unified' case); spherical averaging but
retaining only pairs ($\mathbf{x}$, $r$) for which all distances $r$
lie still within the CDL ('full in' case); along a direction
(line-of-sight, LOS) either perpendicular or parallel to the upstream
flow.  We stress that no density variations or radiative transfer
effects are taken into account.

Given the small inertial range of our structure functions, apparent in
Fig.~\ref{fig:s3_uni_vs_iso_allsim}, top panel, we make use of the ESS
hypothesis and compute linear fits to $Z_{p}$, i.e., to
$\mathrm{log10}(S_{p})$ versus $\mathrm{log10}(S_{3})$. For the 'full
in' case, the quality of the fits is illustrated in
Fig.~\ref{fig:s3_uni_vs_iso_allsim}, bottom panel, for longitudinal
structure functions and $p=2$. In Appendix~\ref{app:sf}, the quality
of the fits is illustrated for all cases considered ('full in',
'unified', different LOSs) for $Z_{5}$ (Fig.~\ref{fig:ess_normXXX})
and computational details are given.  From $Z_{p}$, $p=1$ to $5$, a
best estimate for the co-dimension $C$ is derived for each run (using
Eq.~\ref{eq:ess}). Our demand that the uncertainty of the linear fit
to $Z_{p}$ is smaller than 5\% translates roughly into an uncertainty
of $\pm 0.1$ for $C$ (see Sect.~\ref{sec:sf_basics} and
Fig.~\ref{fig:ess_challenge}).
\subsubsection{CDL structure functions: case 'full in'}
\label{sec:sf_cdl_iso}
The 'full in' case is the most likely to bare similarities, if any,
with 3D periodic box simulations. This because our demand that all 20 rays lie
inside the CDL (see Sect.~\ref{sec:sf_num_comp}) rather disfavors
contributions from the vicinity of the confining shocks with their
(narrow) wiggles, where turbulence is most inhomogeneous and
anisotropic (see Sect.~\ref{sec:2d_slices}). 

\begin{figure}[tbp]
\centerline{\includegraphics[width=9cm,height=4.5cm]{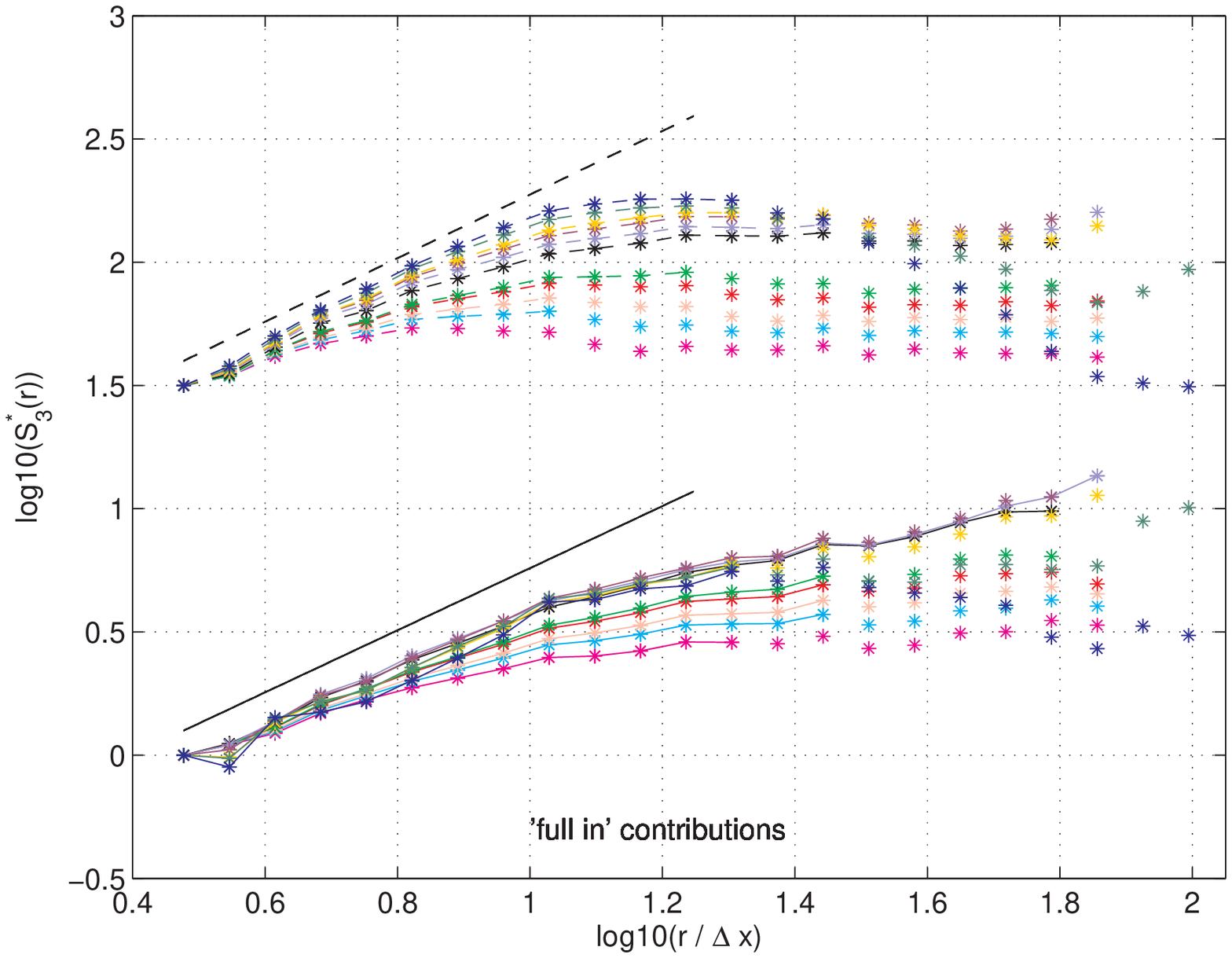}}
\centerline{\includegraphics[width=7cm,height=7.0cm]{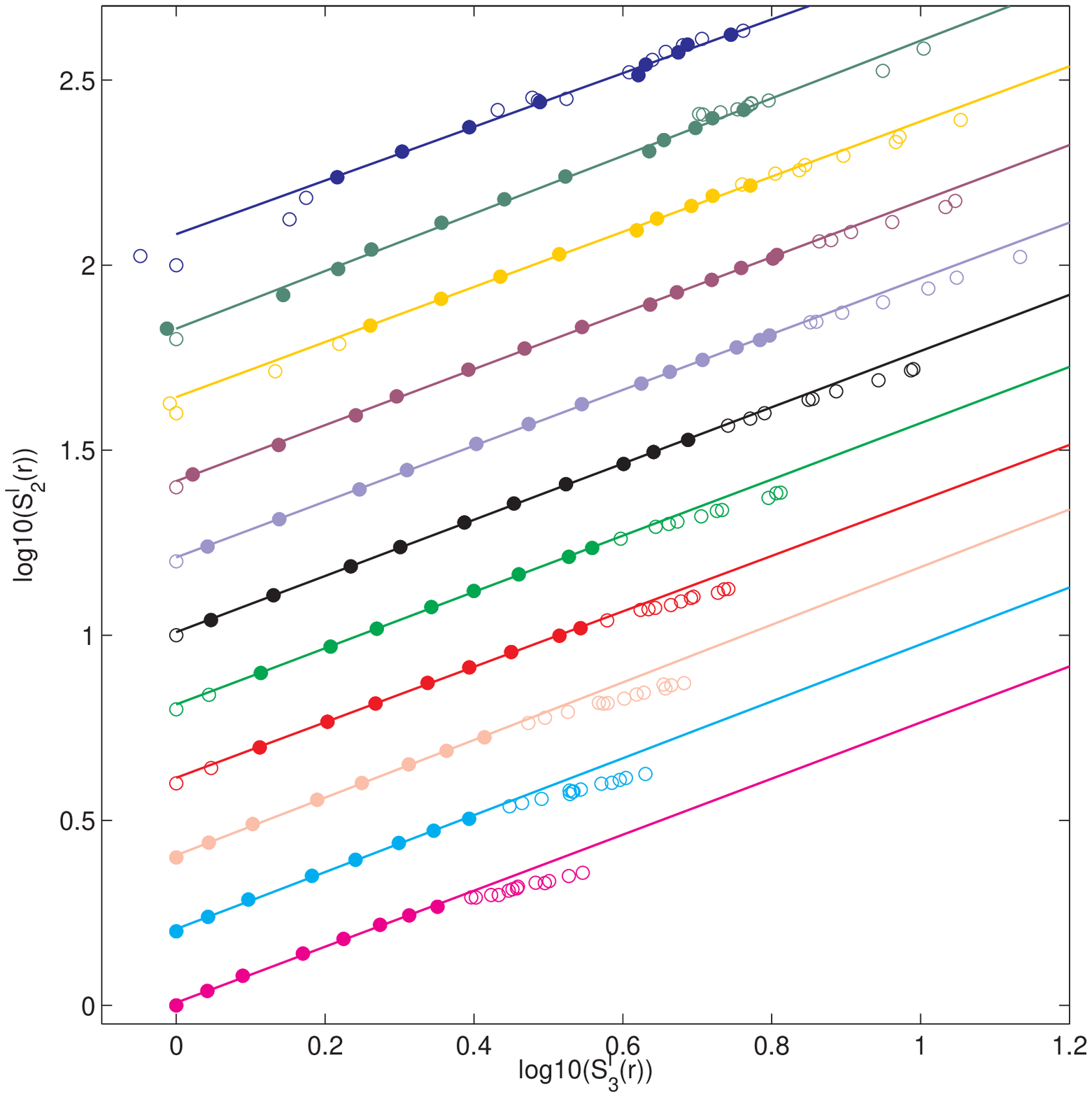}}
\caption{'Full in' case for simulations R*\_2 at $\ell \approx 12$,
  color coding as in Fig.~\ref{fig:dens_mach}. {\bf Top panel:}
  Structure functions $S_{3}^{\perp}(r)$ (upper curves) and
  $S_{3}^{\parallel}(r)$ (lower curves), curves vertically shifted to
  start from one common point. Stars connected by lines indicate data
  used to determine ESS scaling exponents. Also indicated are slopes
  $\zeta_{3}^{\parallel} = 1.26$ and $\zeta_{3}^{\perp} = 1.29$ as
  obtained by~\citet{kritsuk-et-al:07} for 3D periodic box simulations. {\bf
    Bottom panel:} Fit (solid lines) of ESS scaling exponent $Z_{2}^{\parallel}$
  as slope of $\mathrm{log10}(S_{2}^{\parallel})$ versus
  $\mathrm{log10}(S_{3}^{\parallel})$. Filled symbols denote data
  used for the fit, empty symbols denote all data available.}
\label{fig:s3_uni_vs_iso_allsim}
\end{figure}

Looking at the third order structure functions in
Fig.~\ref{fig:s3_uni_vs_iso_allsim}, top panel, five observations may
be made. First, our numerical resolution (256 cells in y-z-direction)
hardly allows for the formation of a proper inertial range. The
approximately linear dependence on $r$ covers less than one order of
magnitude in $r$. Scaling exponents $\zeta_{p}$ (see
Eq.~\ref{eq:spdef}) thus cannot be determined directly with sufficient
quality. Second, this rapid leveling off seems not related to changing
statistics with distance $r$. It takes place at distances $r$ where
the sample size has not yet decreased substantially (see
Fig.~\ref{fig:sf_frac_iso_aniso}, top panel). Third, larger values of
$M_{\mathrm{u}}$ result in shallower slopes of $\mathrm{log10}(S_{3})$
versus $\mathrm{log10}(r)$ that level off earlier. Fourth, the scatter
of third order structure functions due to changing $M_{\mathrm{u}}$ is
larger for $S_{3}^{\perp}$ than for $S_{3}^{\parallel}$, even at small
radii. $S_{3}^{\perp}$ typically also levels off earlier than
$S_{3}^{\parallel}$. Fifth, slopes $\zeta_{3}^{\parallel} = 1.26$ and
$\zeta_{3}^{\perp} = 1.29$ from 3D periodic box simulations at
$M_{\mathrm{rms}}=6$~\citep{kritsuk-et-al:07}, black
lines in Fig.~\ref{fig:s3_uni_vs_iso_allsim}, appear to form an upper
limit to the slab turbulence studied here, which covers a range of
$0.33 < M_{\mathrm{rms}} < 11.5$ (see Table~\ref{tab:list_of_runs}).

ESS scaling exponents are given in Table~\ref{tab:our_ESS_exp}.
Noteworthy are the following points.  Longitudinal structure functions
are consistent with co-dimension $C \approx 1$ or $D \approx 2$, i.e.,
indicating shocks as the most dissipative structures. They may show a
tendency toward larger co-dimension $C$ for larger $M_{\mathrm{u}}$.
The tendency stems to a good part from $Z_{5}$, all other $Z_{p}$ show
no clear dependence on $M_{\mathrm{u}}$.  The fits to $Z_{5}$ look
reasonable (see Fig.~\ref{fig:ess_normXXX}, top left), but firm
conclusions have to wait for better resolved simulation data.
Transverse structure functions display a systematic dependency on
$M_{\mathrm{u}}$: the co-dimension decreases with increasing
$M_{\mathrm{u}}$ from $C \approx 1.2$ for $M_{\mathrm{u}}=2$ to $C
\approx 0.5$ for $M_{\mathrm{u}}=33$. The robustness of this decrease
is supported by the fact that the related increase in $Z_{1}$ is
clearly larger than the 5\% accuracy limit imposed when determining
individual ESS exponents (see Appendix~\ref{app:sf}). 

It is interesting to note that the notion of universal scaling (see
e.g.,~\citet{schmidt-et-al:09}) would imply especially $Z_{2}$ to
increase with $M_{\mathrm{rms}}$ or, equivalently, with
$M_{\mathrm{u}}$.  Our transverse structure functions are compatible
with this expectation.  For the longitudinal structure functions, our
data are inconclusive.  One may speculate that the increase of $Z_{5}$
with $M_{\mathrm{u}}$ rather points toward a decrease of
$Z_{2}$ - a behavior incompatible with universal scaling.

The above findings are robust against increasing CDL size. For R22\_2, the
simulation which we integrated longest, we computed ESS scaling
exponents for CDL sizes $\ell=12, 17, 23, 29,$ and $34$. The resulting
maximum spread of $Z_{p}$ is less than 7\% for $p=1$ and less than 5\%
for $p=2$ to $p=5$. Similar results were obtained for two other
simulations, R11\_2 and R33\_2.
\begin{table}[tbp]
  \caption{Longitudinal (top) and transverse (middle) ESS scaling exponents 
    $Z_{p}$ and best estimates for co-dimension $C$,
    case 'full in' for the different runs at $\ell \approx 12$. Dashes indicate
    fits of insufficient quality. Also given (bottom) are some analytical values following Eq.~\ref{eq:ess}.}
\begin{center}
\begin{tabular}{lccccc} 
\hline
\rule[-2mm]{0pt}{6mm} 
       &$p=1$ &$p=2$ &$p=4$ &$p=5$ & $C$ \\
\hline
R2\_2  & 0.41 & 0.73 & 1.20 &   -  &  -  \\
R4\_2  & 0.45 & 0.78 & 1.14 &   -  &  -  \\
R5\_2  & 0.42 & 0.74 & 1.15 & 1.28 & 0.8 \\
R7\_2  & 0.43 & 0.76 & 1.20 & 1.37 & 0.9 \\
R8\_2  & 0.44 & 0.75 & 1.22 & 1.42 & 1.0 \\ 
R11\_2 & 0.44 & 0.76 & 1.21 & 1.41 & 1.0 \\ 
R16\_2 & 0.45 & 0.76 & 1.20 & 1.38 & 0.9 \\ 
R22\_2 & 0.43 & 0.75 & 1.21 & 1.41 & 1.0 \\ 
R27\_2 & 0.44 & 0.78 & 1.23 & 1.44 & 1.1 \\ 
R33\_2 & 0.42 & 0.77 & 1.23 & 1.48 & 1.2 \\ 
R43\_2 & 0.44 & 0.76 & 1.20 & 1.50 & 1.2 \\ 
\hline
R2\_2  & 0.43 & 0.75 & 1.23 & 1.48 & 1.2 \\
R4\_2  & 0.45 & 0.76 & 1.20 & 1.39 & 0.9 \\
R5\_2  & 0.46 & 0.78 & 1.18 & 1.34 & 0.8 \\ 
R7\_2  & 0.48 & 0.79 & 1.17 & 1.32 & 0.8 \\
R8\_2  & 0.49 & 0.80 & 1.16 & 1.30 & 0.8 \\ 
R11\_2 & 0.52 & 0.82 & 1.13 & 1.26 & 0.7 \\ 
R16\_2 & 0.53 & 0.82 & 1.12 & 1.22 & 0.5 \\
R22\_2 & 0.52 & 0.81 & 1.14 & 1.26 & 0.7 \\ 
R27\_2 & 0.52 & 0.81 & 1.14 & 1.25 & 0.5 \\ 
R33\_2 & 0.52 & 0.81 & 1.14 & 1.24 & 0.5 \\ 
R43\_2 &   -  &   -  &   -  &   -  &  -  \\ 
\hline
  -    & 0.56 & 0.83 & 1.13 & 1.25 & 0.7 \\
  -    & 0.47 & 0.78 & 1.17 & 1.32 & 0.8 \\
  -    & 0.44 & 0.76 & 1.20 & 1.36 & 0.9 \\
  -    & 0.42 & 0.74 & 1.21 & 1.40 & 1.0 \\
  -    & 0.41 & 0.73 & 1.23 & 1.42 & 1.1 \\
  -    & 0.40 & 0.72 & 1.24 & 1.45 & 1.2 \\
\end{tabular}
\end{center}
\label{tab:our_ESS_exp}
\end{table}
\subsubsection{CDL structure functions: beyond the 'full in' case}
\label{sec:sf_cdl_beyond}
\begin{figure}[tbp]
\centerline{\includegraphics[width=9cm,height=4.cm]{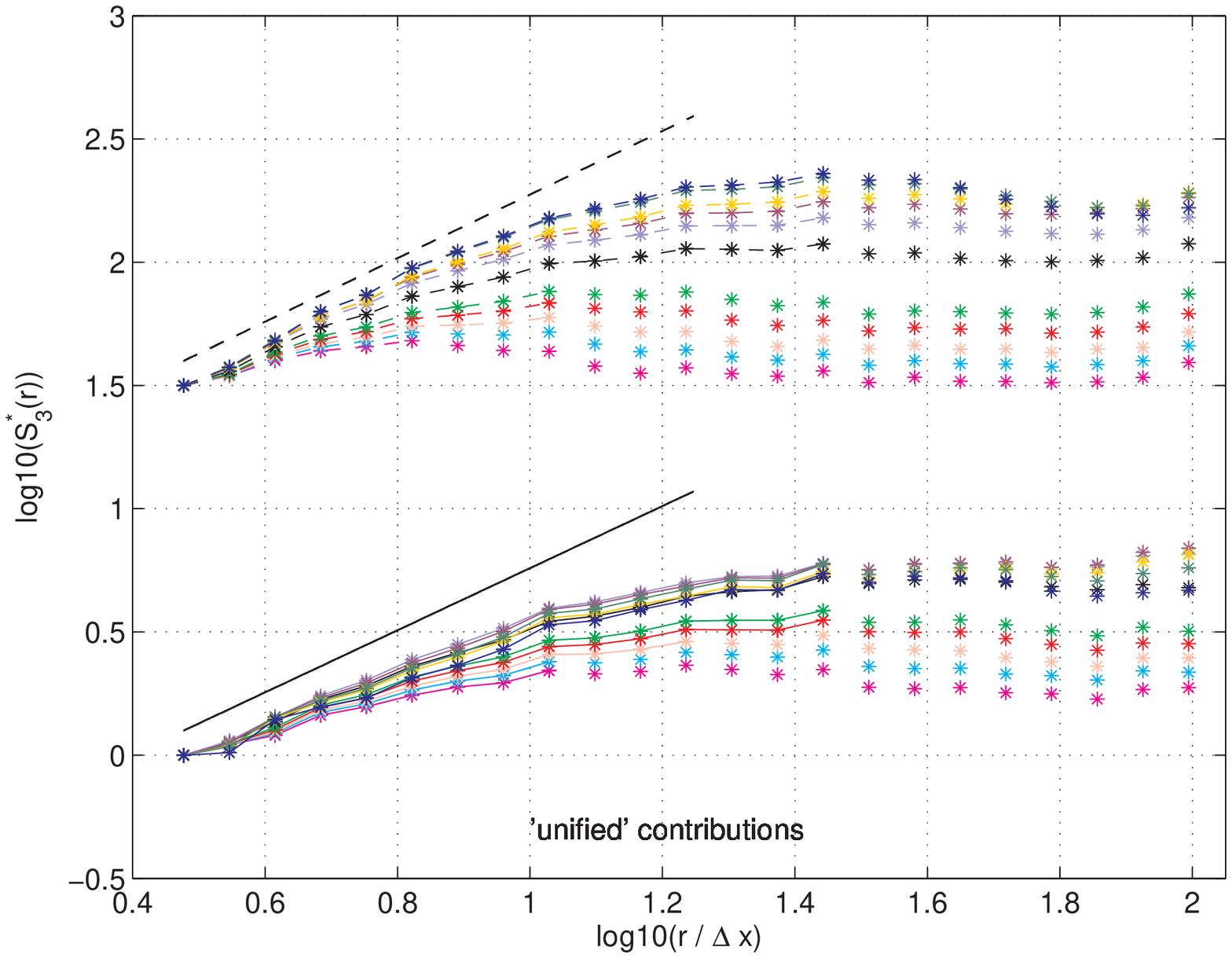}}
\centerline{\includegraphics[width=9cm,height=4.cm]{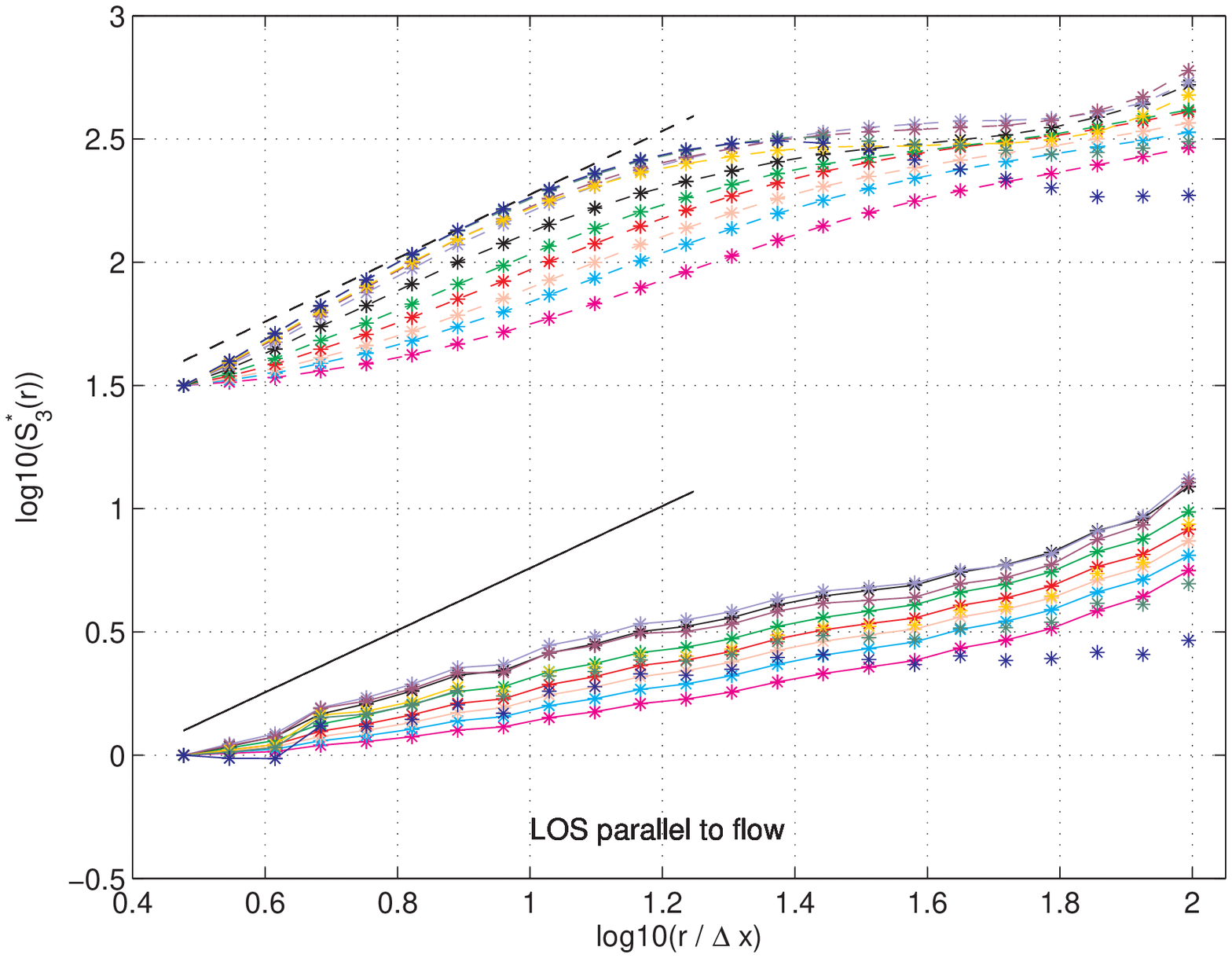}}
\centerline{\includegraphics[width=9cm,height=4.cm]{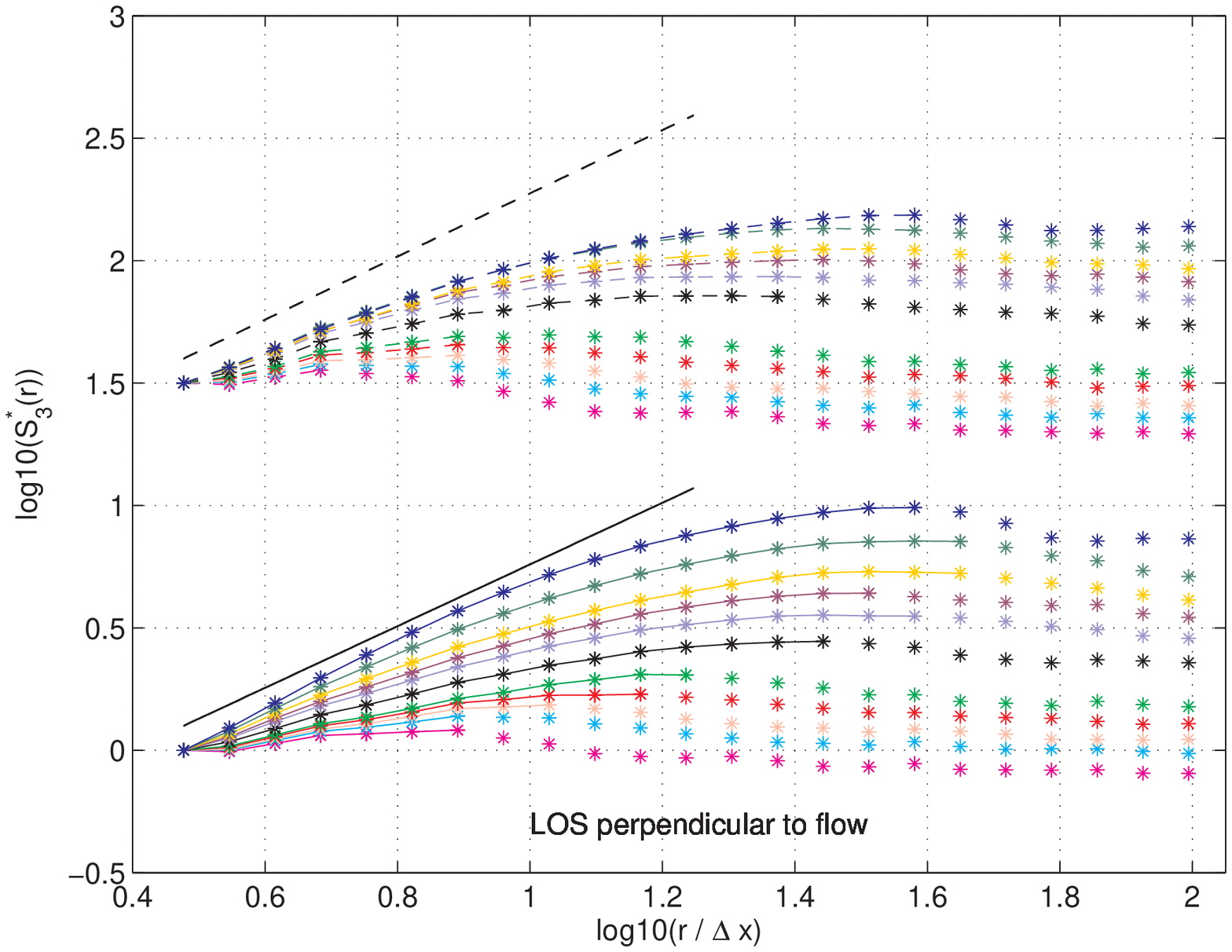}}
\caption{Structure functions $S_{3}^{\parallel}(r)$ (solid) and
  $S_{3}^{\perp}(r)$ (dashed) for the case 'unified' ({\bf top
    panel}), as well as computed along the x-direction ({\bf middle panel})
  and y-direction ({\bf bottom panel}), i.e., parallel and
  perpendicular to the upstream flow, for simulations R*\_2 at $\ell
  = 12$. Color coding as in Fig.~\ref{fig:dens_mach}. Also
  indicated are slopes $\zeta_{3}^{\parallel} = 1.26$ and
  $\zeta_{3}^{\perp} = 1.29$ as obtained by \citet{kritsuk-et-al:07}
  from 3D periodic box simulations. Curves are
  vertically shifted, to start from one common point.}
\label{fig:sf_normXXX}
\end{figure}

The CDL being not strictly periodic and its turbulence asymmetric and
inhomogeneous, structure functions are expected to depend on how the
averages in Eq.~\ref{eq:struc_func_def} are taken. Apart from the
'full in' case presented above, we looked at three more cases:
'unified' and LOSs parallel and perpendicular to the upstream flows
(see Sect.~\ref{sec:sf_num_comp}). For the perpendicular LOS we tested
that results do not depend on whether the y- or z-direction is
analyzed, thus only the y-direction is shown.  Third order structure
functions are shown in Fig.~\ref{fig:sf_normXXX}, ESS scaling
exponents are given in Table~\ref{tab:ESS_exp_aniso}.

The 'unified' case by and large resembles the 'full in' case.  The
structure functions look similar (Figs.~\ref{fig:s3_uni_vs_iso_allsim}
and~\ref{fig:sf_normXXX}, top panel), the quality of ESS fits is
comparable (Fig.~\ref{fig:ess_normXXX}, columns one and two).
Co-dimensions tend to be smaller in the 'unified' case ($0.9 \le C \le
0.4$), indicating a dimension $D>2$ for the most dissipating
structures. This seems plausible as the 'unified' case covers the CDL
more completely than the 'full in' case, especially with regard to the
highly compressible regions close to the confining shocks. As in the
'full in' case, there is a tendency of the co-dimension of longitudinal
(transverse) structure functions to increase (decrease) with
increasing $M_{\mathrm{u}}$.

Structure functions taken along a LOS parallel or perpendicular to the
upstream flow look widely different among themselves
(Fig.~\ref{fig:sf_normXXX}, middle and bottom panel) and with respect
to both the 'unified' and 'full in' case. First, the spread induced by
$M_{\mathrm{u}}$ is larger for the perpendicular LOS than for the
parallel LOS. We speculate that this may be related to our demand that
a ray must not re-enter the CDL (see Sect.~\ref{sec:sf_num_comp}),
which gains relevance as $M_{\mathrm{u}}$ increases, and implies that
regions close to the confining shocks tend to be neglected as $r$
increases.  Second, longitudinal structure functions from a parallel
LOS do not level off but keep increasing over the distance considered.
This seems plausible as with increasing distance the velocity
difference in Eq.~\ref{eq:struc_func_def} approaches the difference
between the two post shock flow velocities at each of the confining
shocks.

Best ESS fits are of reasonable quality for transverse structure
functions and low to intermediate $M_{\mathrm{u}}$
(Fig.~\ref{fig:ess_normXXX}). The concrete values of ESS scaling
exponents given in Table~\ref{tab:ESS_exp_aniso} (middle and right
columns) reveal the crucial role of the orientation of the LOS with
respect to the upstream flow: transverse ESS scaling exponents
indicate co-dimensions $C > 1$ for a LOS parallel to the upstream flow
and $C < 1$ for a perpendicular LOS. For a parallel LOS there is again
a tendency toward smaller co-dimension with increasing
$M_{\mathrm{u}}$.

Best ESS fits for the longitudinal structure functions are typically
of mediocre quality and difficult to interpret. While we give the
corresponding data in the appendix (Table~\ref{tab:ESS_exp_aniso}), we
renounce at their discussion. We only mention two features whose
further analysis may be of interest, but is beyond the scope of the
present paper. First, for the parallel LOS best fits are typically
obtained at intermediate values of
$\mathrm{log10}(S^{\parallel}_{\mathrm{3}})$
(Fig.~\ref{fig:ess_normXXX}, top row, third panel), which translates
to intermediate values of $r$. Second, for perpendicular LOSs there is
a steeper (shallower) linear slope for larger (smaller) values of
$\mathrm{log10}(S^{\parallel}_{\mathrm{3}})$
(Fig.~\ref{fig:ess_normXXX}, top row, fourth panel). As
$M_{\mathrm{u}}$ increases, the shallower slope gradually vanishes and
only the steeper slope prevails.

In summary, our results indicate that the LOS plays a crucial role for
the CDL structure functions and derived quantities, notably the
co-dimension $C$.
\subsection{Numerical resolution}
\label{sec:num_effects}
It is clear from the literature~\citep[e.g.,][]{kritsuk-et-al:07,
  federrath-et-al:10} that the present simulations are at the brink of
the necessary resolution to get physically meaningful results in the
context of isothermal supersonic turbulence. Computational costs limit
concrete numerical resolution studies.

An obvious resolution dependence that can be taken from
Table~\ref{tab:list_of_runs}, and that was already found for the 2D
case~(FW06), is that lower resolution results in higher
$M_{\mathrm{rms}}$ for the same $M_{\mathrm{u}}$ and for the same CDL
size ($\ell \approx 12$). This although there is a tendency toward
smaller driving efficiencies (thus less energy input into the CDL) at
lower resolutions. We can only speculate about potential reasons for
this observed behavior, as to why we find higher resolution to result
in lower $M_{\mathrm{rms}}$ (and not higher $M_{\mathrm{rms}}$, as in
the subsonic case), despite increased energy input.

One potential explanation we can think of, and which we put forward
already in the 2D case: in supersonic turbulence dissipation is
dominated by shocks and with increasing resolution more shocks are
resolved, thus dissipation increases, thus $M_{\mathrm{rms}}$
decreases. This mechanism we would expect to act primarily on the
parallel flow component, as only $M_{\mathrm{rms,\parallel}} >> 1$.

Another possible explanation that comes to mind could be that finer
resolution tends to promote coupling of parallel and transverse
directions. The latter thus may be exploited more efficiently for
energy dissipation, predominantly viscous dissipation as
$M_{\mathrm{rms,\perp}} < 1$. Total energy dissipation (parallel plus
transverse directions) thus may increase with increasing resolution.
Further potential explanations may exist.

The first idea, enhanced dissipation in shocks, is attractive as it
acts predominantly on $M_{\mathrm{rms,\parallel}}$, which dominates
$M_{\mathrm{rms}}$ by far. The second idea, enhanced viscous
dissipation in transverse directions, fits well with our data in that
$M_{\mathrm{rms,\parallel}} / M_{\mathrm{rms,\perp}}$ decreases with
increasing resolution, indicating better coupling of transverse and
parallel directions. Within the frame of our model set up it is not
easy to proof that any of these two ideas are indeed at work,
admittedly leaving us for the time being with an unsatisfying
situation.

It would be interesting to check whether a similar dependence of
$M_{\mathrm{rms}}$ on resolution exists for
driven turbulence in 3D periodic box simulations. This could be done
by monitoring the amount of energy that has to be injected to maintain
a prescribed level of $M_{\mathrm{rms}}$. If the first of the above
potential explanations were true, we would expect that for higher
resolution more energy has to be injected per time to maintain a
prescribed level of turbulence. By contrast, the second of the above
potential explanations may hardly manifest itself in the 3D periodic
box case, as driving there is typically isotropic and thus coupling
among directions is less of an issue than in the present study. The
authors are not aware of any dedicated study in this direction.

Besides $M_{\mathrm{rms}}$, other quantities show some resolution
dependence as well. Clearly increasing with increasing resolution is
$\sigma(\rho)/\rho_{\mathrm{m}}$ (last column of
Table~\ref{tab:list_of_runs}). As $\rho_{\mathrm{m}}$ shows no clear
tendency, this increase is an increase in density variance
with resolution. The increase of peak densities with increasing
resolution is well known~\citep{hennebelle-audit:07,
  kitsionas-et-al:09, federrath-et-al:10} and expected on numerical
grounds.

In summary, we expect our results to essentially hold qualitatively but
to change somewhat quantitatively if we were to go to higher
resolution. In particular, we would expect somewhat lower
$M_{\mathrm{rms}}$ for the same $M_{\mathrm{u}}$ and a larger density
variance.
\section{Discussion}
\label{sec:discussion}
\subsection{3D slabs versus 2D slabs}
\label{sec:3d_vs_2d_slab}
\begin{figure}[tbp]
\centerline{\includegraphics[width=9.5cm,height=5cm]{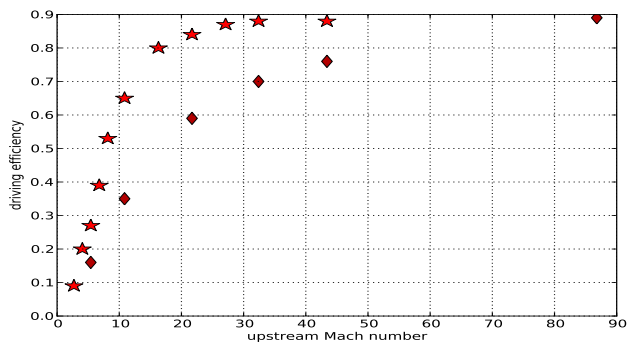}}
\caption{Role of dimensionality for driving efficiency
  $f_{\mathrm{eff}}$ as function of upstream Mach number
  $M_{\mathrm{u}}$. Shown are simulations R*\_2 (red stars) and
  corresponding values for 2D slabs (dark red diamonds, simulations
  R*\_0.2.4 in~FW06).}
\label{fig:feff_2d_3d}
\end{figure}

The relevance of dimensionality for head-on supersonic collision of
isothermal flows was already pointed out in
Sect.~\ref{sec:mean_quantities}: while compression of the collision
zone scales as $\rho_{\mathrm{m}} / \rho_{\mathrm{u}} \propto
M_{\mathrm{u}}^{2}$ in 1D, mean densities of the CDL are largely
independent of $M_{\mathrm{u}}$ in 2D and 3D. Comparing the 3D
numerical results presented here with corresponding 2D
results~(FW06) further shows that for the same upstream conditions the
interaction zone is more turbulent in 3D than in 2D:
$M_{\mathrm{rms}}$ is about 25\% of $M_{\mathrm{u}}$ in 3D, but only
about 20\% in 2D. For the same upstream conditions, driving efficiency
is considerably higher in 3D than in 2D (see
Fig.~\ref{fig:feff_2d_3d}) and the power law dependence of the driving
efficiency on the upstream Mach number is steeper: $\beta_{\mathrm{3}}
= -0.7$ in 2D but $\beta_{\mathrm{3}} = -1.17$ in 3D.

We attribute the larger driving efficiency in 3D (a larger fraction of
the upstream kinetic energy density traverses the confining shocks of
the CDL unthermalized) to the different shock geometry. In 2D, with
the CDL extending to infinity in z-direction, the confining shocks
resemble a corrugated sheet.  In 3D, they look more like a cardboard
egg wrapping. Regions where the confining shocks are perpendicular to
the upstream flow, thus maximum thermalization occurs, are line-like
in the 2D case but point-like in the 3D case (see
Fig.~\ref{fig:eggwrapping}).

Numerical resolution we exclude as an explanation, as 2D simulations
for $M_{\mathrm{u}} = 22$ and three different resolutions yield a
range of $f_{\mathrm{eff}} \in [0.55 , 0.62]$ (FW06), whereas the
corresponding 3D value is $f_{\mathrm{eff}}=0.84$. This for about the
same x-extension of the CDL and about the same coverage (up to a
factor of two) of this extension in terms of grid cells.
\subsection{3D slabs versus 3D periodic boxes}
\label{sec:slab_vs_cube}
The main difference of the simulations presented here to 3D periodic box
simulations lies in the driving of the turbulence - and resulting
consequences. In 3D periodic box models, energy is typically injected in a
controlled, rather homogeneous and isotropic way, although concrete
realizations differ. The 3D slab model may be regarded as the other
extreme: energy input into the turbulent interaction zone only occurs
at the two confining shocks and this in a comparatively uncontrolled
way concerning both the amount of energy and the form of the driving
(compressible / solenoidal and driving wave length).  The kinetic
energy flow entering the interaction zone is predominantly directed
parallel to the upstream flows. Its absolute amount is modulated by
the wiggling of the confining shocks, whose spatial scale in turn
correlates with the spatial extension of the interaction zone.

The different forcing is reflected in the character of the turbulence:
isotropic, homogeneous, and stationary in the case of 3D periodic box models,
anisotropic, inhomogeneous, and with some time evolution for 3D flow
collision zones. We even argue (Sect.~\ref{sec:mach_aniso}) that for
head-on colliding, isothermal flows it is far from trivial, if not
impossible, to reach isotropy while retaining $M_{\mathrm{rms}} >> 1$.
As a corollary, we see some danger in neglecting the concrete large
scale driving of the turbulence and replace it by an isotropic random
driving instead.  For uniform isothermal, head-on colliding
flows as studied here, the anisotropy and inhomogeneity of the
turbulence are a direct consequence of the large scale driving. It
seems plausible that a similar imprint of the large scales on the
small scales is also present in more complex flows, e.g., accretion
onto compact objects~\citep{walder-et-al:10}.  This is not to say that
isothermal isotropic supersonic turbulence cannot exist.  But we
caution that its realization may not be straight forward, at least not
in the context of large scale colliding flows.

Density PDFs deviate from log-normal already in 3D periodic box
simulations of driven isothermal turbulence if the driving has a
substantial compressible component~\citep{schmidt-et-al:09,
  federrath-et-al:10}. It thus seems plausible that our density PDFs
deviate from log-normal. However, the deviations we find are more
pronounced than those seen in compressibly forced 3D periodic box
simulations.  A potential reason could be the inhomogeneity of the
turbulence in the collision zone. The more violent turbulence close to
the confining shocks may be responsible not only for high compressions
but also for pronounced low density voids, as visible in the 2D case,
Fig.~\ref{fig:eggwrapping}, left panel. Density PDFs for individual 2D
slices may shed more light on this issue but were beyond the scope of
the present paper. For molecular clouds, if they indeed result from
colliding flows, the high-density tail we find suggests that current
star formation models based on log-normal density distributions may
severely underestimate high mass star formation, even more than
already postulated in~\citet{schmidt-et-al:09} on the basis of their
simulations.

Regarding ESS scaling exponents, Table~\ref{tab:list_of_ESS_exp} shows
that our values lie well within the range of published values. They
basically range from the theoretical model
by~\citet{she-leveque:94} ($C=2$) to numerical 3D periodic box results
by~\citet{schmidt-et-al:09}. The very low values for $Z_{4}$ and
$Z_{5}$ found in the later study are, however, never reached in our
data. Nevertheless, our data are much closer to these values than 
any other of the listed data.

The range of $Z_{p}$ values we observe results, on the one hand, from
the dependence of $Z_{p}$ on $M_{\mathrm{u}}$. The dependence is
particularly clear for transverse structure functions. Here, $Z_{2}$
increases with $M_{\mathrm{u}}$, in line with the notion of universal
scaling. Longitudinal structure functions show rather inconclusive
results in this respect, possibly contradicting expectations from
universal scaling. Whether isothermal turbulence, in flow collision
zones or 3D periodic box simulations, indeed displays non-universal
scaling properties, possibly due to an additional degree of freedom
related to the large scale forcing as suggested
by~\citet{schmidt-et-al:09}, remains to be seen.

The other reason for the large range of $Z_{p}$ are LOS effects, a
point we take up again in Sect.~\ref{sec:viewingangle}. Here we only
note that parallel and perpendicular LOSs (rows 3 and 4 in
Table~\ref{tab:list_of_ESS_exp}) display nearly complementary ranges
in terms of $Z_{p}$, and that it is mainly the parallel LOS which is
responsible for $Z_{p}$ values close to theoretical values
by~\citet{she-leveque:94}. This kind of LOS effects is likely absent
in 3D periodic box simulations.
\begin{table*}[ht]
  \caption{Comparing ESS scaling exponents $Z_{p}$, $p=1$ to $p=5$,
    following Table~4 in~\citet{schmidt-et-al:09}. Listed are values from this
    work, for longitudinal ($\parallel$) and transverse ($\perp$) structure functions, 
    simulation data by~\citet{boldyrev-et-al2:02} (BNP02), 
    \citet{kritsuk-et-al:07} (KNPW07), \citet{schmidt-et-al:09} (SFHKN09), and theoretical values 
    by~\citet{kolmogorov:41} (K41), \citet{she-leveque:94} (SL94), \citet{boldyrev:02} (B02).}
\begin{center}
\begin{tabular}{lccccc} 
\hline
p                       &  1 & 2 & 3 & 4 & 5 \\
\hline
this work, 'full in' case ($\parallel$) & 0.41 - 0.46 & 0.73 - 0.80 & 1.00 & 1.12 - 1.23 & 1.19 - 1.50 \\
this work, 'full in' case ($\perp$)     & 0.43 - 0.53 & 0.75 - 0.82 & 1.00 & 1.13 - 1.23 & 1.23 - 1.45 \\
this work, perpendicular LOS ($\perp$)  & 0.42 - 0.51 & 0.75 - 0.81 & 1.00 & 1.14 - 1.20 & 1.27 - 1.40 \\
this work, parallel LOS ($\perp$)       & 0.38 - 0.43 & 0.71 - 0.77 & 1.00 & 1.18 - 1.27 & 1.38 - 1.54 \\
BNP02 (MHD 500$^3$)                     &     0.42    &     0.74    & 1.00 &     1.20    &     1.38    \\
KNPW07 (HD 1024$^3$)                    &     0.43    &     0.76    & 1.00 &             &             \\
SFHKN09 (HD 768$^3$)                    &     0.52    &     0.83    & 1.00 &     1.09    &     1.14    \\
K41                                     &     0.33    &     0.67    & 1.00 &     1.33    &     1.67    \\
SL94 ($C=2$)                            &     0.36    &     0.70    & 1.00 &     1.28    &     1.54    \\
B02  ($C=1$)                            &     0.42    &     0.74    & 1.00 &     1.21    &     1.40    \\
\hline
\end{tabular}
\end{center}
\label{tab:list_of_ESS_exp}
\end{table*}
\subsection{Toward real objects?}
\label{sec:disc_obs}
The physical model examined here is clearly too simple for direct
comparison with real world observations.  Nevertheless, we consider it
worthwhile to contemplate on potential real-world implications of two
results, putting them at the same time into perspective by speculating
on effects of some neglected physics. Finally, we reverse the
perspective and ask for which classes of real objects 3D
slab studies may provide useful physical insight.
\subsubsection{Driving the turbulence: high $M_{\mathrm{rms}}$, high $M_{\mathrm{u}}$?}
\label{sec:better_conversion}
The first point we want to discuss more thoroughly is the rather
poor conversion from $M_{\mathrm{u}}$ to $M_{\mathrm{rms}}$, the
latter being only about 10\% to 20\% of the former in the core region
of the CDL. For molecular clouds, where
observations indicate $M_{\mathrm{rsm}} \approx 5 -
50$~\citep{zuckerman-evans:74, mckee-ostriker:07, klessen:11,
  polychroni-et-al:12}, and assuming that molecular clouds result from 
large scale flow collisions, this would imply $M_{\mathrm{u}}
\approx 40 - 500$ within the frame of our model.  

Several questions may be asked. How could such fast flows be produced?
Could inclusion of additional physics within the frame of colliding
flows improve the conversion from $M_{\mathrm{u}}$ to
$M_{\mathrm{rsm}}$? Are driving mechanisms other than colliding flows
compelling? We will mostly dwell on the second question.

The inclusion of radiative cooling may or may not improve conversion
from $M_{\mathrm{u}}$ to $M_{\mathrm{rsm}}$. Simulations where the
cooling limit, and thus the lowest temperature within the CDL, is
equal to the upstream flow temperature show less turbulence (smaller
$M_{\mathrm{rms}}$) than isothermal simulations with the same upstream
Mach number~\citep{walder-folini:00, folini-walder-favre:10}.  If
significant cooling layers form at the confining shocks, the
associated large thermal post-shock pressure will tend to straighten
the confining shocks. A positive feedback results: the more straight
the confining shocks, the more upstream kinetic energy is thermalized
at these shocks, the higher the post-shock temperature, the stronger
the corresponding thermal pressure and the longer the cooling time for
the shock-heated gas. And the less kinetic energy is available for
driving the turbulence.

The situation may be different during the very early collision phase,
when the CDL is still very thin, and especially if the post shock
temperature falls within a temperature range where cooling
instabilities occur, i.e.,  where cooling becomes more and more
efficient as the temperature decreases.  The early (thin) CDL as a
whole becomes very unstable under such
conditions~\citep{walder-folini:00, heitsch-et-al:05,
  folini-walder-favre:10, heitsch-et-al:11}. The situation may also
change if the cooling limit is decidedly below the temperature of the
upstream flow~\citep{pittard-et-al:05}. The CDL then can cool to
temperatures well below the upstream flow temperature, the sound speed
within the CDL decreases and, consequently, the Mach number increases.
Except for this last possibility, it seems rather unlikely that
inclusion of radiative cooling improves the conversion from
$M_{\mathrm{u}}$ to $M_{\mathrm{rsm}}$. Isothermal conditions rather
set an upper limit.

Other physics of potential relevance for $M_{\mathrm{rms}}$ (and
beyond) but not covered by our model include magnetic fields, which
may increase the CDL turbulence by additional energy input
through reconnection or decrease the turbulence as the field guides
the flow direction. More specifically for molecular
clouds, inclusion of self-gravity may augment the turbulence
within the cloud. MHD waves generated by moving accreting cores may
stir the turbulence~\citep{folini-et-al:04}.  Turbulence may also be
driven through jets, winds, and radiation from young or forming stars.

All possibilities just listed to augment $M_{\mathrm{rms}}$ have in
common that they drive the turbulence from within the CDL. For
molecular clouds this may seem in contradiction with observation based
results favoring driving on the scale of the
cloud~\citep{ossenkopf-mac-low:02, heyer-et-al:06, brunt-et-al:09,
  roman-duval-et-al:11}. These results are also somewhat challenging
with regard to our findings. For our CDL, the wiggling of the
confining shocks modulates the incoming flows. It seems plausible that
the spatial scale of this modulation affects, or even sets, the
driving scale of the turbulence. If so, this would imply an energy
injection scale that is smaller than the CDL size, as the spatial
scale of this wiggling is smaller than the spatial extension of the
CDL. The scale of the wiggling increases, however, in proportion to
the CDL size as the CDL grows (see FW06).  Also, our simulations
suggest that spatial modulation of the incoming flow comprises not one
scale but some continuum of scales.

A speculative solution to the above, somewhat conflicting, results and
demands for molecular clouds could be that various driving mechanisms
and associated driving wave lengths co-exist: external large scale and
internal small scale driving. The latter would add to
$M_{\mathrm{rms}}$, thus would help to overcome the poor conversion
from $M_{\mathrm{u}}$ to $M_{\mathrm{rms}}$ and the associated demand
for very high Mach number flows. It also would help to make the
turbulence more isotropic, something we have shown is difficult to
achieve within the frame of isothermal colliding flows.  Large scale
forcing would equally add to $M_{\mathrm{rms}}$ and leave a
characteristic large-scale-driving imprint in the turbulence.
\subsubsection{Viewing angle effects?}
\label{sec:viewingangle}
The second finding that in our opinion deserves discussion is the
strong anisotropy of CDL velocities. It means that line-of-sight
velocities of an observer, and any derived quantities, will depend on
the viewing angle of the CDL. The same CDL can display supersonic or
subsonic line-of-sight velocities, depending on whether the
line-of-sight of the observer is oriented parallel or perpendicular to
the incoming flows. The density Mach number relation suffers from the
same problem. If taken as indicative of the nature of the driving
($b=1/3$ for solenoidal driving and $b=1$ for compressible driving,
from isothermal 3D periodic box simulations,
see~\citet{federrath-et-al:08}), the viewing angle will co-decide
whether one concludes for the same CDL that its driving is rather
compressible or solenoidal.

Equally affected are ESS scaling exponents $Z_{p}$. For transverse
structure functions and derived co-dimensions we have shown that
results depend crucially on the adopted viewing angle: co-dimensions
$C>1$ ($C<1$) are obtained for a line-of-sight parallel
(perpendicular) to the upstream flow. As stressed before, these
results are preliminary in that they do not take into account
radiative transfer, density effects, or projection effects. That such
factors matter is well established in the
literature~\citep{stutzki-et-al:98, brunt-maclow:04, sanchez-et-al:05,
  ossenkopf-et-al:06, esquivel-et-al:07, brunt-et-al:09,
  federrath-et-al:10}. Further analysis of our data in this direction
is envisaged but beyond the scope of the present paper.

Despite this cautionary remark, it is tempting to dwell a moment
longer on the subject and add a concrete illustration of our point
that line-of-sight effects should be considered as one more factor
when interpreting observational data. As an example, consider
simulation R5\_2, whose transverse structure functions are generally
well behaved. In terms of $Z_{p}$ we find $Z_{1} = 0.38$ and
$Z_{5}=1.49$ when looking at the CDL parallel to the upstream flow,
but $Z_{1} = 0.45$ and $Z_{5}=1.37$ when the line-of-sight is
perpendicular to the upstream flow. In the first case, with values
roughly similar to observation based results for the Polaris
Flare~\citep{hily-blant-et-al:08}, we deduce a co-dimension of $C=1.5$
and may conclude that the forcing is predominantly solenoidal and,
possibly, that the turbulence is not too compressible.  If we look at
the same data but from a different viewing angle, now perpendicular to
the upstream flow, we deduce a co-dimension of $C=0.9$ and may
conclude that the turbulence is highly compressible and the forcing
rather compressible than solenoidal.
\subsubsection{3D slabs: a useful concept for real objects?}
\label{sec:useful_concept}
In a number of astrophysical objects, large scale colliding flows play
a key role, i.e., flows with a dominant bulk velocity component on top
of any additional flow structure. In fact, the present study was
motivated by numerical simulations of entire objects by the
authors~\citep[][]{folini-walder-2:00, dumm-et-al:00,
  folini-walder:02, harper-et-al:05, walder-et-al:05,
  walder-folini:08, georgy-et-al:13}.  We advocate that insight into
the physics of such collision zones can be gained from 3D slab
studies. The present study, for example, shows that a richly
structured interaction zone exists even for rather simple physics and
although the colliding flows are essentially reduced to their bulk
properties. Of course, real colliding flows likely harbor more
structure and physics. The relative importance of such additional
complexity may be gradually addressed within the context of 3D slabs.

Taking the perspective of astrophysical objects, 3D slab studies may
be useful for the wind collision zone in massive binaries.  Concrete
questions range from the X-ray and non-thermal emission of such
zones~\citep[e.g.,][]{pittard-et-al-2:10, zhekov:12, de-becker-rauw:13}
to dust formation near periastron passage, presumably in the collision
zone and despite the strong stellar radiation
fields~\citep[][]{tuthill-et-al:99, marchenko-et-al:99,
  cherchneff-et-al:00, folini-walder:02, dougherty-et-al:05,
  williams-et-al:12}. Such studies clearly require more physics than
covered in this paper. Also, massive stellar winds are likely
clumped~\citep[e.g.,][]{owocki-et-al:88, moffat-et-al:88}, which can
affect the turbulence in the collision
zone~\citep[e.g.,][]{walder-folini:02, pittard:07, pittard-et-al:09,
  pittard-et-al:10}. Another large class of objects where 3D slabs
might provide some insight are internal collisions in jets, from young
stars~\citep[e.g.,][]{flower-et-al:03, cunningham-et-al:06,
  cunningham-et-al:11} to high-energy objects, like gamma ray
burst~\citep[][]{mimica-et-al:07, zitouni-et-al:08,
  bosnjak-et-al:09, mimica-et-al:10, granot:12}.  In the later case,
however, relativistic effects are likely relevant. A third context
where 3D slab studies might be useful is star formation on galactic
scales. Turbulence in the wake of flow collision is one suggested
explanation~\citep[e.g.,][]{gabor-bournaud:13} for the observed delay
and potential suppression of star formation for
$z>2$~\citep[e.g.,][]{bouche-et-al:10,weinmann-et-al:12}. The case of
3D slabs and molecular clouds we briefly touched in
Sect.~\ref{sec:better_conversion}. A multi-phase medium is likely
relevant for such studies~\citep[e.g.,][]{hennebelle-perault:99,
  hennebelle-et-al:07, gray-scannapieco:13}. Also, the flows probably
already carry (turbulent) structure, whose effect on the collision
zone remains to be clarified. In fact, we consider it a most
interesting question whether gradual inclusion of such and other
effects allows to recover the 3D periodic box view on molecular cloud
turbulence (see Sect.~\ref{sec:slab_vs_cube}) in the context of 3D
slabs. We are not aware of any such studies.
\section{Summary and conclusions}
\label{sec:conc}
We have performed 3D simulations of head-on colliding isothermal flows
with upstream Mach numbers in the range $2 \le M_{\mathrm{u}} \le 43 $
and presented a first analysis of the turbulence in the resulting
interaction zone, the CDL (cold dense layer). We find that the
characteristics of the CDL turbulence deviate markedly from what is
obtain in corresponding 3D periodic box simulations.

As in 2D, approximate self-similar scaling relations also hold in 3D
for the mean density, root mean square Mach number, driving
efficiency, and energy dissipation in the CDL, albeit with
different numerical constants and a different (steeper) scaling
exponent for the driving efficiency. For the same $M_{\mathrm{u}}$,
the driving efficiency is larger in 3D than in 2D because of
the different shock geometry ('egg wrapping' in 3D, 'corrugated
sheet' in 2D).

Density PDFs are not log-normal but show
fat tails at high densities and a 'two-power-law' composite for low
densities. 

The CDL turbulence is inhomogeneous and anisotropic.  Root mean square
Mach numbers are around 15\% (CDL center) to 25\% (CDL average) of
$M_{\mathrm{u}}$. They are always larger parallel to the upstream flow
than in perpendicular direction, where they never reach supersonic
values. The density variance - Mach number relation inherits this
anisotropy. Isotropization of the flow hardly takes place and we
argue, in fact, that isotropization is hard to achieve while retaining
substantially supersonic root mean square Mach numbers. Even in the
(small scale) central region of the CDL the turbulence remains
anisotropic, even here the imprint of the large scale driving is still
felt.

The viewing angle of the CDL is shown to play a prominent role for
different turbulence characteristics like ESS scaling exponents or
the density variance - Mach number dependence.  ESS scaling exponents
of transverse structure functions imply for the most dissipative
structures a dimension $D<2$ if computed for a line-of-sight parallel
to the upstream flow, but $D>2$ if taken along a line-of-sight
perpendicular to the upstream flow. We suggest to keep this in mind
when interpreting observational data.

Our results show that even our very simple model setup results in a
very richly structured, turbulent interaction zone. The physical
system studied thus opens a perspective on supersonic turbulence that
is rewarding and complementary to 3D periodic box simulations. Under
what conditions both perspectives may be united we consider an
interesting question for the future.  The finite size of the
interaction zone is a challenge with regard to data analysis and
interpretation of results. But this finite size makes the data also
potentially very interesting, as real objects are of finite size as
well, and therefore likely incorporate boundary effects.  Moreover,
bulk flows in real astrophysical objects likely carry some structure
(are inhomogeneous) and require a more elaborate physical description
than the isothermal approach taken here. The assumption of head-on
collision equally needs to be relaxed and the role of numerical
resolution for supersonic turbulence needs investigation.  The
robustness of our findings against these factors remains to be tested.
Nevertheless, we consider the present study a useful contribution for
the physical understanding of complex real objects.
\begin{acknowledgements}
  The authors would like to thank an anonymous referee for his / her
  constructive criticism.  RW and DF acknowledge support from the
  French Stellar Astrophysics Program PNPS.  We acknowledge support
  from the P\^{o}le Scientifique de Mod\'{e}lisation Num\'{e}rique
  (PSMN) and from the Grand Equipement National de Calcul Intensif
  (GENCI), project number x2012046960. The research leading to these
  results has received funding from the European Research Council
  under the European Union's Seventh Framework (FP7/2007-2013)/ERC
  grant agreement no.  320478.
\end{acknowledgements}
%
%
%
%
%
%
%
%
%
\bibliographystyle{apj} 
\bibliography{3d_slabs} 

\begin{thebibliography}{162}
\expandafter\ifx\csname natexlab\endcsname\relax\def\natexlab#1{#1}\fi

\bibitem[{{Agertz} {et~al.}(2009){Agertz}, {Teyssier}, \&
  {Moore}}]{aegertz-et-al:09}
{Agertz}, O., {Teyssier}, R., \& {Moore}, B. 2009, \mnras, 397, L64

\bibitem[{{Aluie}(2011)}]{aluie:11}
{Aluie}, H. 2011, Phys. Rev. Letters, 106, 174502

\bibitem[{{Anninos} \& {Norman}(1996)}]{anninos-norman:96}
{Anninos}, P. \& {Norman}, M.~L. 1996, ApJ, 460, 556

\bibitem[{{Audit} \& {Hennebelle}(2005)}]{audit-hennebelle:05}
{Audit}, E. \& {Hennebelle}, P. 2005, A\&A, 433, 1

\bibitem[{{Audit} \& {Hennebelle}(2010)}]{audit-hennebelle:10}
---. 2010, \aap, 511, A76

\bibitem[{{Ballesteros-Paredes} \& {Mac~Low}(2002)}]{ballesteros-maclow:02}
{Ballesteros-Paredes}, J. \& {Mac~Low}, M. 2002, ApJ, 570, 734

\bibitem[{{Banerjee} \& {Galtier}(2013)}]{banerjee-galtier:13}
{Banerjee}, S. \& {Galtier}, S. 2013, \pre, 87, 013019

\bibitem[{{Benzi} {et~al.}(1993){Benzi}, {Ciliberto}, {Tripiccione}, {Baudet},
  {Massaioli}, \& {Succi}}]{benzi-et-al:93}
{Benzi}, R., {Ciliberto}, S., {Tripiccione}, R., {Baudet}, C., {Massaioli}, F.,
  \& {Succi}, S. 1993, Phys. Rev. E, 48, 29

\bibitem[{{Berger}(1985)}]{berger:85}
{Berger}, M.~J. 1985, Lectures in applied Mathematics, 22, 31

\bibitem[{{Biferale} {et~al.}(2008){Biferale}, {Lanotte}, \&
  {Toschi}}]{biferale-et-al:08}
{Biferale}, L., {Lanotte}, A.~S., \& {Toschi}, F. 2008, Physica D Nonlinear
  Phenomena, 237, 1969

\bibitem[{{Biferale} {et~al.}(2012){Biferale}, {Musacchio}, \&
  {Toschi}}]{biferale-et-al:12}
{Biferale}, L., {Musacchio}, S., \& {Toschi}, F. 2012, Phys. Rev. Letters, 108,
  164501

\bibitem[{{Blondin} \& {Marks}(1996)}]{blondin-marks:96}
{Blondin}, J.~M. \& {Marks}, B.~S. 1996, New {A}stronomy, 1, 235

\bibitem[{{Boldyrev}(2002)}]{boldyrev:02}
{Boldyrev}, S. 2002, ApJ, 569, 841

\bibitem[{{Boldyrev} {et~al.}(2002{\natexlab{a}}){Boldyrev}, {Nordlund}, \&
  {Padoan}}]{boldyrev-et-al2:02}
{Boldyrev}, S., {Nordlund}, {\AA}., \& {Padoan}, P. 2002{\natexlab{a}}, \apj,
  573, 678

\bibitem[{{Boldyrev} {et~al.}(2002{\natexlab{b}}){Boldyrev}, {Nordlund}, \&
  {Padoan}}]{boldyrev-et-al1:02}
---. 2002{\natexlab{b}}, Phys. Rev. Letters, 89, 031102

\bibitem[{Boris {et~al.}(1992)Boris, Grinstein, Oran, \&
  Kolbe}]{boris-et-al:92}
Boris, J., Grinstein, F., Oran, E., \& Kolbe, R. 1992, Fluid. Dynam. Res, 10,
  199

\bibitem[{{Bouch{\'e}} {et~al.}(2010){Bouch{\'e}}, {Dekel}, {Genzel}, {Genel},
  {Cresci}, {F{\"o}rster Schreiber}, {Shapiro}, {Davies}, \&
  {Tacconi}}]{bouche-et-al:10}
{Bouch{\'e}}, N., {Dekel}, A., {Genzel}, R., {Genel}, S., {Cresci}, G.,
  {F{\"o}rster Schreiber}, N.~M., {Shapiro}, K.~L., {Davies}, R.~I., \&
  {Tacconi}, L. 2010, \apj, 718, 1001

\bibitem[{{Bo{\v s}njak} {et~al.}(2009){Bo{\v s}njak}, {Daigne}, \&
  {Dubus}}]{bosnjak-et-al:09}
{Bo{\v s}njak}, {\v Z}., {Daigne}, F., \& {Dubus}, G. 2009, \aap, 498, 677

\bibitem[{{Brunt} {et~al.}(2009){Brunt}, {Heyer}, \& {Mac
  Low}}]{brunt-et-al:09}
{Brunt}, C.~M., {Heyer}, M.~H., \& {Mac Low}, M.-M. 2009, \aap, 504, 883

\bibitem[{{Brunt} \& {Mac Low}(2004)}]{brunt-maclow:04}
{Brunt}, C.~M. \& {Mac Low}, M.-M. 2004, \apj, 604, 196

\bibitem[{{Byrne} {et~al.}(2011){Byrne}, {Xia}, \& {Shats}}]{byrne-et-al:11}
{Byrne}, D., {Xia}, H., \& {Shats}, M. 2011, Phys. Fluids, 23, 095109

\bibitem[{{Cherchneff} {et~al.}(2000){Cherchneff}, {Le Teuff}, {Williams}, \&
  {Tielens}}]{cherchneff-et-al:00}
{Cherchneff}, I., {Le Teuff}, Y.~H., {Williams}, P.~M., \& {Tielens},
  A.~G.~G.~M. 2000, \aap, 357, 572

\bibitem[{{Colella}(1990)}]{colella:90}
{Colella}, P. 1990, Journal of Computional Physics, 87, 171

\bibitem[{{Colella} \& {Woodward}(1984)}]{colella-woodward:84}
{Colella}, P. \& {Woodward}, P.~R. 1984, Journal of Computational Physics, 54,
  174

\bibitem[{{Courant} \& {Friedrichs}(1976)}]{courant-friedrichs}
{Courant}, R. \& {Friedrichs}, K.~O. 1976, Supersonic flow and shock waves (New
  York: Springer-Verlag)

\bibitem[{{Cunningham} {et~al.}(2006){Cunningham}, {Frank}, \&
  {Blackman}}]{cunningham-et-al:06}
{Cunningham}, A.~J., {Frank}, A., \& {Blackman}, E.~G. 2006, \apj, 646, 1059

\bibitem[{{Cunningham} {et~al.}(2011){Cunningham}, {Klein}, {Krumholz}, \&
  {McKee}}]{cunningham-et-al:11}
{Cunningham}, A.~J., {Klein}, R.~I., {Krumholz}, M.~R., \& {McKee}, C.~F. 2011,
  \apj, 740, 107

\bibitem[{{de Avillez} \& {Breitschwerdt}(2007)}]{de-avillez-bretschwerdt:07}
{de Avillez}, M.~A. \& {Breitschwerdt}, D. 2007, \apjl, 665, L35

\bibitem[{{De Becker} \& {Rauw}(2013)}]{de-becker-rauw:13}
{De Becker}, M. \& {Rauw}, F. 2013, \aap, 558, A28

\bibitem[{{Domaradzki}(2010)}]{domaradzki:10}
{Domaradzki}, J.~A. 2010, International Journal of Computational Fluid
  Dynamics, 24, 435

\bibitem[{{Dougherty} {et~al.}(2005){Dougherty}, {Beasley}, {Claussen},
  {Zauderer}, \& {Bolingbroke}}]{dougherty-et-al:05}
{Dougherty}, S.~M., {Beasley}, A.~J., {Claussen}, M.~J., {Zauderer}, B.~A., \&
  {Bolingbroke}, N.~J. 2005, \apj, 623, 447

\bibitem[{{Downes}(2012)}]{downs:12}
{Downes}, T.~P. 2012, \mnras, 425, 2277

\bibitem[{{Dubrulle}(1994)}]{dubrulle:94}
{Dubrulle}, B. 1994, Phys. Rev. Letters, 73, 959

\bibitem[{{Dumm} {et~al.}(2000){Dumm}, {Folini}, {Nussbaumer}, {Schild},
  {Schmutz}, \& {Walder}}]{dumm-et-al:00}
{Dumm}, T., {Folini}, D., {Nussbaumer}, H., {Schild}, H., {Schmutz}, W., \&
  {Walder}, R. 2000, \aap, 354, 1014

\bibitem[{{Elmegreen} \& {Falgarone}(1996)}]{elmegreen-falgarone:96}
{Elmegreen}, B.~G. \& {Falgarone}, E. 1996, \apj, 471, 816

\bibitem[{{Esquivel} {et~al.}(2007){Esquivel}, {Lazarian}, {Horibe}, {Cho},
  {Ossenkopf}, \& {Stutzki}}]{esquivel-et-al:07}
{Esquivel}, A., {Lazarian}, A., {Horibe}, S., {Cho}, J., {Ossenkopf}, V., \&
  {Stutzki}, J. 2007, \mnras, 381, 1733

\bibitem[{{Fan} \& {Wei}(2004)}]{fan-wei:04}
{Fan}, Y.~Z. \& {Wei}, D.~M. 2004, ApJ, 615, L69

\bibitem[{{Federrath} \& {Klessen}(2013)}]{federrath-klessen:13}
{Federrath}, C. \& {Klessen}, R.~S. 2013, \apj, 763, 51

\bibitem[{{Federrath} {et~al.}(2008){Federrath}, {Klessen}, \&
  {Schmidt}}]{federrath-et-al:08}
{Federrath}, C., {Klessen}, R.~S., \& {Schmidt}, W. 2008, \apjl, 688, L79

\bibitem[{{Federrath} {et~al.}(2009){Federrath}, {Klessen}, \&
  {Schmidt}}]{federrath-et-al:09}
---. 2009, ApJ, 692, 364

\bibitem[{{Federrath} {et~al.}(2010){Federrath}, {Roman-Duval}, {Klessen},
  {Schmidt}, \& {Mac Low}}]{federrath-et-al:10}
{Federrath}, C., {Roman-Duval}, J., {Klessen}, R.~S., {Schmidt}, W., \& {Mac
  Low}, M.-M. 2010, A\&A, 512, A81+

\bibitem[{{Flower} {et~al.}(2003){Flower}, {Le Bourlot}, {Pineau des
  For{\^e}ts}, \& {Cabrit}}]{flower-et-al:03}
{Flower}, D.~R., {Le Bourlot}, J., {Pineau des For{\^e}ts}, G., \& {Cabrit}, S.
  2003, \mnras, 341, 70

\bibitem[{{Folini} {et~al.}(2004){Folini}, {Heyvaerts}, \&
  {Walder}}]{folini-et-al:04}
{Folini}, D., {Heyvaerts}, J., \& {Walder}, R. 2004, A\&A, 414, 559

\bibitem[{{Folini} \& {Walder}(2000)}]{folini-walder-2:00}
{Folini}, D. \& {Walder}, R. 2000, \apss, 274, 189

\bibitem[{{Folini} \& {Walder}(2002)}]{folini-walder:02}
{Folini}, D. \& {Walder}, R. 2002, in Astronomical Society of the Pacific
  Conference Series, Vol. 260, Interacting Winds from Massive Stars, ed.
  A.~F.~J. {Moffat} \& N.~{St-Louis}, 605

\bibitem[{{Folini} \& {Walder}(2006)}]{folini-walder:06}
---. 2006, A\&A, 459, 1, fW06

\bibitem[{{Folini} {et~al.}(2010){Folini}, {Walder}, \&
  {Favre}}]{folini-walder-favre:10}
{Folini}, D., {Walder}, R., \& {Favre}, J.~M. 2010, in Astronomical Society of
  the Pacific Conference Series, Vol. 429, Numerical Modeling of Space Plasma
  Flows, Astronum-2009, ed. N.~V. {Pogorelov}, E.~{Audit}, \& G.~P. {Zank}, 9

\bibitem[{{Folini} {et~al.}(2003){Folini}, {Walder}, {Psarros}, \&
  {Desboeufs}}]{folini-et-al:03}
{Folini}, D., {Walder}, R., {Psarros}, M., \& {Desboeufs}, A. 2003, in
  Astronomical Society of the Pacific Conference Series, Vol. 288, Stellar
  Atmosphere Modeling, ed. I.~{Hubeny}, D.~{Mihalas}, \& K.~{Werner}, 433

\bibitem[{{Frisch}(1995)}]{frisch:95}
{Frisch}, U. 1995, {Turbulence} (Cambridge University Press)

\bibitem[{{Gabor} \& {Bournaud}(2013)}]{gabor-bournaud:13}
{Gabor}, J.~M. \& {Bournaud}, F. 2013, \mnras

\bibitem[{{Galtier} \& {Banerjee}(2011)}]{galtier-banerjee:11}
{Galtier}, S. \& {Banerjee}, S. 2011, Phys. Rev. Letters, 107, 134501

\bibitem[{{Garnier} {et~al.}(1999){Garnier}, {Mossi}, {Sagaut}, {Comte}, \&
  {Deville}}]{garnier-et-al:99}
{Garnier}, E., {Mossi}, M., {Sagaut}, P., {Comte}, P., \& {Deville}, M. 1999,
  Journal of Computational Physics, 153, 273

\bibitem[{{Gazol} \& {Kim}(2010)}]{gazol-kim:10}
{Gazol}, A. \& {Kim}, J. 2010, \apj, 723, 482

\bibitem[{{Gazol} \& {Kim}(2013)}]{gazol-kim:13}
---. 2013, \apj, 765, 49

\bibitem[{{Gazol} {et~al.}(2007){Gazol}, {Kim}, {V{\'a}zquez-Semadeni}, \&
  {Luis}}]{gazol-et-al:07}
{Gazol}, A., {Kim}, J., {V{\'a}zquez-Semadeni}, E., \& {Luis}, L. 2007, in
  Astronomical Society of the Pacific Conference Series, Vol. 365, SINS - Small
  Ionized and Neutral Structures in the Diffuse Interstellar Medium, ed.
  M.~{Haverkorn} \& W.~M. {Goss}, 154

\bibitem[{{Georgy} {et~al.}(2013){Georgy}, {Walder}, {Folini}, {Bykov},
  {Marcowith}, \& {Favre}}]{georgy-et-al:13}
{Georgy}, C., {Walder}, R., {Folini}, D., {Bykov}, A., {Marcowith}, A., \&
  {Favre}, J.~M. 2013, \aap, 559, A69

\bibitem[{{Glover} {et~al.}(2010){Glover}, {Federrath}, {Mac Low}, \&
  {Klessen}}]{glover-et-al:10}
{Glover}, S.~C.~O., {Federrath}, C., {Mac Low}, M.-M., \& {Klessen}, R.~S.
  2010, \mnras, 404, 2

\bibitem[{{Gong} \& {Ostriker}(2011)}]{gong-ostriker:11}
{Gong}, H. \& {Ostriker}, E.~C. 2011, \apj, 729, 120

\bibitem[{{Granot}(2012)}]{granot:12}
{Granot}, J. 2012, \mnras, 421, 2467

\bibitem[{{Grauer} {et~al.}(2012){Grauer}, {Homann}, \&
  {Pinton}}]{grauer-et-al:12}
{Grauer}, R., {Homann}, H., \& {Pinton}, J.-F. 2012, New Journal of Physics,
  14, 063016

\bibitem[{{Gray} \& {Scannapieco}(2013)}]{gray-scannapieco:13}
{Gray}, W.~J. \& {Scannapieco}, E. 2013, \apj, 768, 174

\bibitem[{{Gustafsson} {et~al.}(2006){Gustafsson}, {Brandenburg}, {Lemaire}, \&
  {Field}}]{gustafsson-et-al:06}
{Gustafsson}, M., {Brandenburg}, A., {Lemaire}, J.~L., \& {Field}, D. 2006,
  A\&A, 454, 815

\bibitem[{{Hansen} {et~al.}(2011){Hansen}, {McKee}, \&
  {Klein}}]{hansen-et-al:11}
{Hansen}, C.~E., {McKee}, C.~F., \& {Klein}, R.~I. 2011, \apj, 738, 88

\bibitem[{{Harper} {et~al.}(2005){Harper}, {Brown}, {Bennett}, {Baade},
  {Walder}, \& {Hummel}}]{harper-et-al:05}
{Harper}, G.~M., {Brown}, A., {Bennett}, P.~D., {Baade}, R., {Walder}, R., \&
  {Hummel}, C.~A. 2005, \aj, 129, 1018

\bibitem[{{Heitsch} {et~al.}(2005){Heitsch}, {Burkert}, {Hartmann}, {Slyz}, \&
  {Devriendt}}]{heitsch-et-al:05}
{Heitsch}, F., {Burkert}, A., {Hartmann}, L.~W., {Slyz}, A.~D., \& {Devriendt},
  J.~E.~G. 2005, ApJ, 633, L113

\bibitem[{{Heitsch} {et~al.}(2011){Heitsch}, {Naab}, \&
  {Walch}}]{heitsch-et-al:11}
{Heitsch}, F., {Naab}, T., \& {Walch}, S. 2011, \mnras, 415, 271

\bibitem[{{Heitsch} {et~al.}(2006){Heitsch}, {Slyz}, {Devriendt}, {Hartmann},
  \& {Burkert}}]{heitsch-et-al:06}
{Heitsch}, F., {Slyz}, A.~D., {Devriendt}, J.~E.~G., {Hartmann}, L.~W., \&
  {Burkert}, A. 2006, \apj, 648, 1052

\bibitem[{{Heitsch} {et~al.}(2009){Heitsch}, {Stone}, \&
  {Hartmann}}]{heitsch-et-al2:09}
{Heitsch}, F., {Stone}, J.~M., \& {Hartmann}, L.~W. 2009, \apj, 695, 248

\bibitem[{{Hennebelle} \& {Audit}(2007)}]{hennebelle-audit:07}
{Hennebelle}, P. \& {Audit}, E. 2007, \aap, 465, 431

\bibitem[{{Hennebelle} {et~al.}(2007){Hennebelle}, {Audit}, \&
  {Miville-Desch{\^e}nes}}]{hennebelle-et-al:07}
{Hennebelle}, P., {Audit}, E., \& {Miville-Desch{\^e}nes}, M.-A. 2007, \aap,
  465, 445

\bibitem[{{Hennebelle} {et~al.}(2008){Hennebelle}, {Banerjee},
  {V{\'a}zquez-Semadeni}, {Klessen}, \& {Audit}}]{hennebelle-et-al:08}
{Hennebelle}, P., {Banerjee}, R., {V{\'a}zquez-Semadeni}, E., {Klessen}, R.~S.,
  \& {Audit}, E. 2008, A\&A, 486, L43

\bibitem[{{Hennebelle} \& {Chabrier}(2011)}]{hennebelle-chabrier:11}
{Hennebelle}, P. \& {Chabrier}, G. 2011, \apjl, 743, L29

\bibitem[{{Hennebelle} \& {P{\'e}rault}(1999)}]{hennebelle-perault:99}
{Hennebelle}, P. \& {P{\'e}rault}, M. 1999, \aap, 351, 309

\bibitem[{{Heyer} {et~al.}(2006){Heyer}, {Williams}, \&
  {Brunt}}]{heyer-et-al:06}
{Heyer}, M.~H., {Williams}, J.~P., \& {Brunt}, C.~M. 2006, \apj, 643, 956

\bibitem[{{Hily-Blant} {et~al.}(2008){Hily-Blant}, {Falgarone}, \&
  {Pety}}]{hily-blant-et-al:08}
{Hily-Blant}, P., {Falgarone}, E., \& {Pety}, J. 2008, \aap, 481, 367

\bibitem[{{Inoue} \& {Fukui}(2013)}]{inoue-fukui:13}
{Inoue}, T. \& {Fukui}, Y. 2013, ArXiv e-prints

\bibitem[{{Inoue} \& {Inutsuka}(2009)}]{inoue-inutsuka:09}
{Inoue}, T. \& {Inutsuka}, S.-i. 2009, \apj, 704, 161

\bibitem[{{Inoue} \& {Inutsuka}(2012)}]{inoue-inutsuka:12}
---. 2012, \apj, 759, 35

\bibitem[{Jasak \& Weller(1995)}]{jasak-weller:95}
Jasak, H. \& Weller, H. 1995, internal Report, CFD research group, Imperial
  College, London

\bibitem[{{Kaiser} {et~al.}(2000){Kaiser}, {Sunyaev}, \&
  {Spruit}}]{kaiser-et-al:00}
{Kaiser}, C.~R., {Sunyaev}, R., \& {Spruit}, H.~C. 2000, A\&A, 356, 975

\bibitem[{{Kang} {et~al.}(2005){Kang}, {Ryu}, {Cen}, \& {Song}}]{kang-et-al:05}
{Kang}, H., {Ryu}, D., {Cen}, R., \& {Song}, D. 2005, \apj, 620, 21

\bibitem[{{Kitsionas} {et~al.}(2009){Kitsionas}, {Federrath}, {Klessen},
  {Schmidt}, {Price}, {Dursi}, {Gritschneder}, {Walch}, {Piontek}, {Kim},
  {Jappsen}, {Ciecielag}, \& {Mac Low}}]{kitsionas-et-al:09}
{Kitsionas}, S., {Federrath}, C., {Klessen}, R.~S., {Schmidt}, W., {Price},
  D.~J., {Dursi}, L.~J., {Gritschneder}, M., {Walch}, S., {Piontek}, R., {Kim},
  J., {Jappsen}, A.-K., {Ciecielag}, P., \& {Mac Low}, M.-M. 2009, \aap, 508,
  541

\bibitem[{{Klar} \& {M{\"u}cket}(2012)}]{klar-muecket:12}
{Klar}, J.~S. \& {M{\"u}cket}, J.~P. 2012, \mnras, 423, 304

\bibitem[{{Klessen}(2011)}]{klessen:11}
{Klessen}, R.~S. 2011, in EAS Publications Series, Vol.~51, EAS Publications
  Series, ed. C.~{Charbonnel} \& T.~{Montmerle}, 133--167

\bibitem[{{Klessen} \& {Hennebelle}(2010)}]{klessen-hennebelle:10}
{Klessen}, R.~S. \& {Hennebelle}, P. 2010, \aap, 520, A17

\bibitem[{{Kolmogorov}(1941)}]{kolmogorov:41}
{Kolmogorov}, A. 1941, Akademiia Nauk SSSR Doklady, 30, 301, (K41)

\bibitem[{{Konstandin} {et~al.}(2012){Konstandin}, {Girichidis}, {Federrath},
  \& {Klessen}}]{konstandin-et-al:12}
{Konstandin}, L., {Girichidis}, P., {Federrath}, C., \& {Klessen}, R.~S. 2012,
  \apj, 761, 149

\bibitem[{{Kritsuk} {et~al.}(2011{\natexlab{a}}){Kritsuk}, {Nordlund},
  {Collins}, {Padoan}, {Norman}, {Abel}, {Banerjee}, {Federrath}, {Flock},
  {Lee}, {Li}, {M{\"u}ller}, {Teyssier}, {Ustyugov}, {Vogel}, \&
  {Xu}}]{kritsuk-et-al:11}
{Kritsuk}, A.~G., {Nordlund}, {\AA}., {Collins}, D., {Padoan}, P., {Norman},
  M.~L., {Abel}, T., {Banerjee}, R., {Federrath}, C., {Flock}, M., {Lee}, D.,
  {Li}, P.~S., {M{\"u}ller}, W.-C., {Teyssier}, R., {Ustyugov}, S.~D., {Vogel},
  C., \& {Xu}, H. 2011{\natexlab{a}}, ApJ, 737, 13

\bibitem[{{Kritsuk} \& {Norman}(2004)}]{kritsuk-norman:04}
{Kritsuk}, A.~G. \& {Norman}, M.~L. 2004, ApJ, 601, L55

\bibitem[{{Kritsuk} {et~al.}(2006){Kritsuk}, {Norman}, \&
  {Padoan}}]{kritsuk-et-al:06}
{Kritsuk}, A.~G., {Norman}, M.~L., \& {Padoan}, P. 2006, \apjl, 638, L25

\bibitem[{{Kritsuk} {et~al.}(2007{\natexlab{a}}){Kritsuk}, {Norman}, {Padoan},
  \& {Wagner}}]{kritsuk-et-al:07}
{Kritsuk}, A.~G., {Norman}, M.~L., {Padoan}, P., \& {Wagner}, R.
  2007{\natexlab{a}}, ApJ, 665, 416

\bibitem[{{Kritsuk} {et~al.}(2011{\natexlab{b}}){Kritsuk}, {Norman}, \&
  {Wagner}}]{kritsuk-et-al2:11}
{Kritsuk}, A.~G., {Norman}, M.~L., \& {Wagner}, R. 2011{\natexlab{b}}, \apjl,
  727, L20

\bibitem[{{Kritsuk} {et~al.}(2007{\natexlab{b}}){Kritsuk}, {Padoan}, {Wagner},
  \& {Norman}}]{kritsuk-et-al2:07}
{Kritsuk}, A.~G., {Padoan}, P., {Wagner}, R., \& {Norman}, M.~L.
  2007{\natexlab{b}}, in American Institute of Physics Conference Series, Vol.
  932, Turbulence and Nonlinear Processes in Astrophysical Plasmas, ed.
  D.~{Shaikh} \& G.~P. {Zank}, 393--399

\bibitem[{{Krumholz} \& {McKee}(2005)}]{krumholz-mckee:05}
{Krumholz}, M.~R. \& {McKee}, C.~F. 2005, \apj, 630, 250

\bibitem[{{Kurien} \& {Sreenivasan}(2000)}]{kurien-sreenivasan:00}
{Kurien}, S. \& {Sreenivasan}, K.~R. 2000, Phys. Rev. E, 62, 2206

\bibitem[{{Li} \& {Nakamura}(2006)}]{li-nakamura:06}
{Li}, Z.-Y. \& {Nakamura}, F. 2006, \apjl, 640, L187

\bibitem[{{Mac Low}(1999)}]{maclow:99}
{Mac Low}, M.-M. 1999, ApJ, 524, 169

\bibitem[{{Mac~Low} {et~al.}(1998){Mac~Low}, {Klessen}, \&
  {Burkert}}]{maclow-klessen-burkert:98}
{Mac~Low}, M.-M., {Klessen}, R., \& {Burkert}, A. 1998, Phys. Rev. Letters,
  80(3), 2754

\bibitem[{{Marchenko} {et~al.}(1999){Marchenko}, {Moffat}, \&
  {Grosdidier}}]{marchenko-et-al:99}
{Marchenko}, S.~V., {Moffat}, A.~F.~J., \& {Grosdidier}, Y. 1999, \apj, 522,
  433

\bibitem[{{McKee} \& {Ostriker}(2007)}]{mckee-ostriker:07}
{McKee}, C.~F. \& {Ostriker}, E.~C. 2007, \araa, 45, 565

\bibitem[{{Melzani} {et~al.}(2013){Melzani}, {Winisdoerffer}, {Walder},
  {Folini}, {Favre}, {Krastanov}, \& {Messmer}}]{melzani-et-al:13}
{Melzani}, M., {Winisdoerffer}, C., {Walder}, R., {Folini}, D., {Favre}, J.~M.,
  {Krastanov}, S., \& {Messmer}, P. 2013, \aap, 558, A133

\bibitem[{{Mimica} \& {Aloy}(2010)}]{mimica-et-al:10}
{Mimica}, P. \& {Aloy}, M.~A. 2010, \mnras, 401, 525

\bibitem[{{Mimica} {et~al.}(2007){Mimica}, {Aloy}, \&
  {M{\"u}ller}}]{mimica-et-al:07}
{Mimica}, P., {Aloy}, M.~A., \& {M{\"u}ller}, E. 2007, \aap, 466, 93

\bibitem[{{Moffat} {et~al.}(1988){Moffat}, {Drissen}, {Lamontagne}, \&
  {Robert}}]{moffat-et-al:88}
{Moffat}, A.~F.~J., {Drissen}, L., {Lamontagne}, R., \& {Robert}, C. 1988, ApJ,
  334, 1038

\bibitem[{{Molina} {et~al.}(2012){Molina}, {Glover}, {Federrath}, \&
  {Klessen}}]{molina-et-al:12}
{Molina}, F.~Z., {Glover}, S.~C.~O., {Federrath}, C., \& {Klessen}, R.~S. 2012,
  \mnras, 3020

\bibitem[{{Moraghan} {et~al.}(2013){Moraghan}, {Kim}, \&
  {Yoon}}]{moraghan-et-al:13}
{Moraghan}, A., {Kim}, J., \& {Yoon}, S.-J. 2013, \mnras, 432, L80

\bibitem[{{Myasnikov} \& {Zhekov}(1998)}]{myasnikov-zhekov:98}
{Myasnikov}, A.~V. \& {Zhekov}, S.~A. 1998, MNRAS, 300, 686

\bibitem[{{Ntormousi} {et~al.}(2011){Ntormousi}, {Burkert}, {Fierlinger}, \&
  {Heitsch}}]{ntormousi-et-al:11}
{Ntormousi}, E., {Burkert}, A., {Fierlinger}, K., \& {Heitsch}, F. 2011, \apj,
  731, 13

\bibitem[{{Ossenkopf} {et~al.}(2006){Ossenkopf}, {Esquivel}, {Lazarian}, \&
  {Stutzki}}]{ossenkopf-et-al:06}
{Ossenkopf}, V., {Esquivel}, A., {Lazarian}, A., \& {Stutzki}, J. 2006, \aap,
  452, 223

\bibitem[{{Ossenkopf} \& {Mac Low}(2002)}]{ossenkopf-mac-low:02}
{Ossenkopf}, V. \& {Mac Low}, M.-M. 2002, \aap, 390, 307

\bibitem[{{Owocki} {et~al.}(1988){Owocki}, {Castor}, \&
  {Rybicki}}]{owocki-et-al:88}
{Owocki}, S.~P., {Castor}, J.~I., \& {Rybicki}, G.~B. 1988, ApJ, 335, 914

\bibitem[{{Padoan} {et~al.}(2003){Padoan}, {Boldyrev}, {Langer}, \&
  {Nordlund}}]{padoan-et-al:03}
{Padoan}, P., {Boldyrev}, S., {Langer}, W., \& {Nordlund}, {\AA}. 2003, \apj,
  583, 308

\bibitem[{{Padoan} {et~al.}(2004){Padoan}, {Jimenez}, {Nordlund}, \&
  {Boldyrev}}]{padoan-et-al:04}
{Padoan}, P., {Jimenez}, R., {Nordlund}, {\AA}., \& {Boldyrev}, S. 2004,
  Physical Review Letters, 92, 191102

\bibitem[{{Padoan} \& {Nordlund}(1999)}]{padoan-nordlund:99}
{Padoan}, P. \& {Nordlund}, {\AA}. 1999, \apj, 526, 279

\bibitem[{{Padoan} \& {Nordlund}(2002)}]{padoan-nordlund:02}
---. 2002, \apj, 576, 870

\bibitem[{{Padoan} {et~al.}(1997){Padoan}, {Nordlund}, \&
  {Jones}}]{padoan-nordlund-jones:97}
{Padoan}, P., {Nordlund}, A., \& {Jones}, B.~J.~T. 1997, MNRAS, 288, 145

\bibitem[{{Padoan} {et~al.}(2007){Padoan}, {Nordlund}, {Kritsuk}, {Norman}, \&
  {Li}}]{padoan-et-al:07}
{Padoan}, P., {Nordlund}, {\AA}., {Kritsuk}, A.~G., {Norman}, M.~L., \& {Li},
  P.~S. 2007, \apj, 661, 972

\bibitem[{{Pan} {et~al.}(2008){Pan}, {Wheeler}, \& {Scalo}}]{pan-et-al:08}
{Pan}, L., {Wheeler}, J.~C., \& {Scalo}, J. 2008, \apj, 681, 470

\bibitem[{{Panaitescu} {et~al.}(1999){Panaitescu}, {Spada}, \& {M{\' e}sz{\'
  a}ros}}]{panaitescu-et-al:99}
{Panaitescu}, A., {Spada}, M., \& {M{\' e}sz{\' a}ros}, P. 1999, ApJ, 522, L105

\bibitem[{{Parkin} \& {Pittard}(2010)}]{parkin-pittard:10}
{Parkin}, E.~R. \& {Pittard}, J.~M. 2010, \mnras, 406, 2373

\bibitem[{{Passot} \& {V{\'a}zquez-Semadeni}(1998)}]{passot-vazquez:98}
{Passot}, T. \& {V{\'a}zquez-Semadeni}, E. 1998, Physical Review Letters, 58,
  4501

\bibitem[{{Pittard}(2007)}]{pittard:07}
{Pittard}, J.~M. 2007, \apjl, 660, L141

\bibitem[{{Pittard} {et~al.}(2005){Pittard}, {Dobson}, {Durisen}, {Dyson},
  {Hartquist}, \& {O'Brien}}]{pittard-et-al:05}
{Pittard}, J.~M., {Dobson}, M.~S., {Durisen}, R.~H., {Dyson}, J.~E.,
  {Hartquist}, T.~W., \& {O'Brien}, J.~T. 2005, \aap, 438, 11

\bibitem[{{Pittard} {et~al.}(2009){Pittard}, {Falle}, {Hartquist}, \&
  {Dyson}}]{pittard-et-al:09}
{Pittard}, J.~M., {Falle}, S.~A.~E.~G., {Hartquist}, T.~W., \& {Dyson}, J.~E.
  2009, \mnras, 394, 1351

\bibitem[{{Pittard} {et~al.}(2010){Pittard}, {Hartquist}, \&
  {Falle}}]{pittard-et-al:10}
{Pittard}, J.~M., {Hartquist}, T.~W., \& {Falle}, S.~A.~E.~G. 2010, \mnras,
  405, 821

\bibitem[{{Pittard} \& {Parkin}(2010)}]{pittard-et-al-2:10}
{Pittard}, J.~M. \& {Parkin}, E.~R. 2010, \mnras, 403, 1657

\bibitem[{{Polychroni} {et~al.}(2012){Polychroni}, {Moore}, \&
  {Allsopp}}]{polychroni-et-al:12}
{Polychroni}, D., {Moore}, T.~J.~T., \& {Allsopp}, J. 2012, \mnras, 422, 2992

\bibitem[{{Porter} \& {Woodward}(1994)}]{porter-woodward:94}
{Porter}, D.~H. \& {Woodward}, P.~R. 1994, ApJ Supp., 93, 309

\bibitem[{{Price} {et~al.}(2011){Price}, {Federrath}, \&
  {Brunt}}]{price-et-al:11}
{Price}, D.~J., {Federrath}, C., \& {Brunt}, C.~M. 2011, \apjl, 727, L21

\bibitem[{{Quirk}(1994)}]{quirk:94}
{Quirk}, J.~J. 1994, Internat. J. Numer. Methods Fluids, 18, 555

\bibitem[{{Roman-Duval} {et~al.}(2011){Roman-Duval}, {Federrath}, {Brunt},
  {Heyer}, {Jackson}, \& {Klessen}}]{roman-duval-et-al:11}
{Roman-Duval}, J., {Federrath}, C., {Brunt}, C., {Heyer}, M., {Jackson}, J., \&
  {Klessen}, R.~S. 2011, \apj, 740, 120

\bibitem[{{Roman-Duval} {et~al.}(2010){Roman-Duval}, {Jackson}, {Heyer},
  {Rathborne}, \& {Simon}}]{roman-duval-et-al:10}
{Roman-Duval}, J., {Jackson}, J.~M., {Heyer}, M., {Rathborne}, J., \& {Simon},
  R. 2010, \apj, 723, 492

\bibitem[{{S{\'a}nchez} {et~al.}(2005){S{\'a}nchez}, {Alfaro}, \&
  {P{\'e}rez}}]{sanchez-et-al:05}
{S{\'a}nchez}, N., {Alfaro}, E.~J., \& {P{\'e}rez}, E. 2005, \apj, 625, 849

\bibitem[{{Schmidt} {et~al.}(2009){Schmidt}, {Federrath}, {Hupp}, {Kern}, \&
  {Niemeyer}}]{schmidt-et-al:09}
{Schmidt}, W., {Federrath}, C., {Hupp}, M., {Kern}, S., \& {Niemeyer}, J.~C.
  2009, A\&A, 494, 127

\bibitem[{{Seifried} {et~al.}(2011){Seifried}, {Schmidt}, \&
  {Niemeyer}}]{seifried-et-al:11}
{Seifried}, D., {Schmidt}, W., \& {Niemeyer}, J.~C. 2011, \aap, 526, A14

\bibitem[{{She} \& {Leveque}(1994)}]{she-leveque:94}
{She}, Z.-S. \& {Leveque}, E. 1994, Physical Review Letters, 72, 336

\bibitem[{{Stevens} {et~al.}(1992){Stevens}, {Blondin}, \&
  {Pollock}}]{stevens-et-al:92}
{Stevens}, I.~R., {Blondin}, J.~M., \& {Pollock}, A.~M.~T. 1992, ApJ, 386, 265

\bibitem[{{Stone} {et~al.}(1998){Stone}, {Ostriker}, \&
  {Gammie}}]{stone-ostriker-gammie:98}
{Stone}, J.~M., {Ostriker}, E.~C., \& {Gammie}, C.~F. 1998, ApJ, 508, L99

\bibitem[{{Stutzki} {et~al.}(1998){Stutzki}, {Bensch}, {Heithausen},
  {Ossenkopf}, \& {Zielinsky}}]{stutzki-et-al:98}
{Stutzki}, J., {Bensch}, F., {Heithausen}, A., {Ossenkopf}, V., \& {Zielinsky},
  M. 1998, A\&A, 336, 697

\bibitem[{{Tuthill} {et~al.}(1999){Tuthill}, {Monnier}, \&
  {Danchi}}]{tuthill-et-al:99}
{Tuthill}, P.~G., {Monnier}, J.~D., \& {Danchi}, W.~C. 1999, \nat, 398, 487

\bibitem[{{Vazquez-Semadeni}(1994)}]{vazquez-semadeni:94}
{Vazquez-Semadeni}, E. 1994, \apj, 423, 681

\bibitem[{{V{\'a}zquez-Semadeni} {et~al.}(2010){V{\'a}zquez-Semadeni},
  {Col{\'{\i}}n}, {G{\'o}mez}, {Ballesteros-Paredes}, \&
  {Watson}}]{vazquez-semadeni-et-al:10}
{V{\'a}zquez-Semadeni}, E., {Col{\'{\i}}n}, P., {G{\'o}mez}, G.~C.,
  {Ballesteros-Paredes}, J., \& {Watson}, A.~W. 2010, \apj, 715, 1302

\bibitem[{{V{\'a}zquez-Semadeni} {et~al.}(2007){V{\'a}zquez-Semadeni},
  {G{\'o}mez}, {Jappsen}, {Ballesteros-Paredes}, {Gonz{\'a}lez}, \&
  {Klessen}}]{vazquez-semadeni-et-al:07}
{V{\'a}zquez-Semadeni}, E., {G{\'o}mez}, G.~C., {Jappsen}, A.~K.,
  {Ballesteros-Paredes}, J., {Gonz{\'a}lez}, R.~F., \& {Klessen}, R.~S. 2007,
  \apj, 657, 870

\bibitem[{{V{\'a}zquez-Semadeni} {et~al.}(2006){V{\'a}zquez-Semadeni}, {Ryu},
  {Passot}, {Gonz{\'a}lez}, \& {Gazol}}]{vazquez-semadeni-et-al:06}
{V{\'a}zquez-Semadeni}, E., {Ryu}, D., {Passot}, T., {Gonz{\'a}lez}, R.~F., \&
  {Gazol}, A. 2006, ApJ, 643, 245

\bibitem[{{Vishniac}(1994)}]{vishniac:94}
{Vishniac}, E.~T. 1994, ApJ, 428, 186

\bibitem[{{Walder} {et~al.}(2005){Walder}, {Burrows}, {Ott}, {Livne},
  {Lichtenstadt}, \& {Jarrah}}]{walder-et-al:05}
{Walder}, R., {Burrows}, A., {Ott}, C.~D., {Livne}, E., {Lichtenstadt}, I., \&
  {Jarrah}, M. 2005, \apj, 626, 317

\bibitem[{{Walder} \& {Folini}(1996)}]{walder-folini:96}
{Walder}, R. \& {Folini}, D. 1996, A\&A, 315, 265

\bibitem[{{Walder} \& {Folini}(1998)}]{walder-folini:98}
---. 1998, A\&A, 330, L21

\bibitem[{{Walder} \& {Folini}(2000{\natexlab{a}})}]{amaze:00}
{Walder}, R. \& {Folini}, D. 2000{\natexlab{a}}, in Thermal and Ionization
  Aspects of Flows from Hot Stars: Observations and Theory, ed. H.~J. G. L.~M.
  Lamers \& A.~Sapar, ASP Conference Series, 281--285

\bibitem[{{Walder} \& {Folini}(2000{\natexlab{b}})}]{walder-folini:00}
---. 2000{\natexlab{b}}, Ap\&SS, 274, 343

\bibitem[{{Walder} \& {Folini}(2002)}]{walder-folini:02}
{Walder}, R. \& {Folini}, D. 2002, in Astronomical Society of the Pacific
  Conference Series, Vol. 260, Interacting Winds from Massive Stars, ed.
  A.~F.~J. {Moffat} \& N.~{St-Louis}, 595

\bibitem[{{Walder} {et~al.}(2010){Walder}, {Folini}, {Favre}, \&
  {Shore}}]{walder-et-al:10}
{Walder}, R., {Folini}, D., {Favre}, J.~M., \& {Shore}, S.~N. 2010, in
  Astronomical Society of the Pacific Conference Series, Vol. 429, Numerical
  Modeling of Space Plasma Flows, Astronum-2009, ed. N.~V. {Pogorelov},
  E.~{Audit}, \& G.~P. {Zank}, 173

\bibitem[{{Walder} {et~al.}(2008){Walder}, {Folini}, \&
  {Shore}}]{walder-folini:08}
{Walder}, R., {Folini}, D., \& {Shore}, S.~N. 2008, A\&A, 484, L9

\bibitem[{{Wang} {et~al.}(2010){Wang}, {Li}, {Abel}, \&
  {Nakamura}}]{wang-et-al:10}
{Wang}, P., {Li}, Z.-Y., {Abel}, T., \& {Nakamura}, F. 2010, \apj, 709, 27

\bibitem[{{Weinmann} {et~al.}(2012){Weinmann}, {Pasquali}, {Oppenheimer},
  {Finlator}, {Mendel}, {Crain}, \& {Macci{\`o}}}]{weinmann-et-al:12}
{Weinmann}, S.~M., {Pasquali}, A., {Oppenheimer}, B.~D., {Finlator}, K.,
  {Mendel}, J.~T., {Crain}, R.~A., \& {Macci{\`o}}, A.~V. 2012, \mnras, 426,
  2797

\bibitem[{{Williams} {et~al.}(2012){Williams}, {van der Hucht}, {van Wyk},
  {Marang}, {Whitelock}, {Bouchet}, \& {Setia Gunawan}}]{williams-et-al:12}
{Williams}, P.~M., {van der Hucht}, K.~A., {van Wyk}, F., {Marang}, F.,
  {Whitelock}, P.~A., {Bouchet}, P., \& {Setia Gunawan}, D.~Y.~A. 2012, \mnras,
  420, 2526

\bibitem[{{Zhekov}(2012)}]{zhekov:12}
{Zhekov}, S.~A. 2012, \mnras, 422, 1332

\bibitem[{{Zitouni} {et~al.}(2008){Zitouni}, {Daigne}, {Mochkovich}, \&
  {Zerguini}}]{zitouni-et-al:08}
{Zitouni}, H., {Daigne}, F., {Mochkovich}, R., \& {Zerguini}, T.~H. 2008,
  \mnras, 386, 1597

\bibitem[{{Zrake} \& {MacFadyen}(2011)}]{zrake-macfadyen:11}
{Zrake}, J. \& {MacFadyen}, A. 2011, in American Institute of Physics
  Conference Series, Vol. 1358, American Institute of Physics Conference
  Series, ed. J.~E. {McEnery}, J.~L. {Racusin}, \& N.~{Gehrels}, 102--105

\bibitem[{{Zrake} \& {MacFadyen}(2012{\natexlab{a}})}]{zrake-macfadyen-2:12}
{Zrake}, J. \& {MacFadyen}, A. 2012{\natexlab{a}}, ArXiv e-prints

\bibitem[{{Zrake} \& {MacFadyen}(2012{\natexlab{b}})}]{zrake-macfadyen:12}
{Zrake}, J. \& {MacFadyen}, A.~I. 2012{\natexlab{b}}, \apj, 744, 32

\bibitem[{{Zuckerman} \& {Evans}(1974)}]{zuckerman-evans:74}
{Zuckerman}, B. \& {Evans}, II, N.~J. 1974, \apjl, 192, L149

\end{thebibliography}
\begin{appendix}
\section{Derivation of analytical expression for the driving efficiency}
\label{app:feff}
Part of the total (left plus right) upwind kinetic energy flux
density, ${\cal F}_{\mathrm{e_{\mathrm{kin}},u}} =
\rho_{\mathrm{u}}v_{\mathrm{u}}^{3}$, is thermalized at the shocks
confining the CDL.  The remaining part, $\dot{\cal E}_{\mathrm{drv}}$,
drives the turbulence in the CDL. In analogy with the 2D case, we
assume that $\dot{\cal E}_{\mathrm{drv}}$ and ${\cal
  F}_{\mathrm{e_{\mathrm{kin}},u}}$ are related by a function of the
upwind Mach-number alone,
\begin{equation}
\dot{\cal E}_{\mathrm{drv}} = f_{\mathrm{eff}}(M_{\mathrm{u}}){\cal F}_{\mathrm{e_{\mathrm{kin}},u}}.
\label{eq:def_feff}
\end{equation}
We call the function $f_{\mathrm{eff}}$ the driving efficiency.  An
expression for $f_{\mathrm{eff}}$ can be derived by using the jump
conditions for strong, oblique shocks,
\begin{eqnarray}
\rho_{\mathrm{d}}        & = & \rho_{\mathrm{u}} M_{\mathrm{\perp,u}}^{2} 
                           =   \rho_{\mathrm{u}} M_{\mathrm{u}}^{2} \cos^{2}\alpha,\nonumber   \\
v_{\mathrm{\perp,d}}     & = & v_{\mathrm{\perp,u}} M_{\mathrm{\perp,u}}^{-2} 
                           =   \frac{a}{M_{\mathrm{u}} \cos \alpha}, \nonumber  \\
v_{\mathrm{\parallel,d}} & = & v_{\mathrm{\parallel,u}}
                           = a M_{\mathrm{u}} \sin\alpha.
\label{eq:oblique_jump}
\end{eqnarray}
The subscript d denotes downstream quantities, right after shock
passage, the subscripts ${\mathrm{\perp}}$ and ${\mathrm{\parallel}}$
denote flow components perpendicular and parallel to the shock, and
$\alpha$ is the absolute value of the angle between the x-axis and the
normal to the shock.  Using Eq.~\ref{eq:oblique_jump} we obtain
\begin{eqnarray}
\dot{\cal E}_{\mathrm{drv}} & = & \frac{1}{Y Z}\int_{S_{\mathrm{l,r}}} ds 
                                \frac{\rho_{\mathrm{d}} v_{\mathrm{d}}^{2}}{2} 
                                v_{\mathrm{\perp,d}} \nonumber \\
                          & = & \frac{1}{Y Z}\int_{S_{\mathrm{l,r}}} ds 
                                \frac{\rho_{\mathrm{u}} M_{\mathrm{\perp,u}}^{2} 
                                (v_{\mathrm{\parallel,u}}^{2} + v_{\mathrm{\perp,u}}^{2} M_{\mathrm{\perp,u}}^{-4})}{2} 
                                v_{\mathrm{\perp,u}} M_{\mathrm{\perp,u}}^{-2} \nonumber \\
                          & \approx & \frac{1}{Y Z}\int_{S_{\mathrm{l,r}}} ds 
                                \frac{\rho_{\mathrm{u}} v_{\mathrm{\parallel,u}}^{2}}{2} 
                                 v_{\mathrm{\perp,u}} \nonumber \\
                          & = & \frac{\rho_{\mathrm{u}}v_{\mathrm{u}}^{3}}{2 Y Z}\int_{S_{\mathrm{l,r}}} ds 
                                \cos(\alpha) (1 - \cos^{2}(\alpha)) \\
                          & = & \frac{\rho_{\mathrm{u}}v_{\mathrm{u}}^{3}}{2 Y Z}\int_{YZ_{\mathrm{l,r}}} dy dz 
                                (1 - \cos^{2}(\alpha)) ,
\label{eq:edrive_bowshock1}
\end{eqnarray}
where the integrals over $S_{\mathrm{l,r}}$ and $YZ_{\mathrm{l,r}}$
run over the left and right shock and where it was used that $ds
\cos(\alpha) = dy dz$. We further used the shock jump conditions given
in Eq.~\ref{eq:oblique_jump}, that $ v_{\mathrm{d}}^{2} =
v_{\mathrm{\parallel,d}}^{2} + v_{\mathrm{\perp,d}}^{2}$, and that
$v_{\mathrm{\parallel,u}}^{2} = v_{\mathrm{u}}^{2}(1 -
\cos^{2}(\alpha))$.  The term $M_{\mathrm{\perp,u}}^{-4}$ is omitted
as the shocks we observe in our simulations fulfill $\cos \alpha >>
M_{\mathrm{u}}^{-2}$ for the most part.  Analogous to the 2D
case we thus obtain
\begin{equation}
f_{\mathrm{eff}} = \frac{1}{2 Y Z}\int_{Y_{\mathrm{l,r}}} dy dz 
                                (1 - \cos^{2}(\alpha)) \equiv 1 - \cos^{2}(\alpha_{\mathrm{eff}}),
\label{eq:a_feff}
\end{equation}
where we used the midpoint rule. 
\begin{figure}[tbp]
\centerline{\includegraphics[width=8cm,height=8cm]{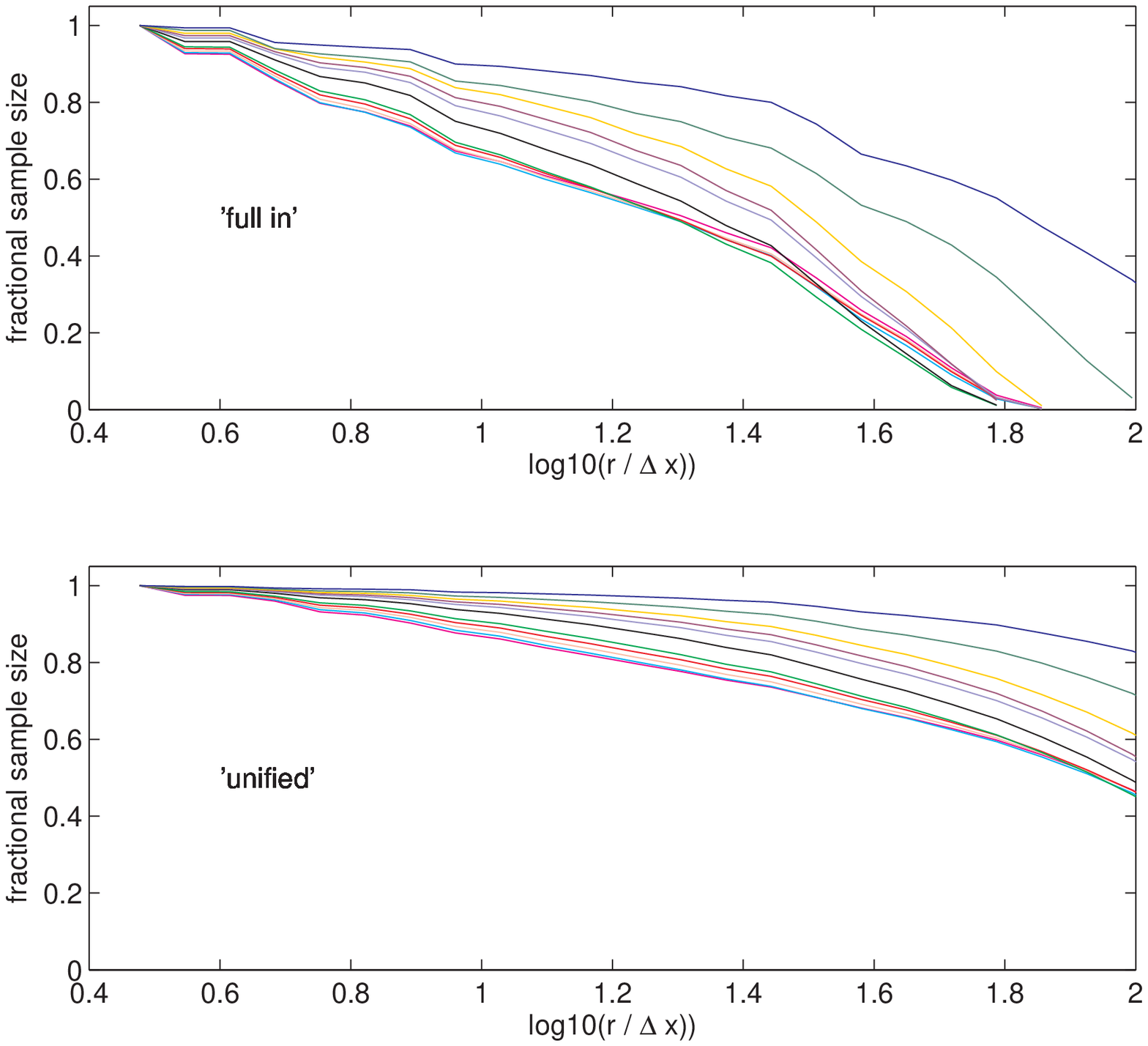}}
\vspace{.5cm}
\centerline{\includegraphics[width=8cm,height=3.5cm]{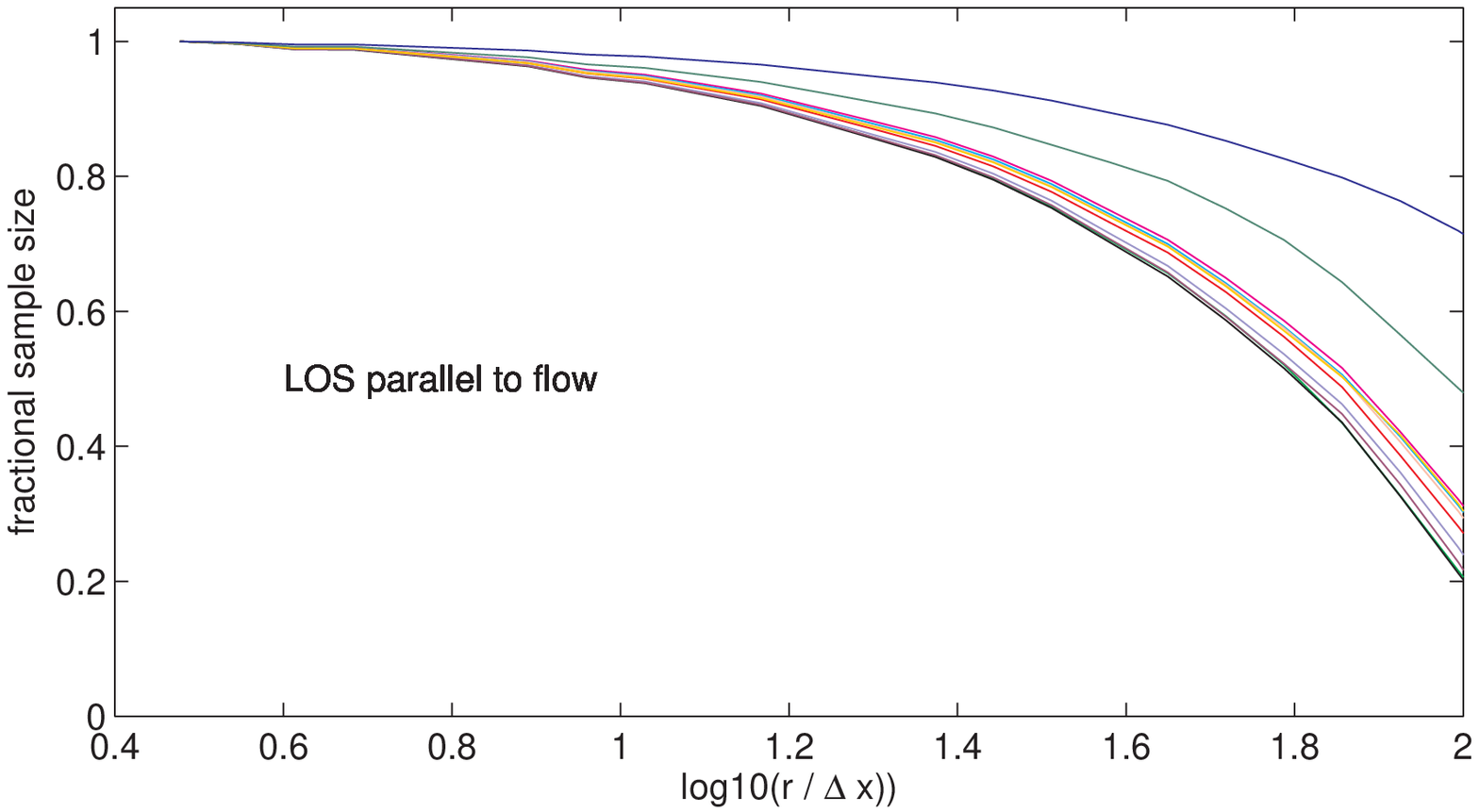}}
\vspace{.5cm}
\centerline{\includegraphics[width=8cm,height=3.5cm]{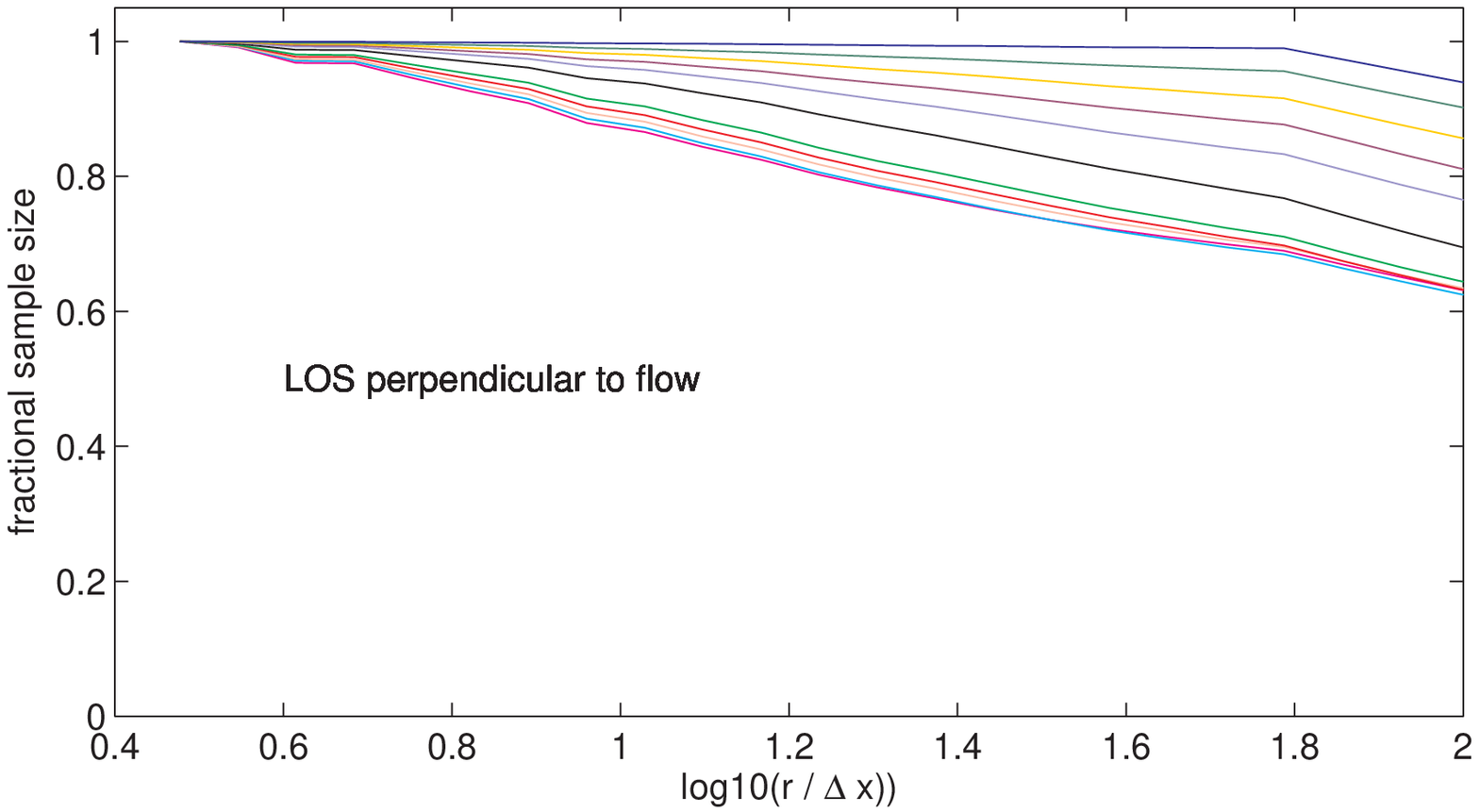}}
\caption{Sample size versus radial distance, normalized to the sample
  size at the smallest radius for cases 'full in' ({\bf first
    panel}), 'unified' ({\bf second panel}), parallel LOS ({\bf third
    panel}), and perpendicular LOS ({\bf fourth panel}) for the
  different runs, all at $\ell \approx 12$.  Color coding as in
  Fig.~\ref{fig:dens_mach}.}
\label{fig:sf_frac_iso_aniso}
\end{figure}
\section{Structure functions from CDL data: computational details and additional results}
\label{app:sf}
\subsection{Computing structure functions from CDL data}
\label{app:sf_num_comp}
To compute $S_{p}(r)$ from our slab data, we consider data of only one
time step, at $\ell \approx 12$. To average over all positions
$\mathbf{x}$, we step through all grid points in the CDL. To average
over all directed distances $\mathbf{r}$, we cast rays at each
position $\mathbf{x}$ into 20 directions, given by the surface normals
of an icosahedron centered at $\mathbf{x}$: $(\pm 1, \pm 1, \pm 1)$,
$(0,\pm 1/\varphi, \pm \varphi)$, $(\pm 1/\varphi, \pm \varphi, 0)$,
and $(\pm \varphi, 0, \pm 1/\varphi)$ with $\varphi = (1 +
\sqrt{5})/2$. We tested the robustness of our results against rotation
of this icosahedron. We step along each ray and compute the
longitudinal and transverse contribution to $S_{p}(r)$. Each ray is
covered by 23 logarithmically spaced steps from 3 $\Delta x$ to 100
$\Delta x$, $\Delta x$ being the grid cell size. Much larger distances
(beyond 128) make no sense because of the periodic domain. Only points
within the CDL may contribute to $S_{p}(r)$.  A ray that once left the
CDL is forbidden to re-enter the CDL at a larger $r$.

As for a given position $\mathbf{x}$ and distance $r$ some rays may
already reach outside the CDL while others still lie within, we
distinguish two cases for each pair ($\mathbf{x}$, $r$): all 20 rays
are still within the CDL ('full in' case) or at least one ray has left
the CDL ('some out' case).  The 'full in' case ascertains that only
symmetric contributions (all 20 directions) enter the mean in
Eq.~\ref{eq:struc_func_def}. The price to pay is a rather rapidly
decreasing and non-uniform sample size: only points $\mathbf{x}$ close
to the center of the CDL contribute to $S_{p}(r)$ for large $r$.
Considering 'full in' and 'some out' contributions in the mean in
Eq.~\ref{eq:struc_func_def} ('unified' case) samples
the CDL as completely as possible, but introduces some asymmetry in
the computation of the mean in Eq.~\ref{eq:struc_func_def}: some
directions are missing for some $\mathbf{x}$ and $r$, especially for
large $r$. The different sample sizes as function of distance $r$ are
illustrated in Fig.~\ref{fig:sf_frac_iso_aniso}. Third order structure
functions are shown in Figs.~\ref{fig:s3_uni_vs_iso_allsim} and~\ref{fig:sf_normXXX}.

While structure functions based on spherical averaging are of
theoretical interest, observation based structure functions are
necessarily based on line-of-sight data. This motivated us to compute
structure functions also along individual one dimensional directions,
parallel or perpendicular to the upstream flow. The direction of
computation may be interpreted as a line-of-sight (LOS), although we
stress that no density variations or radiative transfer effects are
taken into account. Also, our demand that a ray having left the CDL
may not re-enter it, implies for perpendicular LOSs that regions close
to the (wiggled) confining shocks contribute to $S_{p}(r)$ only for
small $r$, whereas large distances are dominated again by
contributions from the central part of the CDL. For a parallel LOS,
the velocity difference in Eq.~\ref{eq:struc_func_def} approaches, as
$r$ increases, the difference between the two post shock flow
velocities at each of the confining shocks.
\subsection{Computation of ESS scaling exponents}
\label{app:sf_ess_comp}
\begin{figure*}[bp]
\centerline{
\includegraphics[width=4.5cm]{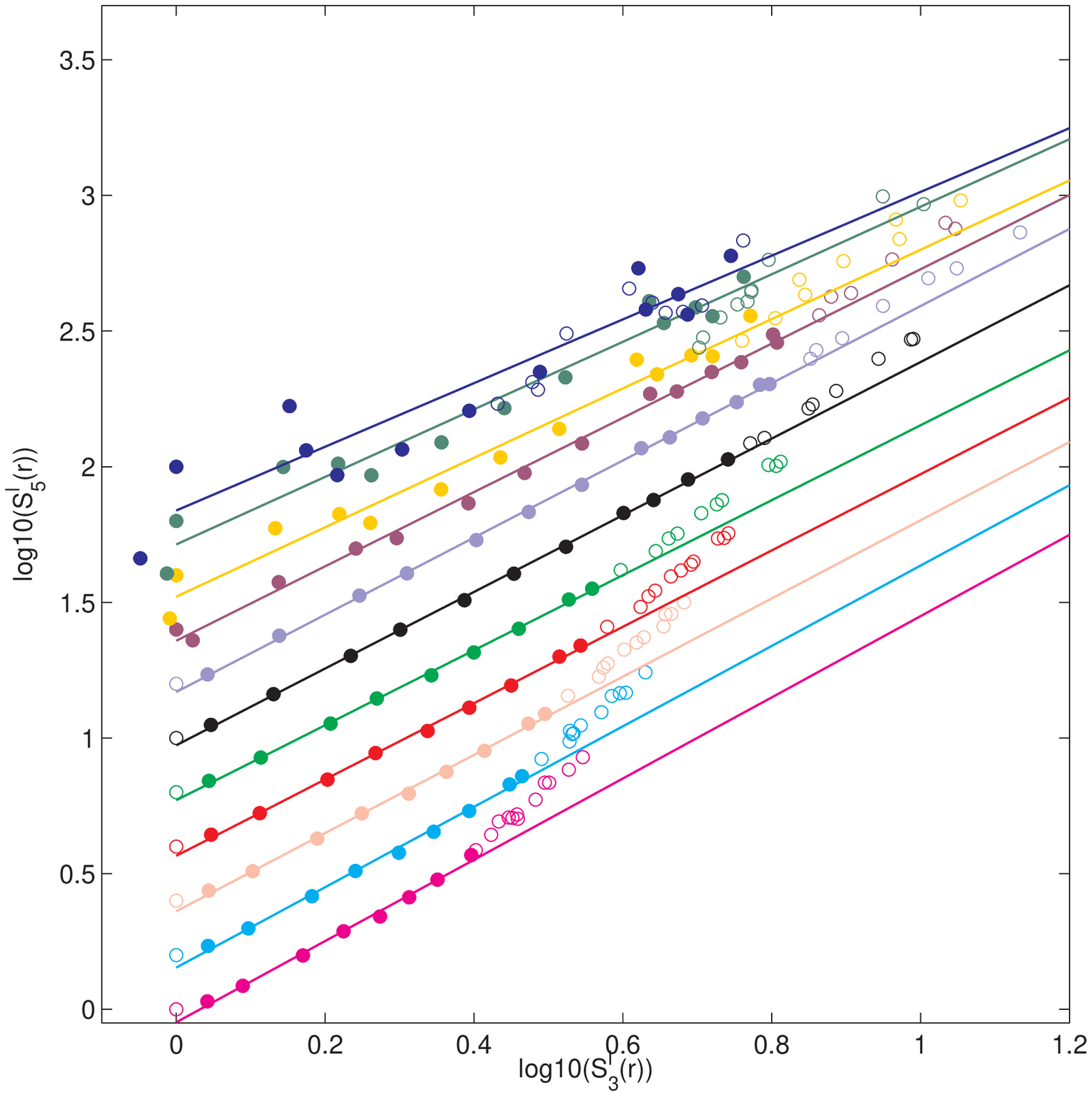}
\includegraphics[width=4.5cm]{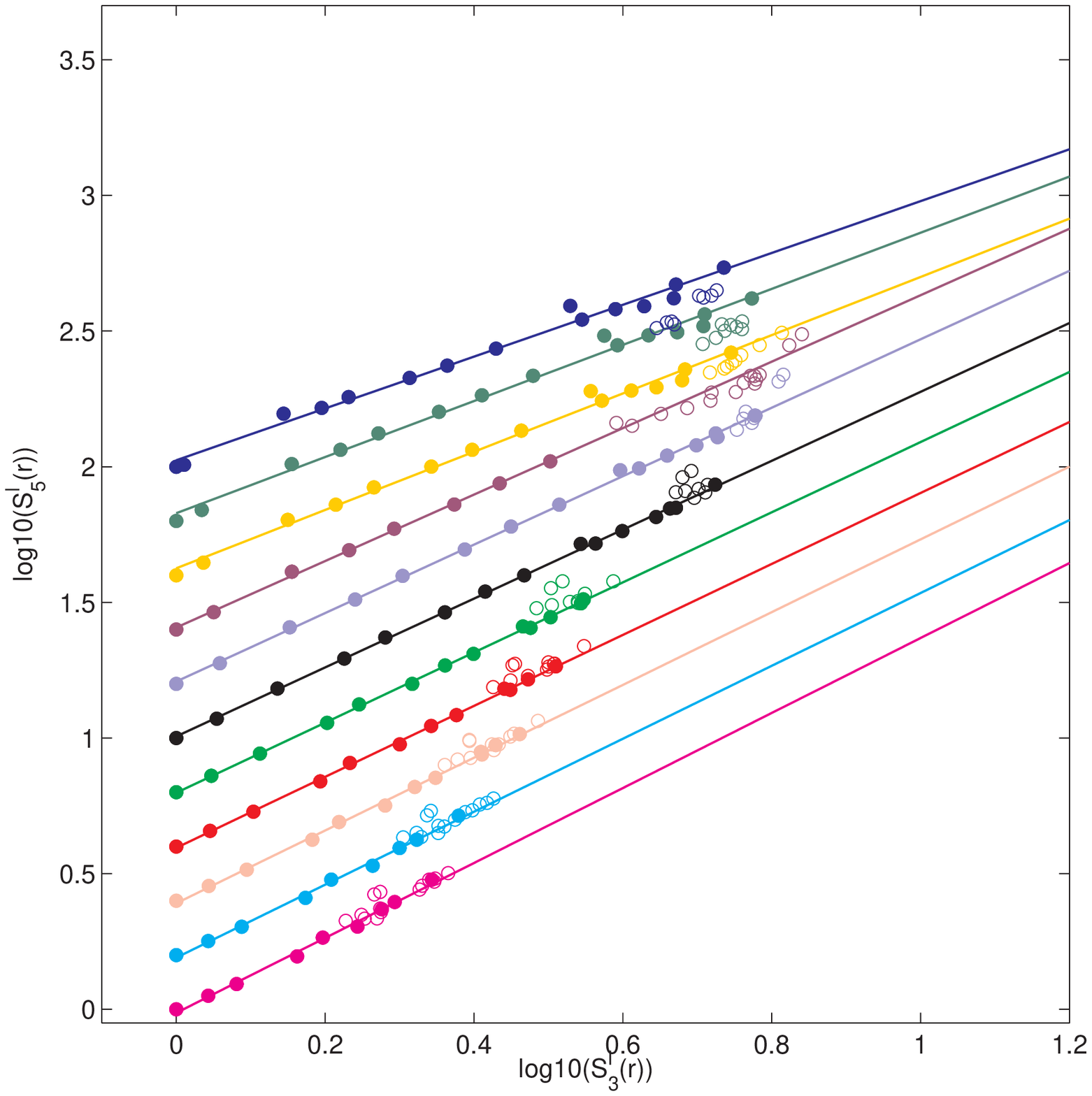}
\includegraphics[width=4.5cm]{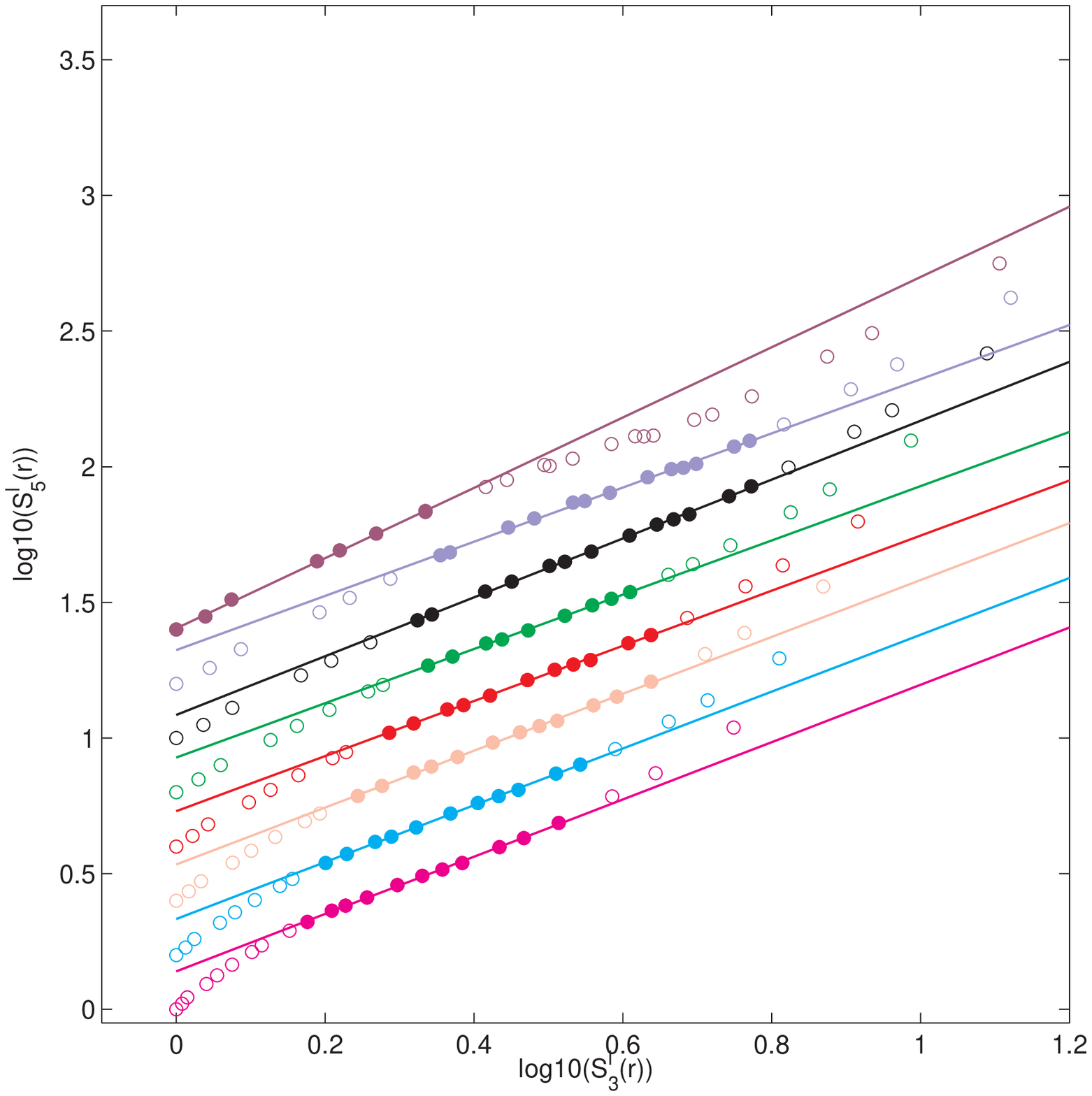}
\includegraphics[width=4.5cm]{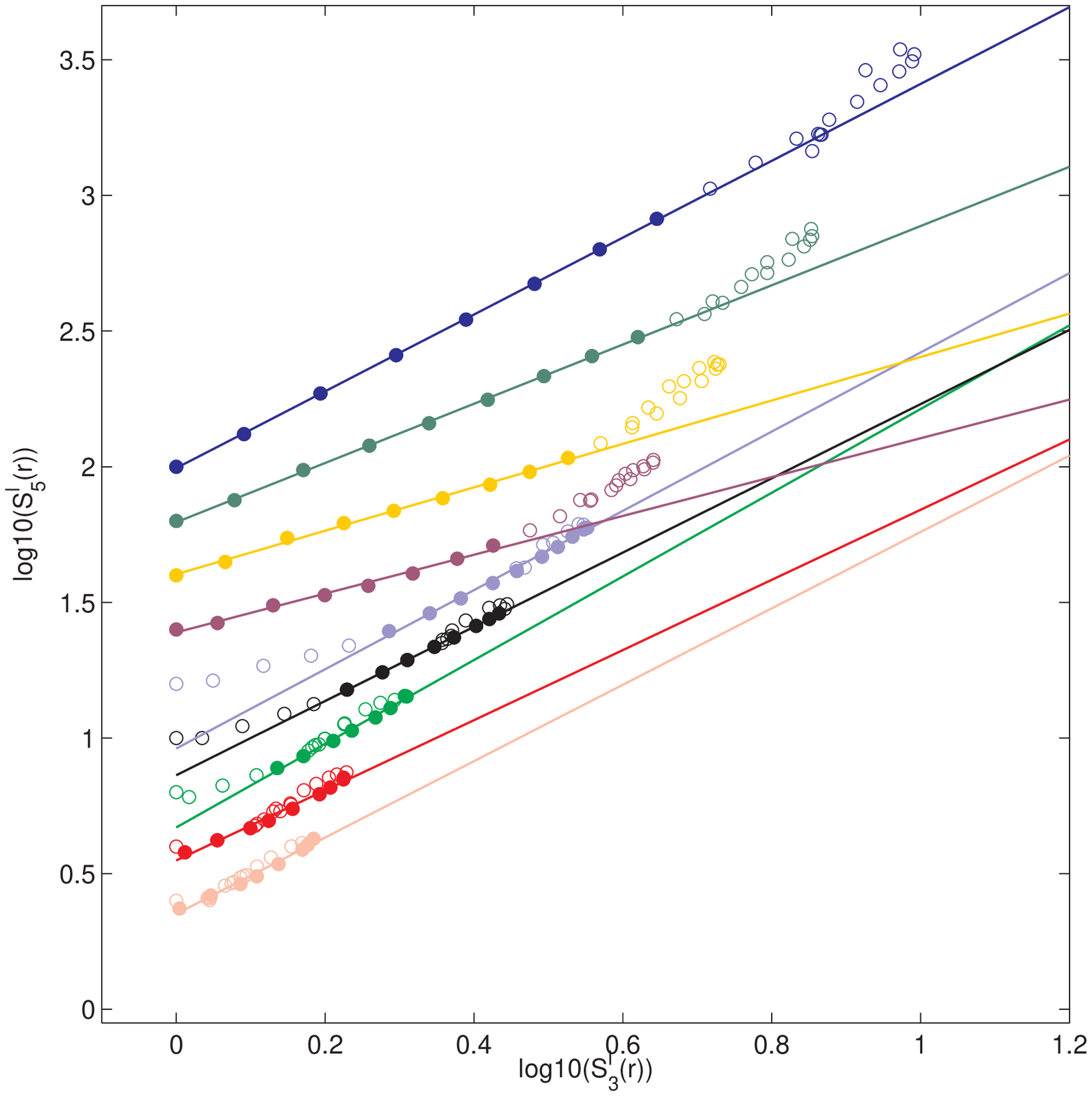}
}
\centerline{
\includegraphics[width=4.5cm]{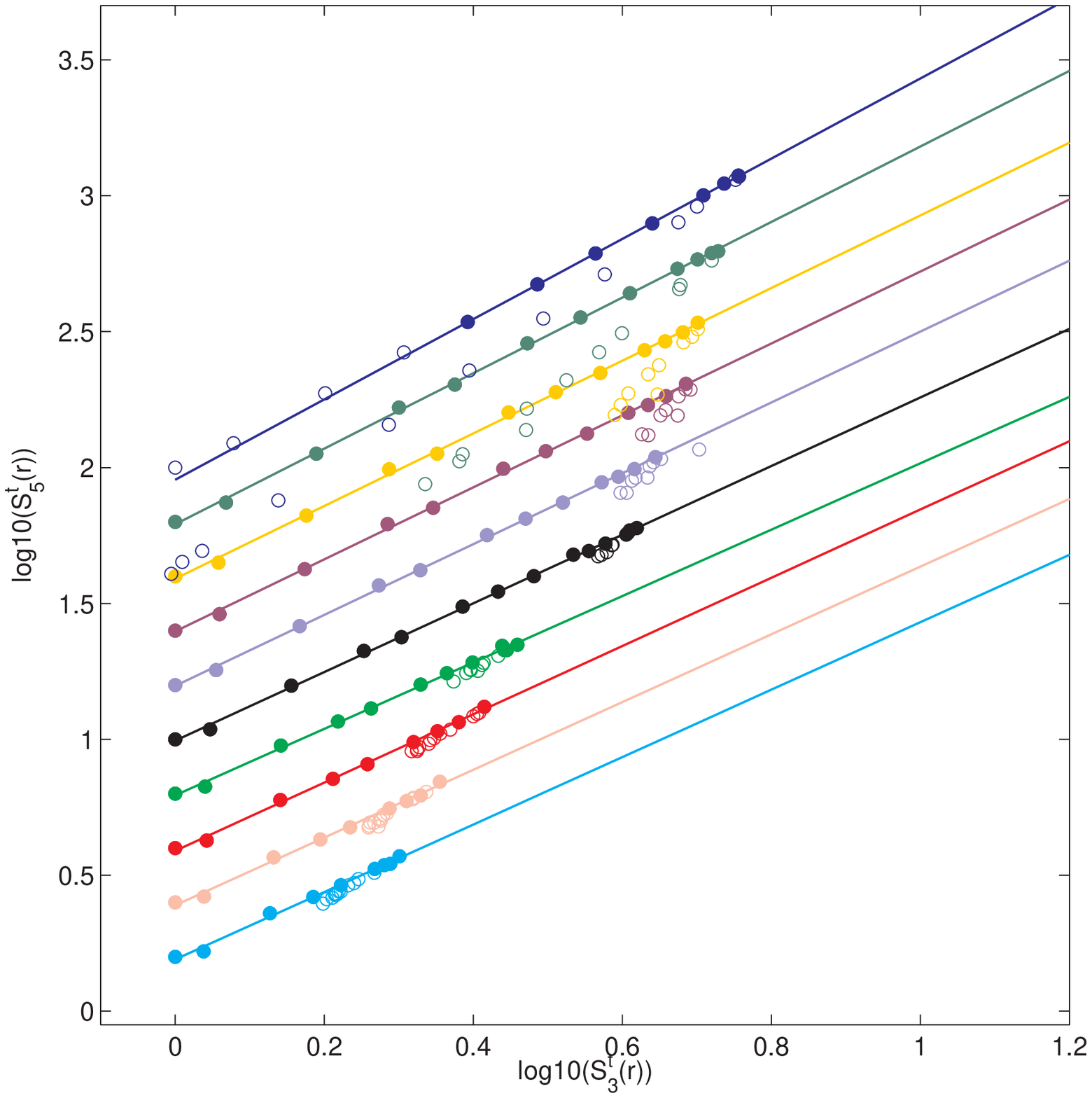}
\includegraphics[width=4.5cm]{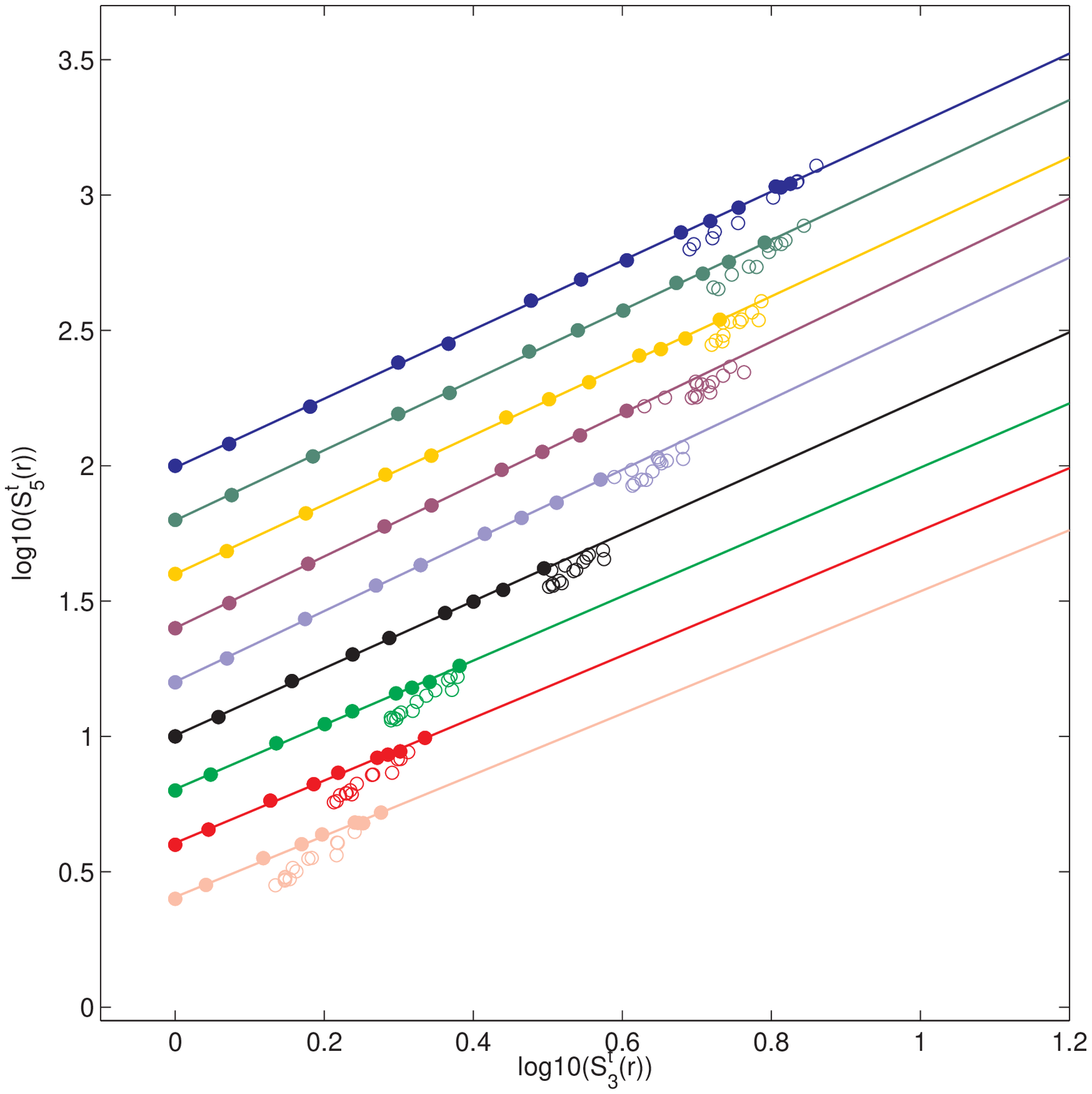}
\includegraphics[width=4.5cm]{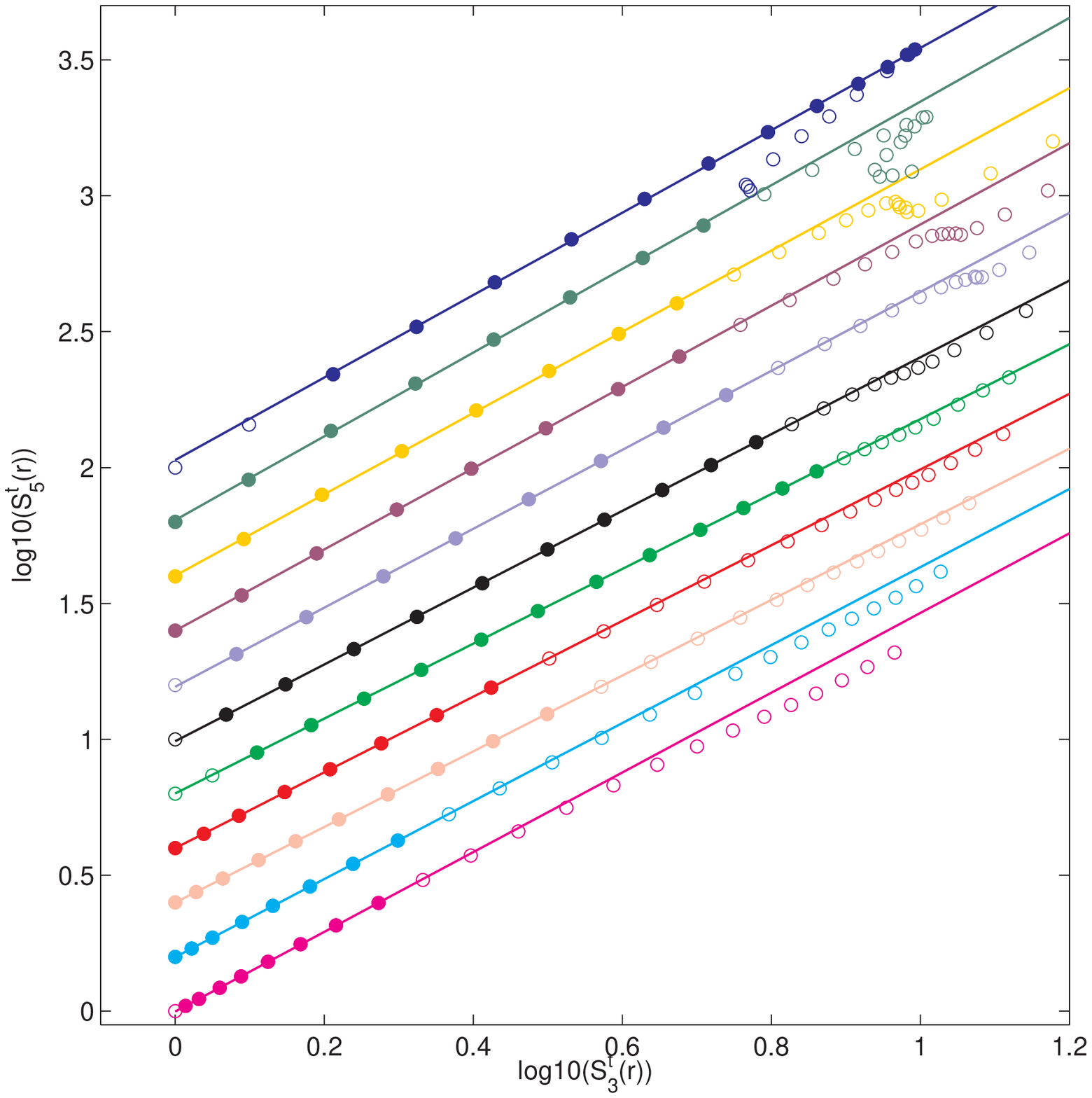}
\includegraphics[width=4.5cm]{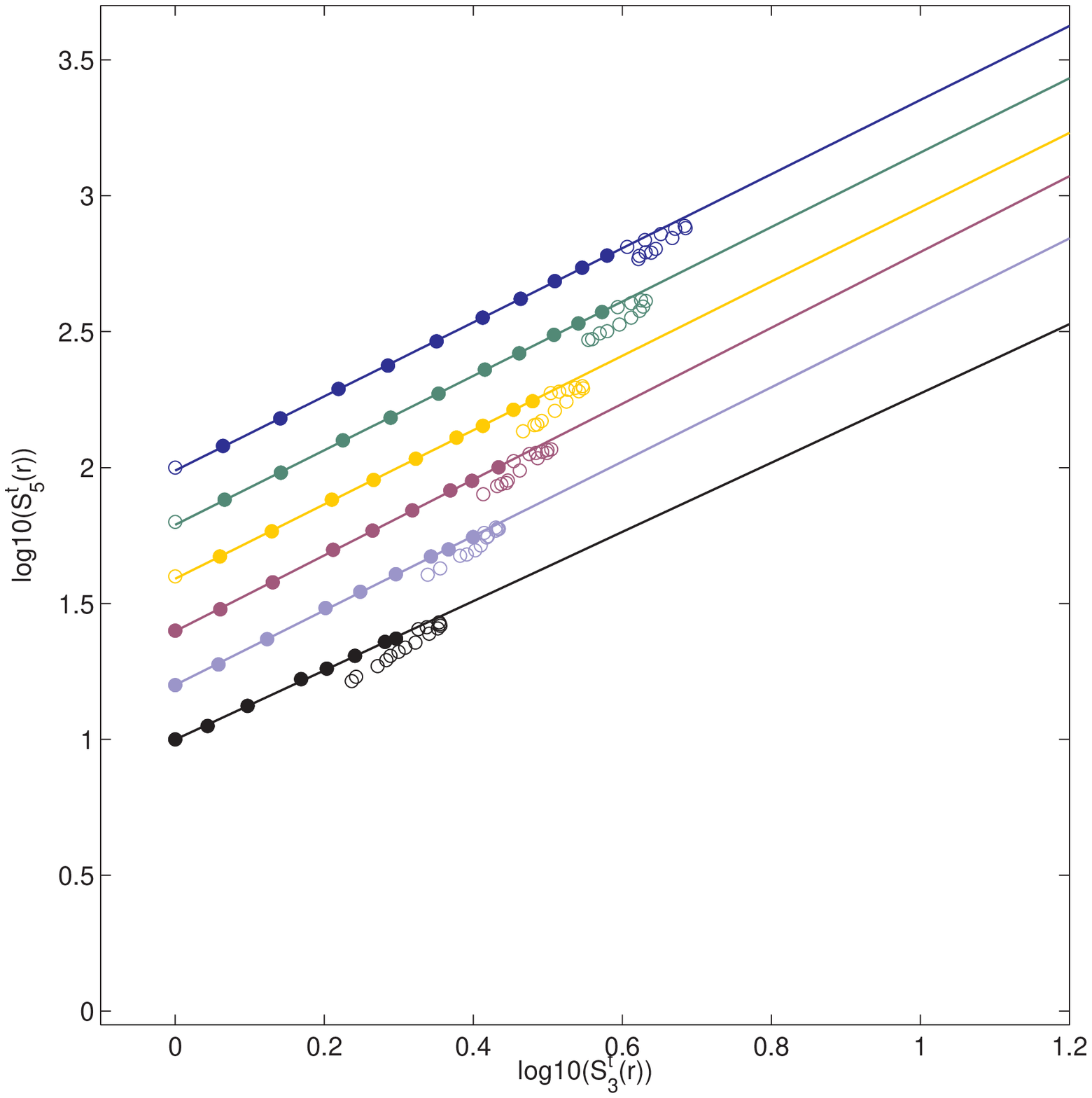}
}
\caption{Fits (solid lines) of ESS scaling exponents $Z_{5}$ as the
  slope of $\mathrm{log10}(S_{5})$ versus $\mathrm{log10}(S_{3})$ for
  longitudinal ({\bf top row}) and transverse ({\bf bottom row})
  structure functions for the cases 'full in' ({\bf first column}),
  'unified' ({\bf second column}), parallel LOS ({\bf third column}), and
  perpendicular LOS ({\bf fourth column}). Filled symbols denote data
  used for the fit, empty symbols denote all data available. Color
  coding as in Fig.~\ref{fig:dens_mach}.}
\label{fig:ess_normXXX}
\end{figure*}

Given the small inertial range of our structure functions, apparent in
Fig.~\ref{fig:s3_uni_vs_iso_allsim}, top panel, we make use of the ESS
hypothesis and compute linear fits to $Z_{p}$, i.e., to
$\mathrm{log10}(S_{p})$ versus $\mathrm{log10}(S_{3})$.  We restrict
the analysis to distances where the third order structure functions
still increase, i.e., we demand that $\mathrm{log10}(S_{3}(r_{i})) >
0.99 \, \mathrm{log10}(S_{3}(r_{i-1}))$ for $i>4$. We further require that
at least eight data points enter the linear fit (corresponding to
roughly half a magnitude in $r$). Within these limits, we search for
the best linear fit (of $Z_{p}$) in terms of relative standard error.
Finally, we reject even the best fit if its relative standard error
exceeds 5\%. While the choice of 5\% is arbitrary, we found the
results to remain essentially unchanged if we repeated the analysis
with a value of 10\%. For the 'full in' case, the quality of the fits
is illustrated in Fig.~\ref{fig:s3_uni_vs_iso_allsim}, bottom panel,
for longitudinal structure functions and $p=2$. In
appendix~\ref{app:sf}, the quality of the fits is illustrated for all
cases considered ('full in', 'unified', different LOSs) for $Z_{5}$
(Fig.~\ref{fig:ess_normXXX}).  From $Z_{p}$, $p=1$ to $5$, and
Eq.~\ref{eq:ess} a best estimate for the co-dimension $C$ is derived
for each run (minimal RMS error).  Note that typically more than the
required 8 points enter each ESS fit and that the best fit is found
for the smallest distances $r$. This situation is typical for nearly
all the situations considered. The only exceptions are ESS for
longitudinal structure functions computed along one dimensional
line-of-sights (see Sect.~\ref{sec:sf_cdl_beyond}).
\subsection{ESS scaling exponents, beyond the 'full in' case}
\label{app:beyon}
\begin{table*}[bp]
  \caption{Longitudinal (top) and transverse (bottom) ESS exponents 
    $Z_{p}$ and best estimates for co-dimension $C$ for the case 'unified' (left columns) and
    line-of-sights parallel (middle columns) and perpendicular (right columns) to the upstream flow, for 
    the different runs at $\ell \approx 12$. Dashes indicate 
    that no fit of sufficient quality could be obtained.}
\begin{center}
\begin{tabular}{lccccccccccccccccc} 
\hline
\rule[-2mm]{0pt}{6mm}
       & \multicolumn{5}{c}{'unified'} & & \multicolumn{5}{c}{'parallel to upstream flow'} & & \multicolumn{5}{c}{'perpendicular to upstream flow'} \\ 
\rule[-2mm]{0pt}{6mm}
       &p=1 &p=2 &p=4 &p=5 & $C$ & \hspace{-.4cm} &p=1 &p=2 &p=4 &p=5 & $C$ & \hspace{-.4cm} &p=1 &p=2 &p=4 &p=5 & $C$ \\
\hline
R2\_2  & 0.59 & 0.93 & 0.97 & 0.95 & 0.6   & &   -  &   -  &   -  &   -  &   -   & & 0.39 & 0.72 & 1.23 & 1.42 & 1.1 \\
R4\_2  & 0.56 & 0.89 & 1.02 & 1.03 & 0.6   & &   -  &   -  &   -  &   -  &   -   & & 0.42 & 0.77 & 1.10 & 1.09 & 0.5 \\
R5\_2  & 0.55 & 0.88 & 1.04 & 1.07 & 0.6   & &   -  &   -  &   -  &   -  &   -   & & 0.46 & 0.82 & 0.97 & 0.80 & 0.3 \\
R7\_2  & 0.51 & 0.83 & 1.11 & 1.22 & 0.5   & & 0.68 & 0.94 & 0.98 & 1.29 & 0.6   & & 0.47 & 0.74 & 0.94 & 0.72 & 0.5 \\
R8\_2  & 0.49 & 0.80 & 1.14 & 1.26 & 0.8   & & 0.62 & 0.90 & 1.02 & 1.00 & 0.3   & & 0.39 & 0.72 & 1.22 & 1.46 & 1.2 \\ 
R11\_2 & 0.50 & 0.81 & 1.15 & 1.27 & 0.8   & & 0.60 & 0.88 & 1.05 & 1.08 & 0.6   & & 0.40 & 0.73 & 1.21 & 1.37 & 1.0 \\ 
R16\_2 & 0.51 & 0.81 & 1.16 & 1.29 & 0.8   & & 0.61 & 0.89 & 1.04 & 1.00 & 0.3   & & 0.43 & 0.76 & 1.23 & 1.54 & 1.7 \\ 
R22\_2 & 0.52 & 0.81 & 1.16 & 1.31 & 0.8   & & 0.61 & 0.89 & 1.04 & 1.02 & 0.3   & & 0.40 & 0.73 & 1.18 & 1.29 & 0.9 \\ 
R27\_2 & 0.51 & 0.81 & 1.18 & 1.34 & 0.8   & & 0.60 & 0.88 & 1.04 & 1.05 & 0.6   & & 0.37 & 0.71 & 1.23 &   -  &   - \\ 
R33\_2 & 0.50 & 0.80 & 1.16 & 1.34 & 0.8   & & 0.58 & 0.87 & 1.05 & 1.05 & 0.6   & &   -  &   -  &   -  &   -  &   - \\ 
R43\_2 & 0.48 & 0.78 & 1.18 & 1.38 & 0.9   & & 0.56 & 0.86 & 1.05 & 1.06 & 0.6   & &   -  &   -  &   -  &   -  &   - \\ 
\hline
R2\_2  & 0.47 & 0.78 & 1.15 & 1.27 & 0.8   & & 0.38 & 0.74 & 1.26 & 1.52 & 1.6   & & 0.42 & 0.75 & 1.20 & 1.36 & 0.9 \\
R4\_2  & 0.48 & 0.79 & 1.16 & 1.29 & 0.8   & & 0.38 & 0.71 & 1.27 & 1.54 & 1.9   & & 0.44 & 0.76 & 1.20 & 1.37 & 0.9 \\
R5\_2  & 0.49 & 0.80 & 1.15 & 1.28 & 0.8   & & 0.38 & 0.71 & 1.26 & 1.49 & 1.5   & & 0.45 & 0.77 & 1.19 & 1.37 & 0.9 \\ 
R7\_2  & 0.49 & 0.80 & 1.17 & 1.32 & 0.8   & & 0.38 & 0.71 & 1.26 & 1.49 & 1.5   & & 0.45 & 0.77 & 1.20 & 1.40 & 0.9 \\
R8\_2  & 0.50 & 0.80 & 1.16 & 1.31 & 0.8   & & 0.39 & 0.72 & 1.24 & 1.45 & 1.2   & & 0.47 & 0.77 & 1.19 & 1.37 & 0.9 \\ 
R11\_2 & 0.53 & 0.82 & 1.13 & 1.24 & 0.5   & & 0.42 & 0.74 & 1.22 & 1.41 & 1.0   & & 0.51 & 0.81 & 1.14 & 1.27 & 0.7 \\ 
R16\_2 & 0.57 & 0.85 & 1.10 & 1.19 & 0.4   & & 0.42 & 0.75 & 1.20 & 1.38 & 1.0   & &   -  &   -  &   -  &   -  &   -  \\
R22\_2 & 0.58 & 0.86 & 1.09 & 1.15 & 0.6   & & 0.43 & 0.75 & 1.21 & 1.39 & 1.0   & &   -  &   -  &   -  &   -  &   -  \\ 
R27\_2 & 0.59 & 0.87 & 1.08 & 1.13 & 0.6   & & 0.42 & 0.75 & 1.21 & 1.39 & 1.0   & &   -  &   -  &   -  &   -  &   -  \\ 
R33\_2 &   -  &   -  &   -  &   -  &   -   & & 0.43 & 0.75 & 1.18 & 1.43 & 1.0   & &   -  &   -  &   -  &   -  &   -  \\ 
R43\_2 &   -  &   -  &   -  &   -  &   -   & & 0.44 & 0.77 & 1.26 & 1.47 & 1.2   & &   -  &   -  &   -  &   -  &   -  \\ 
\end{tabular}
\end{center}
\label{tab:ESS_exp_aniso}
\end{table*}
\end{appendix}

\end{document}